\pdfoutput=1
\documentclass[
twocolumn,
prd, 
aps, 
nofootinbib, 
showpacs, 
superscriptaddress,
floatfix]{revtex4-2}

\usepackage{graphicx}
\usepackage[usenames,dvipsnames]{xcolor}
\usepackage[utf8]{inputenc}
\usepackage[T1]{fontenc}
\usepackage{anyfontsize}
\usepackage{adjustbox}
\usepackage{placeins}
\usepackage{lipsum}
\usepackage{latexsym}
\usepackage{rotating}
\usepackage{mathtools}
\usepackage{amsmath, amssymb, amsthm, amsfonts}
\usepackage{mathrsfs}
\usepackage{bm}
\usepackage{enumerate}
\usepackage{soul}
\usepackage{natbib}
\usepackage{hyperref}
\usepackage{verbatim}
\usepackage{tikz}
\usetikzlibrary{shapes, arrows, positioning}
\tikzstyle{process} = [ rectangle, draw, text centered, minimum height=2em]
\tikzstyle{connector} = [draw, -latex']

\hypersetup{colorlinks=true,citecolor=NavyBlue,linkcolor=NavyBlue,urlcolor=NavyBlue}

\newcommand{\1}{\hspace{1mm}}
\newcommand{\2}{\hspace{2mm}}
\newcommand{\EA}{\text{\ae}}  

\newcommand{\edit}[1]{{\textcolor{black}{{{#1}}}}}

\numberwithin{equation}{section}
\renewcommand{\theequation}{\arabic{section}.\arabic{equation}}

\begin{document}

\title{Gravitational wave constraints on Einstein-\ae ther theory with LIGO/Virgo data}

\author{Kristen Schumacher}
\affiliation{Department of Physics and Illinois Center for Advanced Studies of the Universe, University of Illinois at Urbana-Champaign, Urbana, IL, 61801, USA}

\author{Scott Ellis Perkins}
\affiliation{Department of Physics and Illinois Center for Advanced Studies of the Universe, University of Illinois at Urbana-Champaign, Urbana, IL, 61801, USA}

\author{Ashley Shaw}
\affiliation{Department of Physics and Illinois Center for Advanced Studies of the Universe, University of Illinois at Urbana-Champaign, Urbana, IL, 61801, USA}

\author{Kent Yagi}
\affiliation{Department of Physics, University of Virginia, Charlottesville, Virginia 22904, USA}

\author{Nicol\'as Yunes}
\affiliation{Department of Physics and Illinois Center for Advanced Studies of the Universe, University of Illinois at Urbana-Champaign, Urbana, IL, 61801, USA}

\date{November 27, 2023}

\begin{abstract}
Lorentz symmetry is a fundamental property of Einstein's theory of general relativity that one may wish to test with gravitational wave observations. 
Einstein-\ae ther theory is a model that introduces Lorentz-symmetry breaking in the gravitational sector through an \ae ther vector field, while still leading to second-order field equations. 
This well-posed theory passes particle physics constraints because it modifies directly only the gravitational sector, yet it predicts deviations in the inspiral and coalescence of compact objects. 
We here, for the first time, put this theory to the test by comparing its gravitational wave predictions directly against LIGO/Virgo gravitational wave data.
We first construct a waveform model for Einstein-\ae ther theory, {\tt EA\_IMRPhenomD\_NRT}, through modifications of the general relativity {\tt IMRPhenomD\_NRTidalv2} model (used by the LIGO/VIRGO collaboration). 
This model constructs a response function that not only contains the transverse-traceless polarization, but also additional Einstein-\ae ther (scalar and vectorial) polarizations simultaneously.
We then use the many current constraints on the theory to construct non-trivial priors for the Einstein-\ae ther coupling constants. 
After testing the waveform model, we conduct parameter estimation studies on two gravitational wave events: GW170817 and GW190425. 
We find that these data are not sufficiently informative to place constraints on the theory that are stronger than current bounds from binary pulsar, solar system and cosmological observations. 
This is because, although Einstein-\ae ther modifications include additional polarizations and have been computed beyond leading post-Newtonian order, these modifications are dominated by (already-constrained) dipole effects. 
These difficulties make it unclear whether future gravitational wave observations will be able to improve on current constraints on Einstein-\ae{}ther theory. 
\end{abstract}

\maketitle


\section{Introduction}
Gravitational waves (GWs) are beginning to allow for unprecedented ways to probe the gravitational interaction in regimes in which gravity is strong and highly dynamical. Since the first detection in 2015, there have so far been 90 GW events detected by the LIGO/Virgo Collaboration~\cite{LIGOScientific:2021djp}. These waves originate from compact binary mergers and allow for the study of the astrophysical objects that comprise them and for tests of fundamental physics, such as tests of Einstein's theory of general relativity (GR)~\cite{LIGOScientific:2020tif}. Though this theory has passed every test encountered to date, there are still reasons to believe that it might need to be extended~\cite{Will:2014kxa, Clifton:2011jh}. Thus, it is imperative that GR be tested in previously unexplored regimes.
 
One property of gravity that is especially interesting to compare against experiment is Lorentz invariance. This property is a general principle that states that experiments are independent of the reference frame they are performed in. Though Lorentz violation has already been strongly constrained for matter interactions, violations that couple only to the gravitational sector have not yet been stringently constrained~\cite{Mattingly:2005re, Will:2014kxa}. Furthermore, there are theoretical reasons to believe that Lorentz invariance may not hold at all energies, and that Lorentz violation may be induced by quantum gravity models~\cite{Mattingly:2005re}. All of this provides a good motivation to search for and/or constrain Lorentz violation in the gravitational sector, since any evidence of a violation would be clear evidence of new physics. 

The simplest theory that violates Lorentz symmetry by introducing a single vector field while still leading to second-order equations of motion is Einstein-\ae{}ther theory~\cite{Jacobson:2007veq}. In this theory, spacetime is filled with a congruence of timelike curves, 
the four-velocity of the \ae{}ther field~\cite{Jacobson:2007veq}. This congruence establishes a preferred direction, implying that there is a locally determined state of rest 
and breaking local Lorentz invariance~\cite{Eling:2004dk}. Modifications to the gravitational action in this theory are regulated by four dimensionless coupling constants, which determine the strength of the coupling of the \ae{}ther field four-velocity to the action. Hence, constraining these coupling constants constrains the theory. 

Einstein-\ae{}ther theory has already been constrained with a plethora of astrophysical observations. The most stringent of these constraints comes from the simultaneous observation of GWs and a gamma ray burst from the 2017 binary neutron star (BNS) merger. This event placed tight observational bounds on the speed of the tensor polarization of GWs, immediately restricting one of the coupling constants of Einstein-\ae{}ther theory to be on the order of  $\mathcal{O}(10^{-15})$~\cite{LIGOScientific:2017zic}. The lack of observational evidence for gravitational ``Cherenkov type'' radiation further places tight constraints on the speed of the GWs in Einstein-\ae{}ther theory, which can be related back to the coupling constants~\cite{Elliott:2005va}. Meanwhile, cosmological observations of the abundance of primordial Helium restrict the amount by which the \ae{}ther field can rescale the effective value of Newton's constant that appears in the Friedman equation~\cite{Carroll:2004ai}. Solar system constraints on the preferred frame parameterized post-Newtonian (PN) parameters, due to lunar laser ranging experiments and observations of the solar spin axis, can be translated into constraints on the Einstein-\ae{}ther coupling constants~\cite{Muller:2005sr, solarSystem2}. Finally, in recent work, observations of the damping of the period of binary pulsar and triple systems have further constrained Einstein-\ae{}ther  theory~\cite{Gupta:2021vdj}. However, even after combining all of these constraints there are still large regions of parameter space that are not yet stringently constrained. 
    
The inspiral and merger of compact objects, as observed with GWs, provide a new laboratory in which we may place new constraints on Einstein-\ae{}ther theory, considering the many modifications to GWs in this theory. 
For instance, modifications to the amplitude and phase of quadrupole radiation in this theory can be searched for in GW data~\cite{Hansen:2014ewa, Zhang:2019iim}. Note that the quadrupole correction is partially degenerate with the chirp mass of the binary system, since this also enters at leading-order in a post-Newtonian (PN) expansion\footnote{A PN expansion is one in which all quantities are series-expanded in small velocities and weak-fields~\cite{Blanchet:2013haa}.} of the phase. Similarly, the emission of dipole radiation due to the propagation of vector and scalar modes is another signature of Einstein-\ae{}ther theory (though this particular signature is already well constrained by binary pulsar observations)~\cite{Hansen:2014ewa}. Finally, the mass of strongly gravitating objects is affected by the \ae{}ther field, in a way described by the ``sensitivity'' of objects in this theory~\cite{Yagi:2013qpa,Gupta:2021vdj}. This sensitivity enters the Einstein-\ae ther prediction of the gravitational waveform and it depends on the coupling constants of the theory and the binding energy of the compact objects generating the GWs. If these signatures of Einstein-\ae{}ther theory are not observed in GW data, the coupling constants of the theory can be constrained to smaller and smaller values.

In this paper, we compare the predictions of Einstein-\ae{}ther theory for the GWs emitted in the inspiral of neutron stars (NS) to all LIGO/Virgo data taken during the O1, O2 and O3 observing campaigns to try to place constraints on the coupling constants of the theory. 
To execute this analysis the predictions of Einstein-\ae{}ther theory must first be encoded into a new waveform template that can be directly compared with data. Building off of the {\tt IMRPhenomD} and {\tt IMRPhenomD\_NRTidalv2} waveform templates, we construct a new waveform template we call {\tt EA\_IMRPhenomD\_NRT}. We first update the code we are using, {\tt GW Analysis Tools}~\cite{Perkins:2021mhb}, to be consistent with LALSuite's {\tt IMRPhenomD\_NRTidalv2} waveform template in GR. From there, we add the binary Love relations to the {\tt IMRPhenomD\_NRTidalv2} model so that we can search for the symmetric combination of tidal deformabilities instead of searching for each tidal deformability individually~\cite{Yagi:2015pkc,Yagi:2016qmr,Carson:2019rjx}. Next, we include the C-Love relations into the model to obtain the compactness of each NS, given the tidal deformability, and thus be able to compute the binding energies and the sensitivities in Einstein-\ae{}ther theory~\cite{Yagi:2013bca,Yagi:2013awa, Maselli:2013mva,Carson:2019rjx, Gupta:2021vdj}. Finally, we add the Einstein-\ae{}ther corrections to the waveform model to 1PN order, as computed in~\cite{Zhang:2019iim}, which now explicitly depend only on the coupling constants, the chirp mass, the symmetric mass ratio, the inclination angle, and the tidal deformabilities, leading to the {\tt EA\_IMRPhenomD\_NRT} model. 

Once constructed, we use the new {\tt EA\_IMRPhenomD\_NRT} waveform model to conduct parameter estimation studies with Bayesian inference on the public LIGO/Virgo data. In parameter estimation studies, previous knowledge about the sampling parameters is encoded in their prior and used to determine the correct sampling region of parameter space. Therefore, we begin by constructing a prior for the Einstein-\ae{ther} coupling constants, describing in detail how each of the current constraints on the theory affects the complicated shape of this prior. We further use this prior to motivate our choice of a particular parameterization of the coupling constants. We then test the capabilities of our waveform model by using it to recover synthetic (injected) data for GWs as predicted both in GR and in Einstein-\ae{}ther theory. Finally, we conduct parameter estimation studies on the two BNS mergers so far observed with LIGO: GW170817 and GW190425. 

We find that current LIGO/Virgo data is not sufficiently informative to place constraints on Einstein-\ae{}ther theory that are stronger than other stringent observational bounds
from solar system~\cite{solarSystem2, Muller:2005sr} and binary pulsar~\cite{Gupta:2021vdj} observations.
That is, marginalized posteriors on the Einstein-\ae{}ther coupling parameters from gravitational wave observations are statistically indistinguishable from their priors, even when the latter are enlarged beyond what is allowed by current observational bounds.  
This is because Einstein-\ae{}ther modifications are dominated by dipole radiation (which enter at -1PN relative order in the waveform) and corrections to the binary's orbital energy (which enter at 0PN relative order in the waveform). Dipole effects are already very well constrained by binary pulsar observations, because these binaries are sufficiently widely separated that dipole modifications can become large unless suppressed by the coupling constants. Leading PN order corrections to the orbital energy are highly correlated with the chirp mass, therefore diluting any constraints. 
 
Even though constraints placed with GWs cannot yet surpass those from other experiments, it is possible that future observations with more advanced detectors will be able to better constrain Einstein-\ae{}ther theory.  For instance, previous work predicted that third-generation and space-based GW detectors may place comparable constraints, or improve them by a factor of 2~\cite{Hansen:2014ewa}. This work, however, was carried out in a now ruled-out region of parameter space, before the coincident GW and electromagnetic observation of GW170817, which bounded the speed of GWs to be essentially identical to that predicted in GR.
Additionally, if the sensitivities of black holes (BHs) in Einstein-\ae{}ther theory were calculated, studies with BH binaries or mixed NS/BH binaries could also be considered. Even without these two specific advancements, constraints from GWs will only improve over time as more BNS mergers are observed and constraints are stacked. Thus, our current work serves as an important foundation for how such parameter estimation studies with GWs in Einstein-\ae{}ther theory can be performed in the fourth and fifth observing runs of the LIGO/Virgo collaboration, and in the future with third-generation detectors. Only by carrying out such studies will we be able to determine whether future observations can place competitive bounds on Einstein-\ae{}ther theory relative to binary pulsar and solar system constraints. 

The remainder of this paper is organized as follows. In Sec.\ \ref{sec:background} we give a brief introduction to Einstein-\ae ther theory, describing the coupling constants of the theory and the sensitivities of strongly gravitating objects. Here we justify why these studies can currently only be performed with BNS inspirals. Section\ \ref{sec:GWs} mathematically describes GWs in Einstein-\ae{ther} theory, presenting the Fourier transform of the response function for an L-shaped GW detector, so that we can understand what modifications and extensions had to be made to current waveform template models in Sec.\ \ref{sec:waveform_template} to create and test the new Einstein-\ae ther waveform template, {\tt EA\_IMRPhenomD\_NRT}. To determine what priors to use for parameter estimation, all current constraints on Einstein-\ae ther theory are collected in Sec.\ \ref{sec:current_constraints}. Once we have a prior, the waveform template is tested on injected data in Sec.\ \ref{sec:injections} and finally used on GW data from BNS inspirals in Sec.\ \ref{sec:GW_constraints}. Section \ref{sec:conclusions} discusses our results and potential future work. There are four appendices included to facilitate reproducibility. In Appendix \ref{sec:appendixNRTidal}, we describe in detail the modifications we made to our code to make it consistent with LALSuite's {\tt IMRPhenomD\_NRTidal} waveform model. Appendix~\ref{sec:appendixSensitivities} provides more detail about the sensitivities in Einstein-\ae{}ther theory for the region of parameter space we are considering and justifies why this region cannot be extended. Appendix \ref{sec:appendixCherenkov} gives the exact mathematical expressions used for one of the conditions in the prior, and Appendix \ref{sec:appendixInjecPlots} contains plots that demonstrate the recovery of injected parameters with our waveform template. 

\textit{Conventions:} Greek letters specify spacetime indices, while Latin letters specify spatial indices only. The Einstein summation convention and $c = 1$ is assumed. The gravitational constant $G_N$ is explicitly listed because there are other gravitational constants in Einstein-\ae ther theory and this allows us to keep track of which one is which. Finally, following the conventions of much of the earlier Einstein-\ae ther literature, we use the metric signature $(+,-,-,-)$. 

\section{Einstein-\AE ther Theory}
\label{sec:background}
In this section, we present a brief overview of Einstein-\ae ther theory, following mostly ~\cite{Gupta:2021vdj}. 
We begin by introducing the action and the field equations, and then continue by discussing the sensitivities of compact objects, which play a key role in our GW model. 

\subsection{Einstein-\ae ther Coupling Constants}

The general action of Einstein-\ae ther theory is \cite{Jacobson:2000xp, Jacobson:2013xta} 
\begin{equation}
S = S_{\EA} + S_{\text{mat}} , 
\end{equation}
where $S_{\text{mat}}$ denotes the matter action and $S_{\EA}$ is the gravitational action of Einstein-\ae ther theory:
\begin{equation}
\begin{aligned}
S_{\EA} &= -\frac{1}{16\pi G_{\EA}} \int \sqrt{-g}\hspace{1mm} d^4 x  \left[R + \lambda(U^\mu U_\mu - 1)\right. \\ 
&\2 \left.+ \frac{1}{3} c_\theta \theta^2 + c_\sigma \sigma_{\mu\nu} \sigma^{\mu\nu} + c_\omega \omega_{\mu\nu} \omega^{\mu\nu} + c_a A_\mu A^\mu \right] . 
\end{aligned}
\end{equation}
In this expression, the quantity $G_{\EA}$ is the ``bare'' gravitational constant, related to Newton's constant $G_N$ via 
\begin{equation}
G_N = \frac{G_{\EA}}{1 - (c_a/2)}, 
\end{equation}
$g$ is the determinant of the metric, $R$ is the four dimensional Ricci scalar, $\lambda$ is a Lagrange multiplier that enforces the unit norm of the \ae ther's four-velocity $U^{\mu}$, and $\{c_\theta, c_\sigma, c_\omega, c_a\}$ are dimensionless coupling constants.
In much of the earlier Einstein-\ae ther theory literature, the action was written in terms of different coupling constants, namely $\{c_1, c_2, c_3, c_4\}$. However the constants used here (which were defined in \cite{Jacobson:2013xta}) appear in many of the physical quantities relevant to GWs in  Einstein-\ae ther theory, so they are particularly convenient to us. The two sets of constants can be related to each other through 
\begin{subequations}
\begin{align}
c_\theta &= c_1 + c_3 + 3c_2, \\
c_\sigma &= c_1 + c_3 ,\\
c_\omega &= c_1 - c_3 ,\\
c_a &= c_1 + c_4 .
\end{align}
\end{subequations}
The rest of the terms in the action are the expansion $\theta$, the shear $\sigma_{\mu\nu}$, the vorticity (also called the twist) $\omega_{\mu\nu}$, and the acceleration $A_\mu$ of the \ae ther's four-velocity. These quantities are defined via
\begin{subequations}
\begin{align}
A^\mu &= U^\nu \nabla_\nu U^\mu, \\
\theta &= \nabla_\mu U^\mu, \\
\sigma_{\mu\nu} &= \nabla_{(\nu} U_{\mu)} + A_{(\mu}U_{\nu)} - \frac{1}{3} \theta h_{\mu\nu}, \\
\omega_{\mu\nu} &= \nabla_{[\nu} U_{\mu]} + A_{[\mu}U_{\nu]},
\end{align}
\end{subequations}
with $h_{\mu\nu} = g_{\mu\nu} - U_\mu U_\nu$ a projector to directions orthogonal to the \ae ther's four velocity.

Varying the action with respect to the metric, the \ae ther field, and the Lagrange multiplier (and eliminating this last from the equations) gives the modified Einstein field equations \cite{Gupta:2021vdj}
\begin{equation}
E_{\alpha\beta} \equiv G_{\alpha \beta} - T_{\alpha \beta}^{\EA} - 8\pi G T_{\alpha\beta}^{\text{mat}} = 0 
\end{equation}
and the \ae ther equations
\begin{align}
\text{\AE}_\mu \equiv \left[\nabla_\alpha J^{\alpha\nu}
- \left(c_a - \frac{c_\sigma + c_\omega}{2} \right) A_\alpha \nabla^\nu U^\alpha \right] h_{\mu\nu}  = 0.
\end{align}
In these expressions, $G_{\alpha\beta}$ is the usual Einstein tensor, the matter stress-energy tensor is $T_{\text{mat}}^{\alpha\beta}$,
and the \ae ther stress-energy tensor is 
\begin{align}
T_{\alpha\beta}^{\EA} &= \nabla_\mu \left(J_{(\alpha}^{\2\mu} U_{\beta)} - J^\mu_{\2(\alpha}U_{\beta)} - J_{(\alpha\beta)}U^\mu \right) \nonumber \\
&\2 + \frac{c_\omega + c_\sigma}{2} \left[ \left(\nabla_\mu U_\alpha \right)\left( \nabla^\mu U_\beta \right) - \left(\nabla_\alpha U_\mu \right)\left( \nabla_\beta U^\mu\right) \right] \nonumber\\
&\2 + U_\nu \left(\nabla_\mu J^{\mu\nu} \right)U_\alpha U_\beta \nonumber\\
&\2 - \left(c_a - \frac{c_\sigma + c_\omega}{2} \right) \left[A^2 U_\alpha U_\beta - A_\alpha A_\beta \right] \nonumber\\
&\2 + \frac{1}{2} M^{\sigma \rho}_{\2\2\mu\nu} \nabla_\sigma U^\mu \nabla_\rho U^\nu g_{\alpha\beta}, 
\end{align}
with
\begin{align}
J^{\alpha}_{\2\mu} &\equiv M^{\alpha\beta}_{\2\2\mu\nu} \nabla_\beta U^\nu, \\
M^{\alpha\beta}_{\2\2\mu\nu} &\equiv \left(\frac{c_\sigma + c_\omega}{2}\right) h^{\alpha\beta} g_{\mu\nu} + \left(\frac{c_\theta - c_\sigma}{3} \right) \delta^\alpha_\mu \delta^\beta_\nu \nonumber\\
&\2\2 +\left(\frac{c_\sigma - c_\omega}{2} \right) \delta^\alpha_\nu \delta^\beta_\mu + c_a U^\alpha U^\beta g_{\mu\nu}.
\end{align}

Linearizing these field equations and perturbing about Minkowski space results in propagation equations for the gravitational wave polarization tensor, which can be classified into a transverse-traceless (spin-2) part, a vector (spin-1) part, and a scalar (spin-0) part. Henceforth, we shall refer to these different spins as the tensor, vector and scalar parts respectively of the gravitational wave polarization. The speeds with which these polarizations propagate are given by \cite{Jacobson:2004ts}
\begin{subequations}
\begin{align}
c_T^2 &= \frac{1}{1 - c_\sigma} \label{eqn:tensorSpeed}, \\
c_V^2 &= \frac{c_\sigma + c_\omega - c_\sigma c_\omega}{2c_a (1 - c_\sigma)}, \label{eqn:vectorSpeed} \\
c_S^2 &= \frac{(c_\theta + 2c_\sigma)(1 - c_a/2)}{3c_a(1 - c_\sigma)(1 + c_\theta/2)}. \label{eqn:scalarSpeed}
\end{align}
\label{eqn:polarization_speeds}
\end{subequations}

\subsection{Sensitivities}
\label{subsec:sensitivities}
The \ae{}ther field in Einstein-\ae ther theory couples to matter indirectly via the metric perturbation. In regions where these perturbations are great, as around strongly gravitating bodies, their effect is more important. Hence, the mass of strongly gravitating objects is affected by the \ae ther field. This coupling depends on the relative velocity between the \ae ther field and the object, $\gamma \equiv u_\alpha U^\alpha$, with $u^\alpha$ the four-velocity of the object. In most situations, including the inspiral of two widely separated objects, this quantity $\gamma$ will be small compared to the speed of light. Thus we can Taylor expand the mass of a gravitating body about $\gamma = 1$ \cite{Gupta:2021vdj}:
\begin{equation}
    \mu(\gamma) = \tilde{m} \left[1 + \sigma (1 - \gamma) + \frac{1}{2} \sigma' (1 - \gamma)^2 + ... \right]\,,
\end{equation}
where $\tilde{m}, \sigma,$ and $\sigma'$ are constants. The quantity $\sigma$ is often referred to as the ``sensitivity'' and $\sigma'$ its derivative \cite{Foster:2007gr, Gupta:2021vdj}:
\begin{subequations}
\begin{align}
   \sigma &\equiv -\frac{d \ln \mu(\gamma)}{d \ln \gamma} \Big|_{\gamma = 1}\,, \\
    \sigma' &\equiv \sigma + \sigma^2 + \frac{d^2 \ln \mu(\gamma)}{d(\ln \gamma)^2} \Big|_{\gamma = 1}\,.
\end{align}
\end{subequations}
Computing the equations of motion for a binary system leads to the definition of an ``active'' mass for each object $m_A$, related to the constant $\tilde{m}_A$ via $m_A = (1 + \sigma_A)\tilde{m}_A$. This is done such that the Newtonian limit of Einstein-\ae ther theory agrees with Newtonian gravity, with a rescaled gravitational constant $\mathcal{G}_{AB} = G_N/[(1+\sigma_A)(1+\sigma_B)]$ \cite{Foster:2007gr}.

The sensitivities play a key role in the GWs emitted by binary systems in Einstein-\ae ther theory. This is because not only do they appear in the Hamiltonian (and, therefore, in the equations of motion) of binaries, but they also enter the fluxes of radiation that back-react on the binary, forcing it to inspiral faster than it would otherwise. Unfortunately, the sensitivities of black holes (BHs) have not yet been calculated, but they are known for neutron stars (NSs)\cite{Gupta:2021vdj}. For these objects, the sensitivities range between $10^{-8}$ and $1$ depending on the region of parameter space considered for the coupling constants (see Sec.~\ref{subsec:EA_in_code} and Appendix~\ref{sec:appendixSensitivities} for more detail).
The sensitivities also vary depending on the mass and radius of the NS (and thus on the equation of state (EoS)). Given that we can only model the sensitivity of NSs, henceforth we focus exclusively on GW events produced by binary NS inspirals, namely GW170817 and GW190425.

The calculation of the sensitivity of NSs is highly nontrivial. To solve for this quantity in terms of Einstein-\ae ther parameters, Gupta \emph{et al.}~\cite{Gupta:2021vdj} solved the field equations through linear order in the NS's velocity and extracted the sensitivity from the asymptotic fall off of the metric and \ae ther field. This calculation was done both for (tabulated) realistic EoSs, as well as for the Tolman VII phenomenological EoS. The latter has the advantage of allowing for an analytic solution to the field equations at zeroth-order in velocity, which then renders the calculation of the sensitivities semi-analytical. When compared to the numerical solutions using the other EoSs, the Tolman VII results are highly accurate, and in fact, the sensitivities present an  approximately universal behavior (with less than 3\% variation between EoSs studied) when written in terms of the stellar binding energy $\Omega_A$. 

With this at hand, Gupta \emph{et al.} were able to find an analytic representation of the sensitivities \cite{Gupta:2021vdj}. First, rescaling to a more convenient parameter in the description of GWs, one defines the sensitivities $s_A$ for body $A$ in a binary system via~\cite{Foster:2007gr}
\begin{equation}
    s_A \equiv \frac{\sigma_A}{1 + \sigma_A}.
\label{eqn:s_def}
\end{equation}
Then, carrying out a small binding energy expansion using the Tolman VII EoS, one finds 
\begin{widetext}
\begin{align}
    s_A &= \frac{(3\alpha_1 + 2\alpha_2)}{3} \frac{\Omega_A}{m_A} \nonumber \\
    &\2\2+ \left(\frac{573 \alpha_1^3 + \alpha_1^2 \left(67669 - 764\alpha_2\right) + 96416 \alpha_2^2 + 68 \alpha_1 \alpha_2 \left(9 \alpha_2 - 2632 \right)}{25740 \alpha_1} \right) \frac{\Omega_A^2}{m_A^2} \nonumber\\
    &\2\2+ \frac{1}{656370000 c_\omega \alpha_1^2} \left\{ - 4 \alpha_1^2 \left(\alpha_1 + 8\right) [36773030 \alpha_1^2 - 39543679 \alpha_1 \alpha_2 + 11403314 \alpha_2^2] \right. \nonumber\\
    &\2\2\2\2 + c_\omega [1970100 \alpha_1^5 - 13995878400 \alpha_2^3 - 640 \alpha_1\alpha_2^2 ( - 49528371 + 345040\alpha_2) - 5 \alpha_1^4(19548109 + 788040 \alpha_2)\nonumber\\
    &\2\2\2\2 - 16 \alpha_1^2 \alpha_2 (1294533212 - 29152855\alpha_2 + 212350\alpha_2^2)+ \left. \alpha_1^3 (2699192440 - 309701434\alpha_2 + 5974000\alpha_2^2)] \right\} \frac{\Omega_A^3}{m_A^3} \nonumber \\
    &\2\2 + \mathcal{O} \left(\frac{\Omega_A^4}{m_A^4} \right), \label{eqn:sensitivity}
\end{align}
\end{widetext}
where $\alpha_1$ and $\alpha_2$ are the preferred frame parameterized post-Newtonian parameters for Einstein-\ae ther theory, namely \cite{Foster:2005dk},
\begin{subequations}
\begin{align}
\alpha_1 &= 4 \frac{c_\omega(c_a - 2c_\sigma) + c_a c_\sigma}{c_\omega(c_\sigma - 1) - c_\sigma}, \\
\alpha_2 &= \frac{\alpha_1}{2} + \frac{3(c_a - 2c_\sigma)(c_\theta + c_a)}{(2 - c_a)(c_\theta + 2c_\sigma)},
\end{align}
\label{eqn:alphas}
\end{subequations}
and $\Omega_A/m_A$ is the ratio of the stellar binding energy to the NS mass $m_A$. For the Tolman VII EoS, the compactness of the star, $C := m_A/R_A$, where $R_A$ is the radius of the star, can be expressed in terms of this ratio \cite{Gupta:2021vdj},
\begin{equation}
    C = - \frac{7\Omega_A}{5m_A} + \frac{35819 \alpha_1 \Omega_A^3 }{85800 m_A^3} + \mathcal{O}\left( \frac{\Omega_A^4}{m_A^4} \right)\label{eqn:compact(bindingovermass)}
\end{equation}
for small compactnesses and binding energies. The leading-order terms of the expansion of the sensitivity in Eq.~\eqref{eqn:sensitivity} agrees with that derived by Foster~\cite{Foster:2007gr}.
Inverting this relationship, one finds the binding energy over the mass as a function of compactness
\begin{equation}
    \frac{\Omega_A}{m_A} = -\frac{5}{7} C - \frac{18275 \alpha_1 C^3}{168168} + \mathcal{O}(C^4). \label{eqn:binding_energy}
\end{equation}

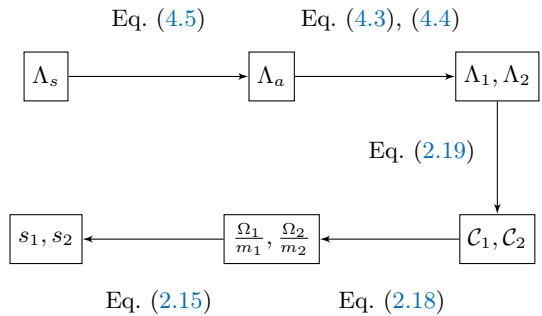
\begin{figure}
\begin{tikzpicture}[node distance = 3cm]
\node[process] (start) {$\Lambda_s$};
\node[process, right of=start](Lambda_a){$\Lambda_a$};
\node[process, right of=Lambda_a](Lambdas){$\Lambda_1, \Lambda_2$};
\node[process, below=1.5cm of Lambdas](compact){$\mathcal{C}_1, \mathcal{C}_2$};
\node[process, left of=compact](binding){$\frac{\Omega_1}{m_1}, \frac{\Omega_2}{m_2}$}; 
\node[process, left of=binding](sensitivity){$s_1, s_2$}; 
\node[draw=none] at (1.5, .75) (binLove){Eq.~\eqref{eqn:binLove}};
\node[draw=none] at (4.45, 0.75) (definitions){Eq.~\eqref{eqn:lambdas},~\eqref{eqn:lambdaa}};
\node [draw=none] at (5, -1) (CLove){Eq.~\eqref{eqn:CLove}};
\node [draw=none] at (4.6, -3) (bindingenergyofC){Eq.~\eqref{eqn:binding_energy}};
\node [draw=none] at (1.5, -3) (sofbindingenergy){Eq.~\eqref{eqn:sensitivity}};

\path [connector] (start) -- (Lambda_a);
\path [connector] (Lambda_a) -- (Lambdas); 
\path [connector] (Lambdas) --(compact); 
\path [connector] (compact) -- (binding); 
\path [connector] (binding) -- (sensitivity); 
\end{tikzpicture}
\caption{A flow chart of computing sensitivities from the parameter sampled on (symmetric tidal deformability, $\Lambda_s$). We use the binary Love relations, the C-Love relations, the Tolmann VII EoS, and the equation for sensitivities as a function of the binding energy to mass ratio. These sensitivities will then be used in the waveform as described in section~\ref{subsec:responsefunc}.}
\label{fig:code_logic_cartoon}%
\end{figure}

Our Einstein-\ae ther waveform model relies on knowing the sensitivities $s_A$, but as shown in Eqs.~\eqref{eqn:sensitivity} and~\eqref{eqn:binding_energy}, these depend ultimately on the compactness. We can relate the compactness of each star to their tidal deformabilities as follows. First, following previous work on nuclear astrophysics with GWs~\cite{Veitch:2014wba, LIGOScientific:2017zic, LIGOScientific:2018hze, LIGOScientific:2020aai}, we will sample the GW likelihood by varying the symmetric tidal deformability $\Lambda_s = (\Lambda_1 + \Lambda_2)/2$ (among many other parameters). From $\Lambda_s$, we can obtain $\Lambda_a = (\Lambda_2 - \Lambda_1)/2$ using the binary Love relations~\cite{Yagi:2013awa, Carson:2019rjx}, 
and from $\Lambda_s$ and $\Lambda_a$ we can easily obtain $\Lambda_1$ and $\Lambda_2$.  
Now, from the latter two quantities, we will obtain the compactness  through the approximately universal C-Love relations \cite{Yagi:2013awa, Carson:2019rjx} 
\begin{equation}
    C_A(\Lambda_A) = 0.2496 \Lambda_A^{-1/5} \frac{1 + \sum_{i=1}^3 a_i \Lambda_A^{-i/5}}{1 + \sum_{i=1}^3 b_i \edit{\Lambda_A^{-i/5}}}, \label{eqn:CLove}
\end{equation}
where the fitting coefficients are 
\begin{align}
    a_i &= \{-919.6, 330.3, -857.2\}, \\
    b_i &= \{-383.5, 192.5, -811.1\}.
\end{align}
From the compactness, we can then evaluate the stellar binding energy, and from that, the sensitivities. 
The logic is outlined in Fig.~\ref{fig:code_logic_cartoon}. 

The binary Love and C-Love relations feature heavily in the construction of our waveform model, but they are known to only be \textit{approximately} EoS insensitive. In fact, their variability is about  $10\%$~\cite{Chatziioannou:2018vzf}. One can include this variability in Bayesian parameter estimation, and then marginalize over it, as done for example in~\cite{Chatziioannou:2018vzf}. We will here not include it, however, because the statistical error in the extraction of the symmetric tidal deformability dominates over any systematic error introduced by this variability, as shown in~\cite{Carson:2019rjx}, at least in the current GW detector era. 

\section{GWs in Einstein-\ae ther Theory}
\label{sec:GWs}
In this section, we review the work of~\cite{Zhang:2019iim} and \cite{Schumacher:inPrep} to construct expressions for the GW polarizations of Einstein-\ae ther theory for a quasi-circular inspiraling binary composed of non-spinning NSs. We then present the Fourier transform of the response function in explicit form, ready for use in parameter estimation and data analysis. 

\subsection{GW Polarizations in Einstein-\ae ther theory}

Following the example of many other studies~\cite{Zhang:2019iim, Foster:2006az, Yagi:2013ava, Jacobson:2000xp}, we begin by considering linear perturbations to a background Minkowski metric, $\eta_{\mu\nu} = \text{diag}(-1, 1, 1, 1)$, and linear perturbations to a stationary \ae ther field: 
\begin{equation}
    h_{\mu\nu} = g_{\mu\nu} - \eta_{\mu\nu}, \hspace{5mm} w^0 = U^0 -1, \hspace{5mm} w^i = U^i.
\end{equation}
The one-form $h_{0i}$ and the vector $w^i$ can be uniquely decomposed into irreducible transverse and longitudinal pieces, while the spatial components of the metric perturbation $h_{ij}$ can be uniquely decomposed into a transverse traceless tensor, a transverse vector, and transverse and longitudinal traces~\cite{Foster:2006az}:
\begin{subequations}
\begin{align}
    w_i &= \nu_i + \nu_{,i}, \\
    h_{0i} &= \gamma_i + \gamma_{,i}, \\
    h_{ij} &= \phi_{ij} + 2\phi_{(i,j)} + \frac{1}{2} P_{ij}[f] + \phi_{,ij}
\end{align} 
\label{eqn:perturbation_decomp}%
\end{subequations}
where the quantity $P_{ij} := \delta_{ij} \Delta - \partial_i \partial_j$ is a transverse differential operator, the quantity $\Delta := \delta^{ij} \partial_i \partial_j$ is the flat-space spatial Laplacian, and $F := \Delta f$ is a scalar. The transverse vector and tensor fields here satisfy the divergence-free condition,
\begin{equation}
    \partial^i \gamma_i = \partial^i \nu_i = \partial^i \phi_i = 0, \hspace{5mm}
    \partial^j \phi_{ij} = 0,
\end{equation}
and the field $\phi_{ij}$ is also traceless, $\phi_i^{\2i} = 0$. Note that we also make the conventional gauge choice, $\phi_i = 0$ and $\nu = \gamma = 0$~\cite{Foster:2006az}.

With these convenient decompositions in hand, we would like to use the formula for GW polarizations in generic modified theories of gravity provided by~\cite{Chatziioannou:2012rf}. However, that work made the implicit assumption that all polarizations of the GW travel at the same speed, specifically the speed of light, and this assumption does not hold for Einstein-\ae ther theory. We extended the work of~\cite{Chatziioannou:2012rf} in~\cite{Schumacher:2023jxq} to accommodate for theories that allow for modes with different and arbitrary speeds. In that work, we also explicitly computed the expressions for GW polarizations in Einstein-\ae ther theory by inserting Eqs.~\eqref{eqn:perturbation_decomp} and~\eqref{eqn:polarization_speeds} into our general formula. We found that
\begin{subequations} 
\begin{align}
&h_+ = \frac{1}{2} \phi_{ij} e^{ij}_+, 
&h_\times = \frac{1}{2} \phi_{ij} e^{ij}_\times,\\
&h_b = \frac{1}{2} F, &h_L = (1 + 2\beta_2) h_b, \\
&h_X = \frac{1}{2} \beta_1 \nu_i e^i_X, &h_Y = \frac{1}{2} \beta_1 \nu_i e^i_Y,
\end{align}
\label{eqn:h_N_ours}
\end{subequations}
where 
\begin{align}
    \beta_1 &= -\frac{2c_\sigma}{c_V},\label{eqn:beta1}\\
    \beta_2 &= \frac{c_a - 2c_\sigma}{2c_a (1 - c_\sigma) c_S^2}, \label{eqn:beta2}
\end{align}
and $e_+^{ij} = e_X^i e_X^j - e_Y^i e_Y^j$ and $e_\times^{ij} = e_X^i e_Y^j + e_Y^i e_X^j$ are combinations of basis vectors, defined in the orthogonal basis for GWs propagating in the $e_Z$ direction:
\begin{subequations}
\begin{align}
    e_X &= \left(\cos\vartheta \cos \varphi, \cos\vartheta \sin \varphi, -\sin\vartheta\right), \\
    e_Y &= \left(-\sin\vartheta, \cos\varphi, 0 \right), \\
    e_Z &= \left(\sin \vartheta \cos\varphi, \sin\vartheta \sin\varphi, \cos\vartheta   \right).
\end{align}
\end{subequations}
Equation~\eqref{eqn:beta2} corrects a small minus sign error in~\cite{Zhang:2019iim} that has since been addressed. Aside from that, our results in Eq.~\eqref{eqn:h_N_ours} agree with those of Eq.\ (3.28) in~\cite{Zhang:2019iim}, which were computed in a different way (i.e.~starting from the time-like geodesic deviation equation and working with the linearized Riemann tensor in terms of the metric perturbation; see \cite{Zhang:2019iim} for more details).

\begin{figure}[t]
    \includegraphics[width=\linewidth]{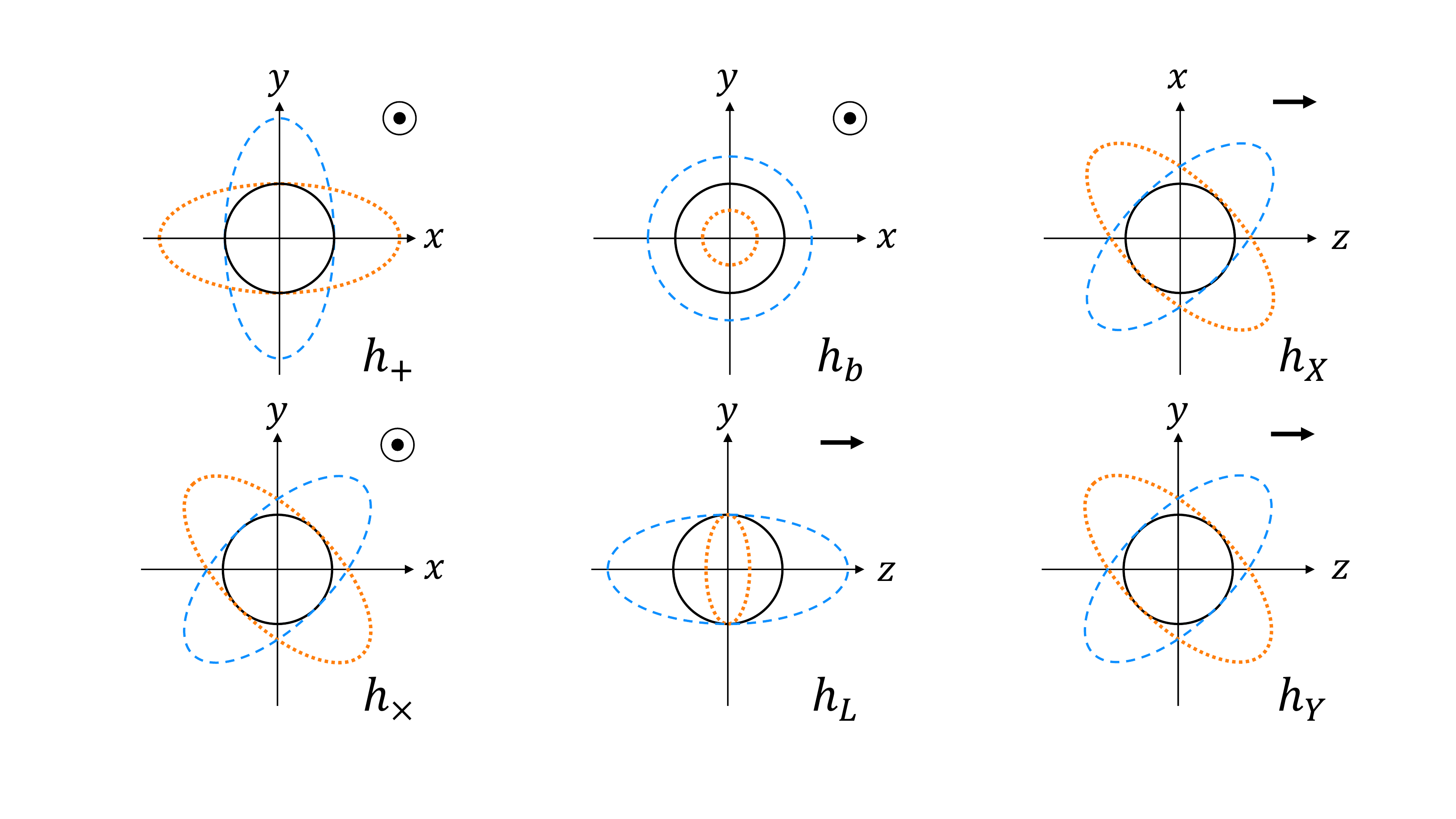}
    \caption{The oscillation of a ring of test particles when each of the six possible polarizations of a GW in Einstein-\ae ther theory passes through, propagating in the $z$-direction. The solid black line represents the ring at times $\omega t = 0, \pi$, the dashed blue line represents the ring at time $\omega t = \pi/2$, and the dotted orange line shows $\omega t = 3\pi/2$.}
    \label{fig:ModesTogether}
\end{figure}

An intuitive understanding of these different GW polarizations in Einstein-\ae ther theory can be gleaned from considering their impact on a ring of test particles. 
In modified theories of gravity, the most general GW has up to six polarization modes. This includes two each of tensor, vector, and scalar type. The two tensor polarizations, $h_+$ and $h_\times$, are the plus and cross modes familiar from GR. The two vector polarizations, $h_X$ and $h_Y$, are labeled for the plane in which they would make a ring of test particles oscillate for a wave propagating in the $z$-direction (see Fig.\ \ref{fig:ModesTogether}). Finally, the two scalar polarizations, $h_b$ and $h_L$, are called the breathing and longitudinal modes for the way in which they would make a ring of test particles oscillate in and out or longitudinally along the direction of propagation (again see Fig.\ \ref{fig:ModesTogether}). 

We continue to follow~\cite{Zhang:2019iim} to compute the GW polarizations that appear in Eq.~\eqref{eqn:h_N_ours} specifically for a binary system. We will not repeat that calculation here, but the details can be found in~\cite{Zhang:2019iim}. That paper assumes that the detectors are far away from the source and solves the linearized Einstein-\ae ther field equations to derive expressions for $\phi_{ij}, \nu_i, \gamma_i$ and $F$ in terms of the Einstein-\ae ther coupling constants, the mass quadrupole moment, the trace-free mass quadrupole moment, the renormalized versions of these quantities, the renormalized mass dipole moment and the renormalized current quadrupole moment. Reference~\cite{Zhang:2019iim} then focuses on two non-spinning compact objects in a quasi-circular orbit to find expressions for these multipolar moments in terms of typical binary system parameters (for example, the binary chirp mass and orbital frequency of the system). Unlike previous work, Ref.~\cite{Zhang:2019iim} allows the center of mass of the binary system to not be comoving with the \ae ther, essentially letting their relative velocity be nonzero, $V^i \neq 0$.  We will again choose to set $V^i = 0$ since we know it must be $V^i \approx \mathcal{O}(10^{-3})$, given the peculiar velocity of our own galaxy relative to the cosmic microwave background, and we consider this to be negligible compared to the other Einstein-\ae ther modifications~\cite{Yagi:2013ava, Foster:2007gr}. 

\subsection{The Response Function}
\label{subsec:responsefunc}
\allowdisplaybreaks[4]
Parameter estimation on actual data from advanced LIGO, advanced Virgo, or KAGRA requires the Fourier transform $\tilde{h}(f)$ of the response function $h(t)$ for an L-shaped GW detector. From \cite{PoissonAndWill}, we can write the latter as
\begin{equation}
h(t) = \sum_N F_N(\theta, \phi, \psi) h_N(t),
\end{equation}
where $N \in \{+, \times, b, L, X,Y\}$ and $F_N(\theta, \phi, \psi)$ are the angle pattern functions, which depend on the polar, azimuthal and polarization angles ($\theta$, $\phi$, and $\psi$, respectively)~\footnote{Figure\ 2 of~\cite{Yunes:2013dva} and Figure\ 11.5 of~\cite{PoissonAndWill} illustrate how these angles relate the orientation of the detector and the source.}:
\begin{subequations}
\begin{align}
F_+ &\equiv \frac{1}{2} \left(1 + \cos^2\theta \right)\cos2\phi \cos2\psi \nonumber\\
&\2\2- \cos\theta \sin2\phi \sin2\psi, \\
F_\times &\equiv \frac{1}{2} \left(1 + \cos^2\theta \right)\cos2\phi \sin2\psi \nonumber\\
&\2\2 + \cos\theta \sin2\phi \cos2\psi, \\
F_b &\equiv -\frac{1}{2} \sin^2\theta \cos2\phi, \\
F_L &\equiv \frac{1}{2} \sin^2\theta \cos2\phi, \\
F_X &\equiv -\sin\theta \left(\cos\theta \cos2\phi \cos\psi - \sin2\phi \sin\psi \right), \\
F_Y &\equiv -\sin\theta \left(\cos\theta \cos2\phi \sin\psi + \sin2\phi \cos\psi \right).
\end{align}
\label{eqn:angularPatternFuncs}
\end{subequations}

Through the stationary phase approximation (SPA), one can compute the Fourier transform of the response function, namely 
\begin{equation}
    \tilde{h}(f) = \int h(t)e^{2i \pi ft} dt. \label{eqn:fourier}
\end{equation}
Doing so, we have reproduced Eq.\ 5.7 of~\cite{Zhang:2019iim}, and then collect terms by the $F_N$ functions of Eq.~\eqref{eqn:angularPatternFuncs}. We also choose to separate contributions to these expressions from the $\ell = 2$ and $\ell = 1$ orbital harmonics. We do so because the $\ell =1 $ harmonics are multiplied by an overall amplitude factor that depends on the coupling constants and that is of ${\cal{O}}(10^{-5})$ relative to the overall amplitude of the $\ell=2$ harmonic, when one saturates current constraints. 
Ultimately, we arrive at
\begin{equation}
    \tilde{h}(f) = \sum_{N} \sum_{\ell = 1,2} \tilde{h}_{N,\ell} (f) F_N,
    \label{eqn:response_func_sum}
\end{equation}
with the expressions for $\tilde{h}_{N,\ell}$ given by\footnote{To use these expressions in the {\tt IMRPhenomD} waveform model, we need to convert to the convention of that paper, which defined the Fourier transform as $\tilde{h}(f) = \int h(t) e^{-2i\pi f t} dt$~\cite{Husa:2015iqa}, instead of as in Eq.~\eqref{eqn:fourier}. To transform these expressions to those used in the code, one can simply take $i \rightarrow -i$.} 
\begin{widetext}
\begin{align}
    \tilde{h}_{(+,2)} (f) &= A_{(2)}(f) \left[(1 + \cos^2\iota) \right] e^{i\Psi_{(2)}} e^{-i2\pi f D_L(1 - c_T^{-1})}, \label{eqn:hplus}\\
    \tilde{h}_{(\times,2)} (f) &= A_{(2)}(f) \left[2i \cos \iota \right] e^{i\Psi_{(2)}} e^{-i2\pi f D_L(1 - c_T^{-1})}, \label{eqn:hcross}\\
    \tilde{h}_{(b,2)} (f) &= A_{(2)}(f) \left[ \frac{1}{2 - c_{a}} \left(3c_{a} (Z - 1)-\frac{ 2 \mathcal{S}}{c_S^2} \right) \sin^2 \iota \right] e^{i\Psi_{(2)}} e^{-i2 \pi f D_L\left(1 - c_{S}^{-1} \right)}, \label{eqn:l2hb}\\
    \tilde{h}_{(L,2)} (f) &= a_{bL}\tilde{h}_{(b,2)} (f)  ,\\
    \tilde{h}_{(X,2)} (f) &=   A_{(2)}(f) \left[ \frac{\beta_1}{c_\sigma + c_\omega - c_\sigma c_\omega} \frac{1}{2c_V} \left( \mathcal{S} - \frac{c_{\sigma}}{1 - c_{\sigma}} \right) \sin(2\iota) \right] e^{i\Psi_{(2)}} e^{-i2 \pi f D_L\left(1 - c_{V}^{-1} \right)},\\
    \tilde{h}_{(Y,2)} (f) &= A_{(2)}(f) \left[  \frac{i\beta_1}{c_\sigma + c_\omega - c_\sigma c_\omega} \frac{1}{c_V} \left( \mathcal{S} - \frac{c_{\sigma}}{1 - c_{\sigma}} \right) \sin(\iota) \right] e^{i\Psi_{(2)}}  e^{-i2 \pi f D_L\left(1 - c_{V}^{-1} \right)},\label{eqn:l2hY} \\
    \tilde{h}_{(b,1)} (f) &= A_{(1)}(f) \left[ \frac{2i}{(2 - c_{a})c_S} \sin\iota \right] e^{i \Psi_{(1)}} e^{-i2 \pi f D_L\left(1 - c_{S}^{-1} \right)}, \label{eqn:l1hb}\\
    \tilde{h}_{(L,1)} (f) &= a_{bL}\tilde{h}_{(b,1)} (f)  ,\\
    \tilde{h}_{(X,1)} (f) &=   A_{(1)}(f) \left[ \frac{ i \beta_1}{c_\sigma + c_\omega - c_\sigma c_\omega} \cos \iota  \right] e^{i \Psi_{(1)}} e^{-i2 \pi f D_L\left(1 - c_{V}^{-1} \right)},\\
    \tilde{h}_{(Y,1)} (f) &=  A_{(1)}(f) \left[ - \frac{\beta_1}{c_\sigma + c_\omega - c_\sigma c_\omega}  \right] e^{i \Psi_{(1)}} e^{-i2 \pi f D_L\left(1 - c_{V}^{-1} \right)},\label{eqn:l1hY}
\end{align}
\end{widetext}
with common amplitude and phase functions given by 
\begin{widetext}
\begin{align}
    A_{(2)}(f) &= - \frac{1}{2} \sqrt{\frac{5\pi}{48}} \sqrt{\frac{(2-c_{a})}{(1 - s_1)(1 - s_2)}} \frac{1}{D_L} G_N^2 \bar{\mathcal{M}}^{2} \kappa_3^{-1/2} \left(G_N \pi \bar{\mathcal{M}} f \right)^{-7/6} \left[1 - \frac{1}{2} \left(G_N \pi \bar{\mathcal{M}}  f \right)^{-2/3} \eta^{2/5} \epsilon_x \right], \label{eqn:l2amp} \\
    \Psi_{(2)} (f) &= \frac{3}{64} \frac{(1 - s_1)(1 - s_2)}{(2 - c_{a})} \kappa_3^{-1}  \left(G_N \pi \bar{\mathcal{M}} f \right)^{-5/3} \left[1 - \frac{4}{7} \left(G_N \pi \bar{\mathcal{M}} f \right)^{-2/3} \eta^{2/5} \epsilon_x \right] + 2\pi f \bar{t}_c - 2 \Phi(t_c) - \frac{\pi}{4},\label{eqn:l2phase} \\
    A_{(1)}(f) &= - \frac{1}{4}\sqrt{\frac{5\pi}{48}} \sqrt{\frac{2 - c_{a}}{(1 - s_1)(1 - s_2)}} \frac{1}{D_L} \Delta s G_N^2 \bar{\mathcal{M}}^{2} \kappa_3^{-1/2} \eta^{1/5} \left(G_N \pi \bar{\mathcal{M}} f \right)^{-3/2} \left[1 - \frac{1}{2} \left(2 G_N \pi \bar{\mathcal{M}} f \right)^{-2/3} \eta^{2/5} \epsilon_x \right], \label{eqn:l1amp}\\
    \Psi_{(1)} (f) &= \frac{3}{128} \frac{(1 - s_1)(1 - s_2)}{(2 - c_{a})} \kappa_3^{-1}  \left(2G_N \pi \bar{\mathcal{M}} f \right)^{-5/3} \left[1 - \frac{4}{7} \left(2 G_N \pi \bar{\mathcal{M}} f \right)^{-2/3} \eta^{2/5} \epsilon_x \right] + 2\pi f \bar{t}_c - \Phi(t_c) - \frac{\pi}{4}. \label{eqn:l1phase}
\end{align}
\end{widetext}
Note that the $\ell =1$ harmonic only affects the additional non-GR polarizations.

The quantities in these expressions that we have not yet explicitly defined are given in~\cite{Zhang:2019iim}, but we repeat their definitions here for completeness: 
\begin{subequations}
\begin{align}
    a_{bL} &= 1 + 2\beta_2, \\
    \bar{t_c} &= t_c + D_L, \\
    \mathcal{M} &= (m_1 + m_2) \eta^{3/5}, \label{eqn:chirpmass} \\
    Z &= \frac{(\alpha_1 - 2\alpha_2) (1 - c_\sigma)}{3(2c_\sigma - c_a)}, \label{eqn:Z}\\
    \epsilon_x &= \frac{5\Delta s^2}{32 \kappa_3}  \mathcal{C}, \label{eqn:epsilon_x}\\
    \Delta s &= s_1 - s_2, \\
    \kappa_3 &= \mathcal{A}_1 + \mathcal{A}_2 \mathcal{S} + \mathcal{A}_3 \mathcal{S}^2 \label{eqn:kappa3},
\end{align}
\end{subequations}
where 
\begin{subequations}
\begin{align}
    \mathcal{S} &= s_1 \mu_2 + s_2\mu_1, \label{eqn:mass_weighted_sensitivity}\\
    \mu_A &= \frac{m_A}{(m_1 + m_2)}, \\
    \eta &= \frac{m_1 m_2}{(m_1 + m_2)^2}, \label{eqn:eta}
\end{align}
\end{subequations}
and
\begin{subequations}
\begin{align}
    \mathcal{A}_1 &= \frac{1}{c_T} + \frac{2 c_{a}c_{\sigma}^2}{(c_\sigma + c_\omega - c_\sigma c_\omega)^2 c_V} + \frac{3 c_{a} (Z - 1)^2}{2(2 - c_{a}) c_S}, \label{eqn:A1}\\
    \mathcal{A}_2 &= - \frac{2c_{\sigma}}{(c_\sigma + c_\omega - c_\sigma c_\omega) c_V^3} - \frac{2(Z - 1)}{(2 - c_{a})c_S^3}, \\
    \mathcal{A}_3 &= \frac{1}{2c_{a} c_V^5} + \frac{2}{3c_{a}(2 - c_{a})c_S^5}, \\
    \mathcal{C} &= \frac{4}{3 c_{a} c_V^3} + \frac{4}{3 c_{a} (2 - c_{a})c_S^3}. \label{eqn:curlyC}
\end{align}
\end{subequations}
For convenience, we also have defined a new quantity 
\begin{equation}
    \bar{\mathcal{M}} = (1-s_1)(1-s_2)\mathcal{M}\bar{\mathcal{M}} = (1-s_1)(1-s_2)\mathcal{M}. \label{eqn:sensitivity_chirpmass}
\end{equation}
Now that we have the mathematical expressions for the Fourier transform of the response function separated out into these convenient pieces, corresponding to the $\ell =2$ and $\ell =1$ contributions to each of the different polarizations of the GW, we can implement them in a waveform model, as we shall describe in the next section. 

\section{An Einstein-\AE ther Waveform Template}
\label{sec:waveform_template}
To compare gravitational wave predictions from Einstein-\ae ther theory directly with data, we need an Einstein-\ae ther waveform model. This section starts with a basic description of {\tt GW Analysis Tools} ({\tt GWAT}), the code used for this analysis. Next we follow~\cite{Dietrich:2019kaq} and update {\tt GWAT} to incorporate the {\tt IMRPhenomD\_NRTidalv2} model for binary NS mergers. Finally, we describe the additions that were made to the {\tt GWAT} implementation of the {\tt IMRPhenomD\_NRTidalv2} model to create the {\tt EA\_IMRPhenomD\_NRT} model, which is capable of modeling coalescing NSs in Einstein-\ae ther theory. Throughout this section, we compare output from our code to previous work to demonstrate its functionality and validity.

\subsection{{\tt GWAT} Implementation of BBH Waveform Models in GR}
\label{subsec:GWATBaseline}

The code used for the parameter estimation analysis that will be presented in this paper was built off of {\tt GWAT}, a set of tools for statistical studies in GW science developed by Scott Perkins and collaborators at the University of Illinois Urbana Champaign~\cite{Perkins:2021mhb}. This software allows the user to select different waveform templates and perform parameter estimation on binary BH systems using Bayesian inference (for a review of how parameter estimation is done in GW science, see e.g.~\cite{Veitch:2014wba}). To gather independent samples for the posterior, {\tt GWAT} uses a Markov Chain Monte Carlo (MCMC) sampler, aided by parallel tempering and a mix of jump proposals. For example, for the un-tempered
chains, 30\% of jumps are proposed with differential evolution and 70\% of jumps are proposed along the eigenvectors of the Fisher matrix.

{\tt GWAT} contains several waveform templates available for use, but for the purposes of this paper, we started development from the {\tt IMRPhenomD} model \cite{Husa:2015iqa, Khan:2015jqa}. 
This waveform is defined in GR with an 11-dimensional parameter space, spanned by the parameter vector $\vec{\theta} = \{\alpha', \sin\delta, \psi, \cos\iota, \phi_{\text{ref}}, t_c, D_L, \mathcal{M}, \eta, \chi_1, \chi_2\}$, where $\alpha'$ and $\delta$ are the right ascension and declination angles of the binary in the sky, $\psi$ is the polarization angle with respect to Earth-centered coordinates, $\iota$ is the inclination angle of the binary, $\phi_{\text{ref}}$ is the phase at a reference frequency ($f_{\text{ref}}$, chosen to be consistent with LALSuite), $t_c$ is the time of coalescence, $D_L$ is the luminosity distance, $\mathcal{M}$ is the chirp mass of the binary, as defined in Eq.~\eqref{eqn:chirpmass}, $\eta$ is the symmetric mass ratio, as defined in Eq.~\eqref{eqn:eta}, and $\chi_1 \1(\chi_2)$ is the dimensionless spin of the heavier (lighter) object. 
The dimensionless astrophysical parameters are sampled from uniform priors in the following regions: $\alpha' \in [0, 2\pi],\1 \sin\delta \in [-1,1],\1 \psi \in [0,\pi],\1 \cos\iota \in [-1, 1],\1 \phi_{\text{ref}} \in [0, 2\pi],\1 \chi_1 \in [-.01, .01],\1 \chi_2 \in [-.01, .01]$. 
The dimensionful astrophysical parameters have the following priors: $t_c$ has a flat prior that is restricted to be within 0.1 seconds of the trigger time of the event, $D_L$ is sampled uniformly in the volume defined by the range $[5, 300]$ Mpc, 
and instead of using a prior uniform in $\mathcal{M}$ and $\eta$, we use a prior uniform in $m_1$ and $m_2$ in the range $[1, 2.5]M_\odot$ for NSs.

\subsection{Extending the {\tt GWAT} Implementation of BBH Waveform Models to BNS inspirals in GR}
\label{subsec:NRTmodifications}

As mentioned in Sec.~\ref{sec:background}, constraints on Einstein-\ae ther theory can currently be studied only with signals from BNS inspirals. Thus, as a first step, the {\tt GWAT} implementation of the  {\tt IMRPhenomD} model has to be extended to include finite-size BNS effects. This extension requires modifications to the GW amplitude and phase, which we implemented following the {\tt IMRPhenomD\_NRTidalv2} model \cite{Dietrich:2018uni}. The exact form of these modifications can be found in Appendix~\ref{sec:appendixNRTidal}, but in essence, they are characterized by the mass-weighted tidal deformability $\tilde{\Lambda}$, which is defined by \cite{Wade:2014vqa}, 
\begin{align}
    \tilde{\Lambda} &= \frac{8}{13} \left[ \left(1 + 7\eta - 31 \eta^2 \right)(\Lambda_1 + \Lambda_2) \right. \nonumber \\
    &\left.\2+ \sqrt{1 - 4\eta} \left(1 + 9\eta - 11 \eta^2\right)(\Lambda_1 - \Lambda_2) \right].
    \label{eqn:mass_weighted_tidal}
\end{align}
Therefore, in addition to the BH astrophysical parameters of Sec.~\ref{subsec:GWATBaseline}, $\vec{\theta}$ must now also include the tidal deformabilities of each NS, $\Lambda_1$ and $\Lambda_2$, increasing the dimensionality of the parameter space to 13. Another important modification is the smooth filtering of the signal at the end of the inspiral, which is accomplished with a Plank taper function. This is implemented to avoid including the merger phase of the BNS coalescence, whose phenomenological analytic description does not yet exist and which would otherwise be present because the {\tt IMRPhenomD} model includes merger and ringdown. 

To compare our code with {\tt LALSuite}, we first generated 100 different random combinations of source parameters, and then we computed their respective waveforms in {\tt GWAT} and in {\tt LALSuite}. We then calculated the relative fractional difference between the amplitudes computed with both codes
\begin{equation}
    \frac{A_{LAL} - A_{GW}}{A_{avg}} = \frac{2(A_{LAL} - A_{GW})}{A_{LAL} + A_{GW}},
\end{equation}
where $A_{LAL}$ is the amplitude calculated by {\tt LALSuite} and $A_{GW}$ is the amplitude calculated by {\tt GWAT}. The difference in the phase computed by the two programs was calculated via $\Psi_{LAL} - \Psi_{GW}$. 
The relative amplitude and phase differences are below $0.001\%$ and constant across frequency, which will thus not affect our parameter estimation studies. 

As we explained in Sec.~\ref{subsec:sensitivities}, however, the Einstein-\ae ther modifications to the waveform model will require knowledge of the sensitivites, which are functions of the compactness, and through the Love-C relations, functions of the individual tidal deformabilities. To extract the individual tidal deformabilities, we will use the binary Love relations \cite{Yagi:2013awa, Carson:2019rjx}. The symmetric and antisymmetric combinations of the NS tidal deformability \cite{Yagi:2013awa, Carson:2019rjx}: 
\begin{align}
    \Lambda_s &= \frac{\Lambda_2 + \Lambda_1}{2}, \label{eqn:lambdas}\\
    \Lambda_a &= \frac{\Lambda_2 - \Lambda_1}{2}\,, \label{eqn:lambdaa}
\end{align}
can be related to each other through nearly EoS-insensitive relations $\Lambda_a = \Lambda_a(\Lambda_s)$. The most recent incarnation of this relation is\footnote{Note that the exponent on $\Lambda_s$ in Eq.~\eqref{eqn:binLove} is negative, which corrects a small typo in Ref.~\cite{Carson:2019rjx} that those authors also corrected recently.} 
\begin{equation}
    \Lambda_a = F_n(q) \frac{1 + \sum_{i=1}^{3}\sum_{j=1}^{2} b_{ij} q^j \Lambda_s^{-i/5}}{1 +\sum_{i=1}^{3}\sum_{j=1}^{2} c_{ij} q^j \Lambda_s^{-i/5} } \Lambda_s^\alpha,
    \label{eqn:binLove}
\end{equation}
where $F_n(q)$ is the Newtonian limiting-control factor, $q$ is the mass ratio with $m_2 \leq m_1$, and $\{n,\alpha\}$ are constants, given by
\begin{subequations}
\begin{align}
    F_n(q) &= \frac{1 - q^{10/(3 - n)}}{1 + q^{10/(3 - n)}}, \qquad 
    q = \frac{m_2}{m_1}, \\
    n &= 0.743, \qquad
    \alpha = 1,
\end{align}
\end{subequations}
while the coefficients $\{b_{ij},c_{ij}\}$ are given by
\begin{align}
    b_{ij} &= \begin{bmatrix}
    -14.40 & 14.45 \\
    31.36 & -32.25 \\
    -22.44 & 20.35
    \end{bmatrix}, \\
    c_{ij} &= \begin{bmatrix}
    -15.25 & 15.37 \\
    37.33 & -43.20 \\
    -29.93 & 35.18
    \end{bmatrix}\,,
\end{align}
which were obtained by fitting 100 EoSs that obey physical constraints~\cite{Carson:2019rjx}.

Using the binary Love relations, we can then sample the waveform on all astrophysical parameters plus just $\Lambda_s$, reducing the dimensionality of the parameter space to 12. Moreover, from the sampled value of $\Lambda_s$, we can also compute $\Lambda_a$ from the binary Love relations, and from these two quantities, we can recover $\Lambda_1$ and $\Lambda_2$. All of this, however, requires that we choose a prior for $\Lambda_s$. We here choose an uniform prior in $(10, 10^4)$. However, for the set of EoSs used to generate the binary Love relations, $\Lambda_s$ and $q$ are also related by the approximate inequality,
\begin{equation}
    q \geq 1.2321 - 0.124616\ln(\Lambda_s)\,,
\end{equation}
which can be obtained by fitting data from Ref. \cite{Carson:2019rjx}. Therefore, any point that does not satisfy the above constraint does not pass the prior and is rejected. 

\begin{figure}
    \centering
    \includegraphics[width=\linewidth]{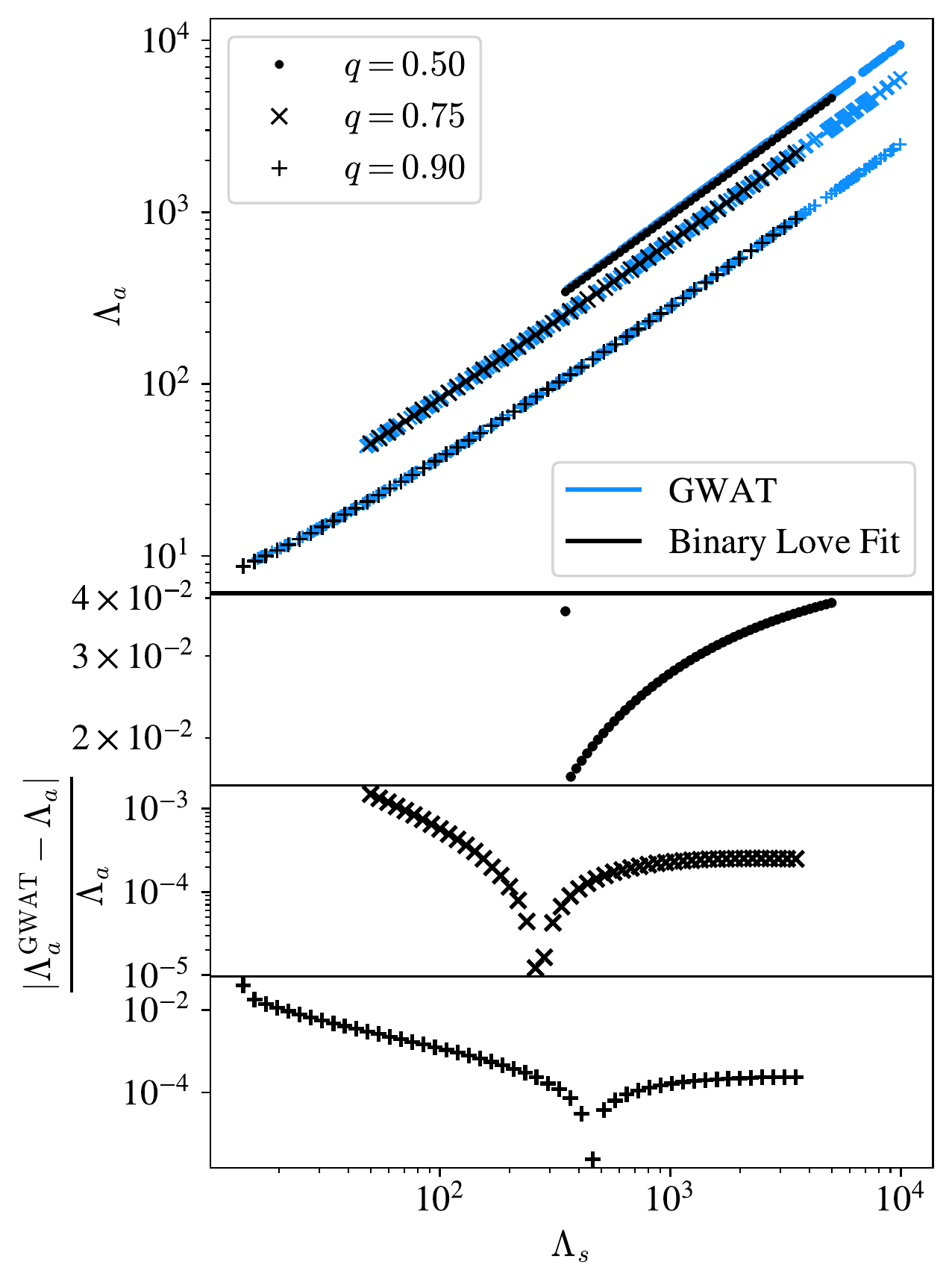}
    \caption{Comparison between the binary Love relation implemented in {\tt GWAT} (points in blue) and that computed in \cite{Carson:2019rjx} (points in black) for three different values of $q$. Beneath, the relative fractional differences (for $q=0.50, \1 q=0.75,$ and $q=0.90$ respectively) demonstrate that the {\tt GWAT} implementation is correct. }
    \label{fig:binLove}
\end{figure}

To validate our {\tt GWAT} implementation of the binary Love relations, we computed $\Lambda_a(\Lambda_s, q)$ for three different values of $q = \{ 0.5, 0.75, 0.9\}$ and 250 randomly generated values of $\Lambda_s$ each. Figure\ \ref{fig:binLove} compares our results to the data published in \cite{Carson:2019rjx}. Observe that the relative fractional difference is below 5\% in all cases, which confirms that our implementation is correct. 

Given the agreement between our code and previous work, we conclude that our {\tt GWAT} implementation of {\tt IMRPhenomD\_NRT} can successfully perform parameter estimation for BNS inspirals, sampling on the symmetric tidal deformability. 

\subsection{Extending the {\tt GWAT} Implementation of BNS Waveform Models in GR to Einstein-\ae ther Theory}
\label{subsec:EA_in_code}

With the GR groundwork in place, we now implement Einstein-\ae ther modifications to the {\tt IMRPhenomD\_NRT} model, thus generating the {\tt EA\_IMRPhenomD\_NRT} model. We will describe here what these modifications are and how we will implement them in {\tt GWAT}. 

As we discussed in Sec.~\ref{subsec:responsefunc} the Einstein-\ae ther modifications to the inspiral part of coalescence include corrections to the amplitude and phase of the Fourier transform of the plus and cross GW polarizations, as well as the introduction of 
the Fourier transform of the four additional GW polarizations present in this theory (Eqs.~\eqref{eqn:hplus}--\eqref{eqn:l1phase}). 
We extend the  {\tt IMRPhenomD\_NRT} model by introducing these modifications to the inspiral portion of coalescence. 
\edit{Beginning at the} merger, a Planck taper function takes the amplitude of the response function to zero, ending the waveform model, because both the {\tt IMRPhenomD\_NRT} and the {\tt EA\_IMRPhenomD\_NRT} models do not include the merger or post-merger portions of coalescence \edit{for NSs}\footnote{\edit{The EoS of NSs is not yet known, so there are different possible outcomes of a binary NS merger including stable NSs, hyper-massive NSs, supra-massive NSs, and BHs\cite{Gao:2015xle}. Thus, any model of the merger or post-merger portions of coalescence is not accurate for NS binaries.}}. 

Since the {\tt EA\_IMRPhenomD\_NRT} model is new, there does not yet exist any other code infrastructure that has implemented Einstein-\ae ther modifications to a coalescence model. We therefore implemented it all within the {\tt GWAT} code as follows. 
Given a point in the 16-dimensional parameter space of 
\begin{align*}
    \vec{\theta} &= \{\alpha', \sin\delta, \psi, \cos\iota, \phi_{\text{ref}}, t_c, D_L, \mathcal{M}, \eta, \chi_1, \chi_2, \Lambda_s,\\
    &\hspace{8mm}c_a, c_\theta, c_\omega, c_\sigma\},
\end{align*}
the code first computes sensitivities, since they play a prominent role in all of the Einstein-\ae{}ther modifications discussed above. The logic for this calculation is outlined in Fig.~\ref{fig:code_logic_cartoon} and proceeds as follows.
From the symmetric combination of the tidal deformabilities $\Lambda_s$, the code uses the binary Love relations to find the antisymmetric combination of the tidal deformabilities $\Lambda_a$, and from these two quantities, the individual tidal deformabilities $\Lambda_1$ and $\Lambda_2$ (see discussion in Sec.~\ref{subsec:NRTmodifications}). From the latter two, the code uses the C-Love relations to compute the individual compactnesses $C_1$ and $C_2$ (see Sec.~\ref{subsec:sensitivities}). Finally, from the compactnesses and the Einstein-\ae ther coupling constants, the code computes the sensitivities $s_1$ and $s_2$ (see  Eqs.~\eqref{eqn:sensitivity} and \eqref{eqn:binding_energy}).

\begin{figure*}
    \centering
    \includegraphics[width=0.49\linewidth]{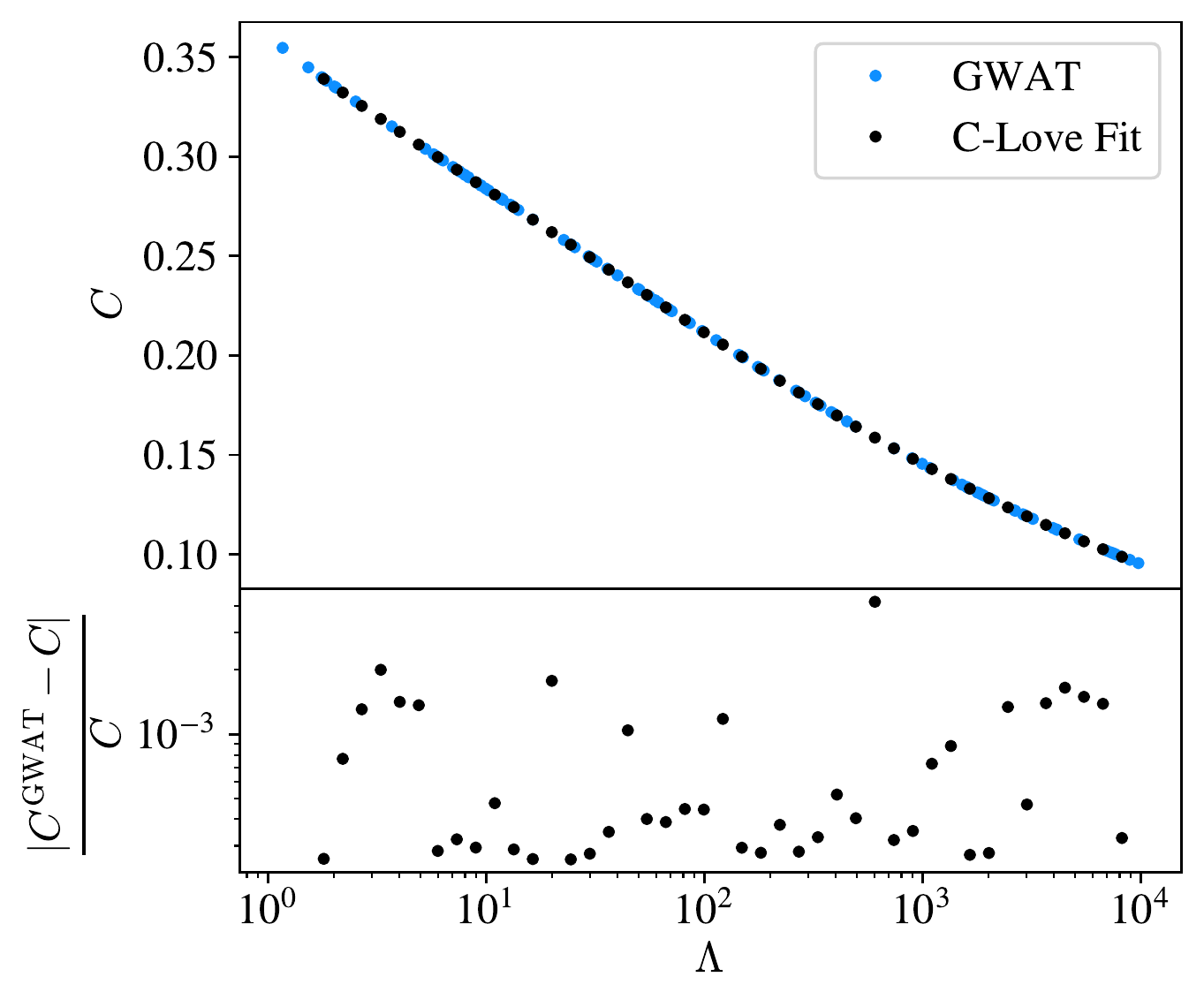}
    \includegraphics[width=0.49\linewidth]{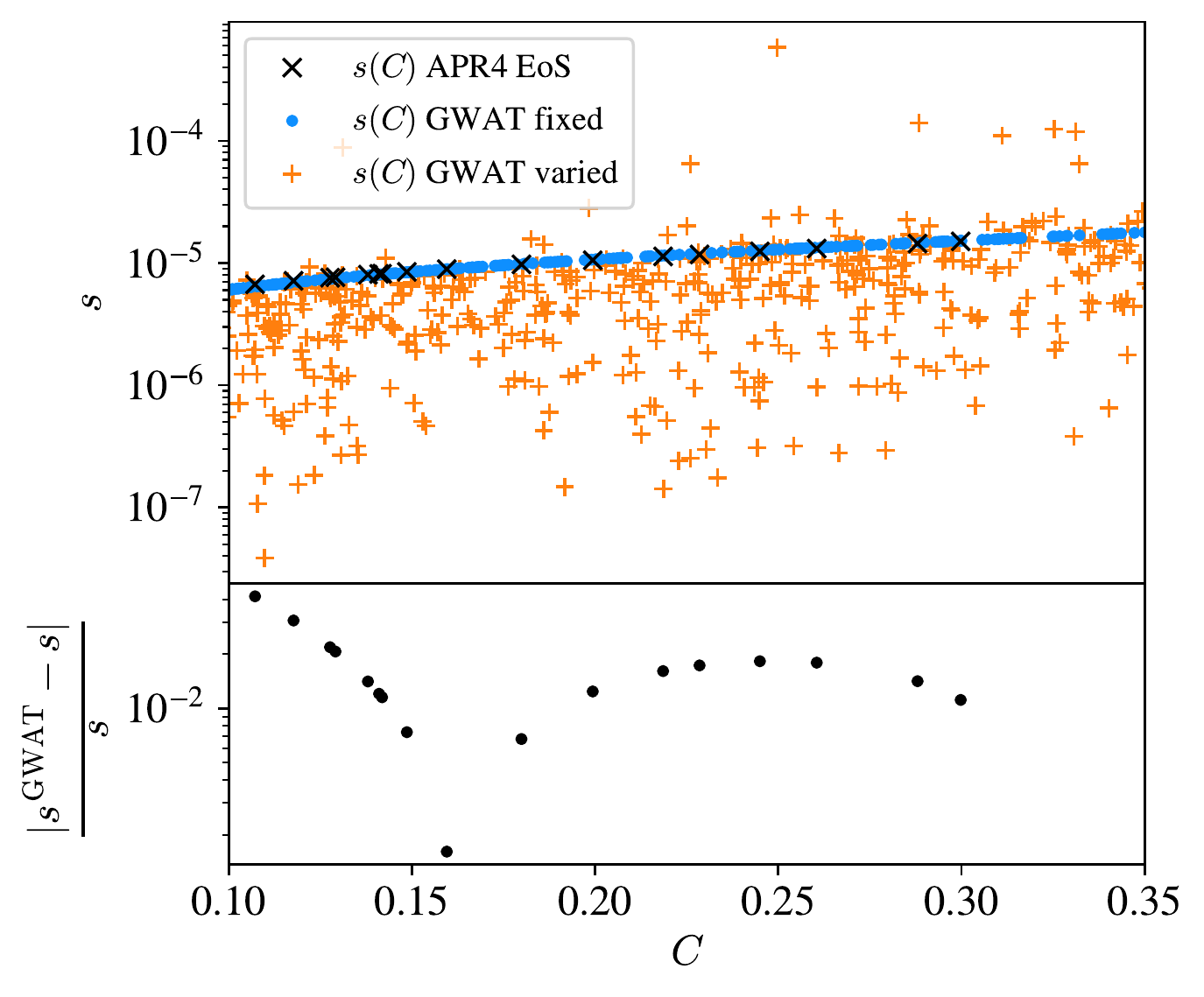}
    \caption{Left: Compactness as a function of $\Lambda$ computed by {\tt GWAT} for 100 random combinations of source parameters and compared to data from~\cite{Carson:2019rjx}. The relative fractional difference between these two data sets is plotted below and serves as a test of the C-Love relations in our code.
    Right: Comparing the sensitivity as a function of compactness computed by {\tt GWAT} with that published in~\cite{Gupta:2021vdj}. For direct comparison, we follow the example of~\cite{Gupta:2021vdj} and fix the Einstein-\ae{}ther coupling constants to $\{c_a, c_\theta, c_\omega, c_\sigma\} = (10^{-4}, 4\times10^{-7}, 10^{-4}, 0)$, plotted in blue. The relative fractional difference between these points and those from~\cite{Gupta:2021vdj} is shown below. Though these points are computed using different EoSs (APR4 in~\cite{Gupta:2021vdj} and Tolmann VII  in {\tt GWAT}), they differ by less than 5\% for realistic values of compactness for NSs. We also compute sensitivity as a function of compactness varying the Einstein-\ae{}ther parameters in the full range of parameter space allowed by current constraints (for a description of this allowed region, see Sec.~\ref{sec:current_constraints}). These points are plotted in orange and represent the typical values of sensitivity we expect to appear in the waveform.}
    \label{fig:CLove_sC}
\end{figure*}

For validation purposes, the inverse of the Love-C relation, $C(\Lambda)$ as computed by the {\tt GWAT} implementation, is shown in the left panel of Fig.\ \ref{fig:CLove_sC} for 100 random tidal deformabilities (ranging between 1 and $10^4$). 
Comparing this to the data from \cite{Carson:2019rjx}, we can compute the relative fractional difference, shown in the left-bottom panel of Fig.\ \ref{fig:CLove_sC}. Observe that the relative fractional difference is at most $0.5\%$, due mostly to interpolation error.

The s-C relation, $s(C)$, as computed by the {\tt GWAT} implementation, is plotted in the right panel of Fig.~\ref{fig:CLove_sC}. First, for direct comparison to~\cite{Gupta:2021vdj}, we fix the Einstein-\ae{}ther coupling constants to $\{c_a, c_\theta, c_\omega, c_\sigma \} = (10^{-4}, 4 \times 10^{-7}, 10^{-4}, 0)$ and compute sensitivity for 250 random values of compactness. 
As before, the relative fractional difference between the {\tt GWAT} sensitivities and that of the original paper are shown in the right-bottom panel of Fig.~\ref{fig:CLove_sC}. Observe again the relative fractional difference is at most $5\%$, once more validating our implementation\footnote{Note that though there is good agreement in the range of compactnesses relevant for this study, this agreement does not hold in the small $C$ limit. As described in Sec.~\ref{subsec:sensitivities}, the sensitivity calculation depends on the Tolmann VII EoS. While this analytic EoS is physically reasonable for realistic NS compactnesses, the justification for this model breaks down for very small $C$.}. 
We then compute sensitivity for 500 random values of compactness when the Einstein-\ae{}ther parameters, $\{c_a, c_\theta, c_\omega, c_\sigma\}$, are also varied.  
These coupling constants are randomly drawn from the complicated region of parameter space allowed by current constraints on the theory (described in detail in Sec.~\ref{sec:current_constraints}).  
Note the wide range of sensitivities possible for a single compactness when these coupling constants are varied. 
Furthermore, Appendix~\ref{sec:appendixSensitivities} discusses the magnitude of sensitivities in a wider region of parameter space that will become useful later.

Once the sensitivities have been evaluated, we can then proceed to evaluate all of the other Einstein-\ae ther quantities that appear in the Fourier transform of the response function. Explicitly, this includes the quantities $\{c_S, c_V, \beta_1, Z, \mathcal{S}, \mathcal{A}_1, \mathcal{A}_2, \mathcal{A}_3, \mathcal{C}, \kappa_3, \epsilon_x, \bar{\mathcal{M}}\}$ as defined in Eqs.~\eqref{eqn:scalarSpeed},\eqref{eqn:vectorSpeed},\eqref{eqn:beta1}, \eqref{eqn:Z}--\eqref{eqn:mass_weighted_sensitivity}, and \eqref{eqn:A1}--\eqref{eqn:sensitivity_chirpmass}. With these Einstein-\ae{}ther quantities computed, the response function can be put together by first evaluating the amplitude and phase of each of the GW polarizations on a frequency array, and then linearly combining the product of the latter with the antenna patterns. 
    
We want to take advantage of the full machinery of {\tt IMRPhenomD\_NRT} that has already been successfully implemented in {\tt GWAT}. Thus, we promote the chirpmass, $\mathcal{M}$, to the Einstein-\ae{}ther scaled version, $\bar{\mathcal{M}}$ (Eq.~\eqref{eqn:sensitivity_chirpmass}), everywhere in {\tt IMRPhenomD\_NRT}. We then use this waveform template to compute the amplitude, $A_{NRT}(f)$, and phase, $\Psi_{NRT}(f)$, of the plus and cross GW polarizations such that
\begin{align}
    \tilde{h}_{+,NRT} (f) = A_{NRT}(f) (1 + \cos^2\iota) e^{i \Psi_{NRT}(f)}, 
    \\
    \tilde{h}_{\times,NRT}(f) = A_{NRT}(f) (2i\cos\iota) e^{i \Psi_{NRT}(f)}.
\end{align}
This introduces uncontrolled remainders at higher orders. However, since the Einstein-\ae{}ther waveform has not yet been computed to those orders, it is reasonable to use the ``promoted'' {\tt IMRPhenomD\_NRT} version for higher order terms. Note that the {\tt EA\_IMRPhenomD\_NRT} waveform template \textit{is only accurate to 0PN}\footnote{\edit{For the PN accuracy of the {\tt IMRPhenomD\_NRT} model, see~\cite{Dietrich:2019kaq, Kawaguchi:2018gvj, Husa:2015iqa, Khan:2015jqa}.}}. 

Now we are ready to construct the amplitude and phase of all of the different GW polarizations in Einstein-\ae{}ther theory in {\tt GWAT} for {\tt EA\_IMRPhenomD\_NRT}. We will do this by adding the appropriate corrections to the already computed $A_{NRT}$ and $\Psi_{NRT}$. First for the plus and cross modes, 
\begin{align}
    \tilde{h}_{+,EA} (f) = A_{EA}(f) (1 + \cos^2\iota) e^{i \Psi_{EA}(f)}, 
    \\
    \tilde{h}_{\times,EA}(f) = A_{EA}(f) (2i\cos\iota) e^{i \Psi_{EA}(f)},
\end{align}
where 
\begin{align}
    A_{EA} (f) &= A_{NRT}(f) + A_{(2)}(f) - A_{0PN}(f), \label{eqn:AEA}\\
    \Psi_{EA}(f) &= \Psi_{NRT}(f) + \Psi_{(2)}(f) - \Psi_{0PN}(f) + \Psi_{c_N}(f).\label{eqn:PsiEA}
\end{align}
$A_{NRT}$ and $\Psi_{NRT}$ are the amplitude and phase computed by {\tt IMRPhenomD\_NRT} as described above. $A_{(2)}$ and $\psi_{(2)}$ are given in Eqs.~\eqref{eqn:l2amp} and~\eqref{eqn:l2phase}. $A_{0PN}$ and $\Psi_{0PN}$ are the 0PN contributions present in both {\tt IMRPhenomD\_NRT} and $A_{(2)}, \Psi_{(2)}$ respectively that are subtracted off so as not to be double counted. Explicitly, 
\begin{align}
    A_{0PN}(f) &= -\sqrt{\frac{5\pi}{96}} \frac{1}{D_L} G_N^2 \bar{\mathcal{M}}^2 \left( G_N \pi \bar{\mathcal{M}}f \right)^{-7/6},\\
    \Psi_{0PN}(f) &= \frac{3}{128} \left(G_N \pi \bar{\mathcal{M}}f \right)^{-5/3} + 2\pi f \bar{t}_c \nonumber \\
    &\2\2 - 2 \Phi(t_c) - \frac{\pi}{4}.
\end{align}
Finally, $\Psi_{c_N}$ is a term that depends on the speed of the GW polarization,
\begin{equation}
    \Psi_{c_N} \equiv -2\pi f D_L(1 - c_N^{-1}) \label{eqn:psicN}
\end{equation}
for $N \in \{T, S, V\}$. Since the plus and cross modes are tensor polarizations, Eqs.~\eqref{eqn:hplus} and~\eqref{eqn:hcross} show that the $\Psi_{c_N}$ term should be $-2\pi f D_L(1 - c_T^{-1})$. 

{\tt EA\_IMRPhenomD\_NRT} similarly computes the other terms in the response function that come from the second harmonic of the orbital period ($\tilde{h}_{N,2}$ with $N \in \{b, L, X, Y\}$ from Eqs.~\eqref{eqn:l2hb}-\eqref{eqn:l2hY}). For example, following Eq.~\eqref{eqn:l2hb},
\begin{align}
    \tilde{h}_{(b,2)} &= A_{EA} \left[ \frac{1}{2 - c_{a}} \left(3c_{a} (Z - 1)-\frac{ 2 \mathcal{S}}{c_S^2} \right) \sin^2 \iota \right] e^{i\Psi_{EA}},
\end{align}
where $A_{EA}$ and $\Psi_{EA}$ are defined as in Eqs.~\eqref{eqn:AEA} and~\eqref{eqn:PsiEA}, with $\Psi_{c_N} = -2\pi f D_L(1 - c_S^{-1})$. For each of the two scalar modes, $\tilde{h}_{(b,2)}$ and $\tilde{h}_{(L,2)}$, $\Psi_{c_N}$ depends on the scalar speed, $c_S$, and likewise for each of the two vector modes $\tilde{h}_{(X,2)}$ and $\tilde{h}_{(Y,2)}$, $\Psi_{c_N}$ depends on the vector polarization speed, $c_V$.

For the terms that come from the first harmonic of the orbital period ($\tilde{h}_{N,1}$ with $N \in \{b, L, X, Y\}$ from Eqs.~\eqref{eqn:l1hb}-\eqref{eqn:l1hY}), {\tt EA\_IMRPhenomD\_NRT} computes a new amplitude and phase, $A_{EA,1}$ and $\Psi_{EA,1}$. Since there is no $\ell=1$ component of amplitude in {\tt IMRPhenomD\_NRT}, $A_{EA,1}$ is simply equivalent to $A_{(1)}$ as defined in Eq.~\eqref{eqn:l1amp}. Meanwhile, 
\begin{align}
    \Psi_{EA,1}(f) &= \Psi_{NRT}(f/2) + \Psi_{(1)}(f) - \Psi_{0PN,1}(f) \nonumber \\
    &\2\2+ \Psi_{c_N}(f)
\end{align}
where 
\begin{align}
     \Psi_{0PN,1}(f) &= \frac{3}{256} \left(2 G_N \pi \bar{\mathcal{M}}f \right)^{-5/3} + 2\pi f \bar{t}_c \nonumber\\
     &\2\2- \Phi(t_c) - \frac{\pi}{4}
\end{align}
and $\Psi_{c_N}$ is defined the same as in the $\ell=2$ case (Eq.~\eqref{eqn:psicN}). 

Finally, {\tt EA\_IMRPhenomD\_NRT} linearly combines each $\tilde{h}_{N,\ell}$ with the appropriate antenna pattern function, $F_{N}$, to construct the full waveform. 
In the limit that the Einstein-\ae ther coupling constants go to zero\footnote{Setting the coupling constants identically to zero can lead to {\tt nan}s in the code because of the many instances of {\tt nans} in the mathematical expressions due to $0/0$ numerical problems. In order to take the GR limit without introducing {\tt nan}s, we set the coupling constants to very small values: $c_a = 1.0\times10^{-30}, c_\theta = 2\times 10^{-30}, c_\omega = 2\times 10^{-30}, c_\sigma = 0$.}, {\tt EA\_IMRPhenomD\_NRT} reduces to {\tt IMRPhenomD\_NRT}. We demonstrate this by comparing the two waveform templates for 100 randomly generated combinations of source parameters, varying each of the parameters in the 16-dimensional parameter space \textit{except for} the Einstein-\ae{}ther coupling constants, which are fixed to small values. 
We draw these parameters from the same priors described in Secs.~\ref{subsec:GWATBaseline} and~\ref{subsec:NRTmodifications}. The relative fractional difference in the amplitude and the difference in the phase are below 0.001\% and constant across frequency. Hence, we conclude that our Einstein-\ae ther waveform template is consistent with GR in the limit that the coupling constants go to zero.

\section{Current Constraints on the theory}
\label{sec:current_constraints}
Several theoretical and experimental results have placed constraints on Einstein-\ae ther theory and its coupling constants. In this section, we discuss the most stringent constraints so that we can use them to construct non-trivial priors for each of the Einstein-\ae ther parameters in two separate parameterizations of the theory. We also explain why the second parameterization is more convenient for analysis of GW data and will be used throughout the rest of this work. 
\subsection{Summary of Existing Constraints}
\label{subsec:summary-of-constraints}

Let us begin with theoretical constraints. In order to avoid gradient instabilities and ghosts, the squared speed of the GW polarizations must be positive \cite{Jacobson:2004ts, Garfinkle:2011iw},
\begin{equation}
c_T^2 > 0,\hspace{5mm} c_V^2 > 0,\hspace{5mm} c_S^2 > 0.
\label{eqn:positiveSpeeds} 
\end{equation}
Furthermore, if we consider a plane wave solution of the linearized field equations with wave vector $(k_0, 0, 0, k_3)$, the energy densities of the different modes  
\cite{Eling:2005zq, Foster:2006az}
\begin{subequations}
\begin{align}
\mathcal{E}_{T} &= \frac{1}{8\pi G} k_3^2 |A|^2, \\
\mathcal{E}_{V} &= \frac{1}{8\pi G} k_3^2 |A|^2 \frac{c_\sigma + c_\omega (1 - c_\sigma)}{1 - c_\sigma}, \label{eqn:vectorEnergyDensity} \\
\mathcal{E}_{S} &= \frac{1}{8\pi G} k_3^2 |A|^2 c_a (2 - c_a), \label{eqn:scalarEnergyDensity}
\end{align}
\end{subequations}
must be positive. Since $c_T^2 > 0 \Rightarrow (1 - c_\sigma) > 0$, Eqs.~\eqref{eqn:vectorEnergyDensity} and \eqref{eqn:scalarEnergyDensity} immediately imply
\begin{subequations}
\begin{align}
&c_\omega \geq -\frac{c_\sigma}{1 - c_\sigma}, \\
0 \leq &c_a \leq 2,
\end{align}
\label{eqn:positiveEnergyCond}%
\end{subequations}
respectively. We refer to Eqs.~\eqref{eqn:positiveSpeeds} and \eqref{eqn:positiveEnergyCond} together as the \textit{stability conditions}, since they are both required to have stable Einstein-\ae ther GWs. 

Now we turn to constraints on the Einstein-\ae ther parameters due to experimental results. The most stringent of these constraints comes from the simultaneous observation of GWs from a NS binary merger and the corresponding short gamma ray burst, GW170817 and GRB170817A. This event placed observational bounds on the speed of the tensor polarizations of GWs: $-3 \times 10^{-15} < c_T - 1 < 7 \times 10^{-16}$ \cite{LIGOScientific:2017zic}. Given the simple dependence of $c_T^2$ on $c_\sigma$, these observations restrict $c_\sigma \approx \mathcal{O}(10^{-15})$. Thus, we will henceforth set $c_\sigma = 0$, dramatically simplifying many of the expressions and reducing the total parameter space from 16 to 15. 

Another observational bound on Einstein-\ae{}ther theory derives from the observation of high-energy cosmic rays. In Einstein-\ae ther theory, the amount of energy atmospheric cosmic rays have is higher than that in GR because GWs and \ae ther field excitations can endow cosmic rays with more energy through a gravitational ``Cherenkov type'' process \cite{Elliott:2005va}. By considering the amount of energy observed in high energy cosmic rays, one can place an upper limit on how efficient this Cherenkov process can be, further constraining the coupling constants of Einstein-\ae ther theory. This was done separately for tensor-like, vector-like, and scalar-like excitations, assuming that all speeds $c_N$ (with $N = T, V, S$) are subluminal. 
The constraints obtained in \cite{Elliott:2005va} with these assumptions are very strict and we will refer to them hereafter as the \textit{Cherenkov constraints}. They are often summarized in the literature (~\cite{Sarbach:2019yso, Gupta:2021vdj} and others) as\footnote{\edit{Note also that $c_N > 1$ is allowed. This does not violate causality in Lorentz-violating theories such as Einstein-\ae{}ther theory.}} 
\begin{equation}
    c_N^2 \gtrsim 1 - \mathcal{O}(10^{-15}) \,,
    \label{eqn:Cherenkov}
\end{equation}
because the constraints give very strict conditions on $\{c_a, c_\theta, c_\omega, c_\sigma\}$ that must be satisfied if $c_N^2 < 1$. It is very challenging, \textit{though not impossible}, to pick a point in parameter space that satisfies the latter. For a more careful summary of what the constraints are and how we applied them in our code, see Appendix~\ref{sec:appendixCherenkov}. 

Another constraint on Einstein-\ae{}ther theory derives from Big Bang Nucleosynthesis (BBN). The Lorentz-violating \ae ther field of Einstein-\ae ther theory rescales the effective value of Newton's constant that appears in the Friedman equation \cite{Carroll:2004ai, Mattingly:2001yd, Jacobson:2007veq}, 
\begin{equation}
    G_{\text{cosmo}} = \frac{G_N(1 - c_a/2)}{1 + c_\theta/2}.
    \label{eqn:Gcosmo}
\end{equation}
However, observations of primordial $^4$He from BBN restrict \cite{Carroll:2004ai}
\begin{equation}
    \Big| \frac{G_{\text{cosmo}}}{G_N} - 1 \Big| \lesssim \frac{1}{8}.
\end{equation}
Inserting Eq.~\eqref{eqn:Gcosmo} into this requirement and simplifying leads to the two inequalities 
\begin{subequations}
\begin{align}
    c_\theta + \frac{8c_a}{7} &\lesssim \frac{2}{7}, \label{eqn:BBNgeneralpos}\\
    c_\theta + \frac{8c_a}{9} &\gtrsim -\frac{2}{9}.
    \label{eqn:BBNgeneralneg}
\end{align}
\end{subequations}
This constraint becomes simpler in certain regions of parameter space, as will be described in the next section.

There are three more experimental constraints that should be discussed here, all of which lead to bounds on the preferred-frame PN parameters $\alpha_1$ and $\alpha_2$, which were defined in Eq.~\eqref{eqn:alphas}. With the constraint that $c_\sigma = 0$, these parameters simplify to \begin{subequations}
\begin{align}
\label{eq:alpha1_simped}
\alpha_1 &= -4 c_a , \\
\alpha_2 &= -2 c_a + \frac{3c_a(c_\theta + c_a)}{(2 - c_a)c_\theta}. 
\end{align}
\end{subequations}
Two of the constraints arise from solar system observations. The first one comes from the close alignment of the solar spin axis with the total angular momentum vector of the solar system, which restricts \cite{solarSystem2}
\begin{equation}
    |\alpha_2| \lesssim 4 \times 10^{-7}.\label{eqn:solarsystem1}
\end{equation} 
The second one comes from lunar laser ranging observations, which bound $  -1.6 \times 10^{-4}<\alpha_1 < 2 \times 10^{-5}$ to (1-$\sigma$)~\cite{Muller:2005sr}; for simplicity, this bound can be conservatively stated as 
\begin{equation} 
|\alpha_1| \lesssim 10^{-4} \label{eqn:solarsystem2}
\end{equation} 
as done in several previous papers \cite{Will:2014kxa, Sarbach:2019yso, Gupta:2021vdj}. This choice will not affect our results (as discussed later). The bounds in Eqs.~\eqref{eqn:solarsystem1} and~\eqref{eqn:solarsystem2} will be referred to as \textit{solar system constraints}. 
 Finally, combining these constraints with observations of the damping of the orbital period of certain binary pulsars and the triple binary pulsar places the even tighter bound \cite{Gupta:2021vdj}: 
\begin{equation}
-1.6 \times 10^{-5} \lesssim \alpha_1 \lesssim 4.6 \times 10^{-6}
     \label{eqn:binarypulsarconstraint}
\end{equation} 
to 1-$\sigma$ uncertainty.

\subsection{Priors on ($\mathbf{c_a,c_\theta,c_\omega}$) from existing constraints}
\label{subsec:c_prior}
Now that we have introduced all of the main constraints on the theory in the previous subsection, let us now study how they lead to a prior on the coupling constant parameter space of Einstein-\ae ther theory. One way to do so is via \textit{rejection sampling} of the constraints, i.e.~to evaluate a given constraint millions of times by sampling uniformly on $\{c_a, c_\theta, c_\omega \}$ and rejecting those choices of these parameters that violate the given constraint. We will start by sampling each of these parameters in the arbitrarily chosen region $[-3,3]$ and show how the parameter space shrinks with the addition of constraints. 

\begin{figure*}
\includegraphics[width=0.49\linewidth,clip=true]{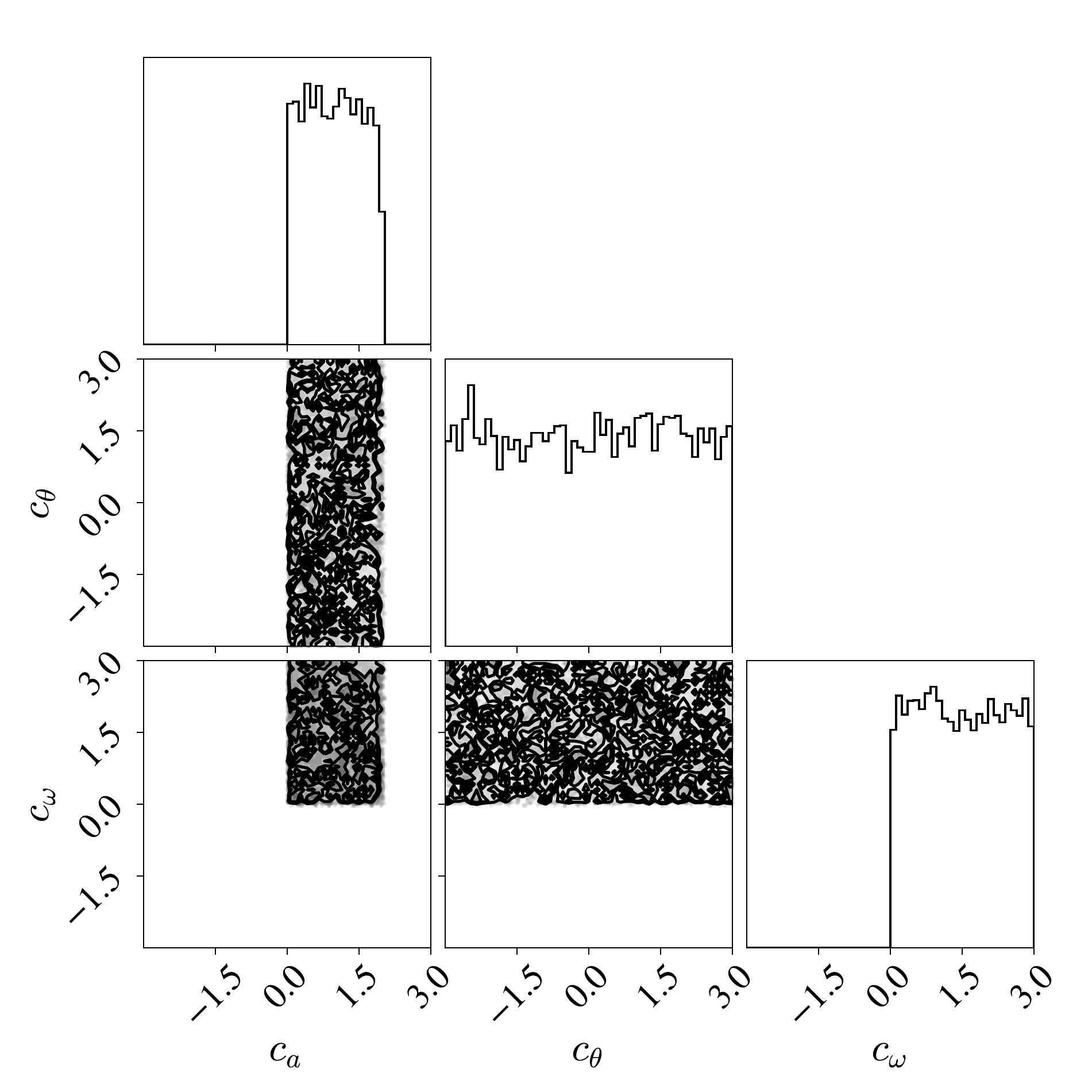}
\includegraphics[width=0.49\linewidth,clip=true]{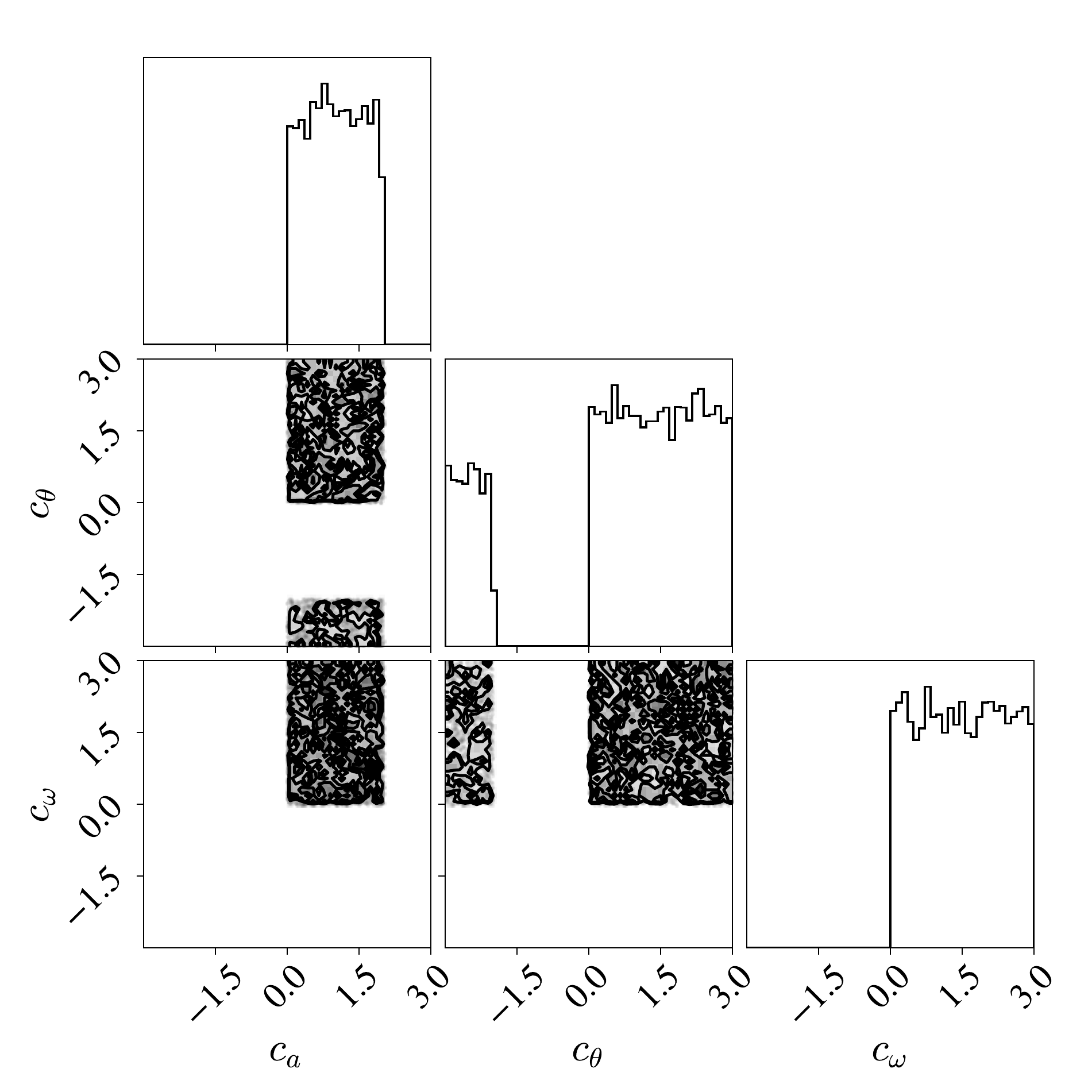} \\
\includegraphics[width=0.49\linewidth,clip=true]{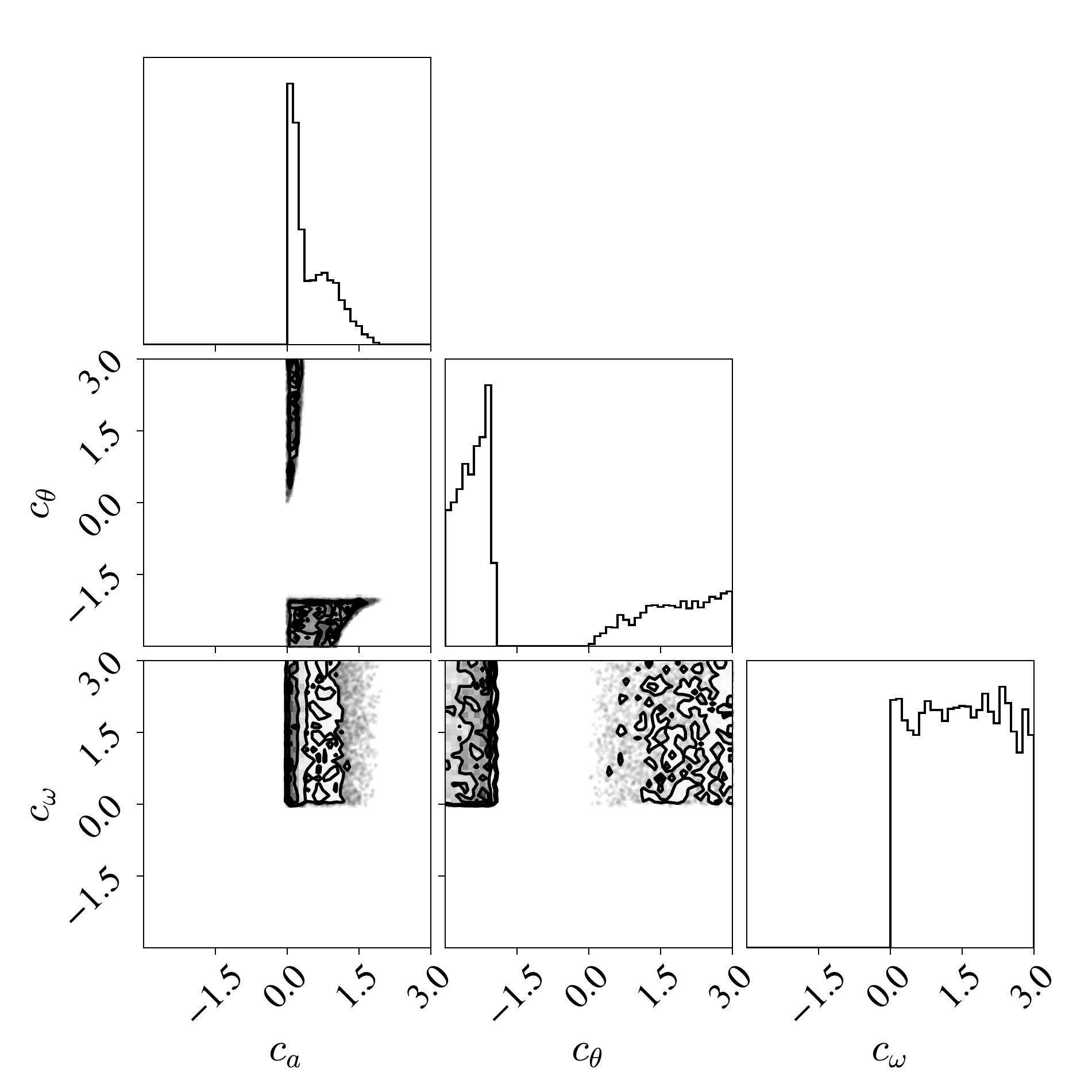}
\includegraphics[width=0.49\linewidth,clip=true]{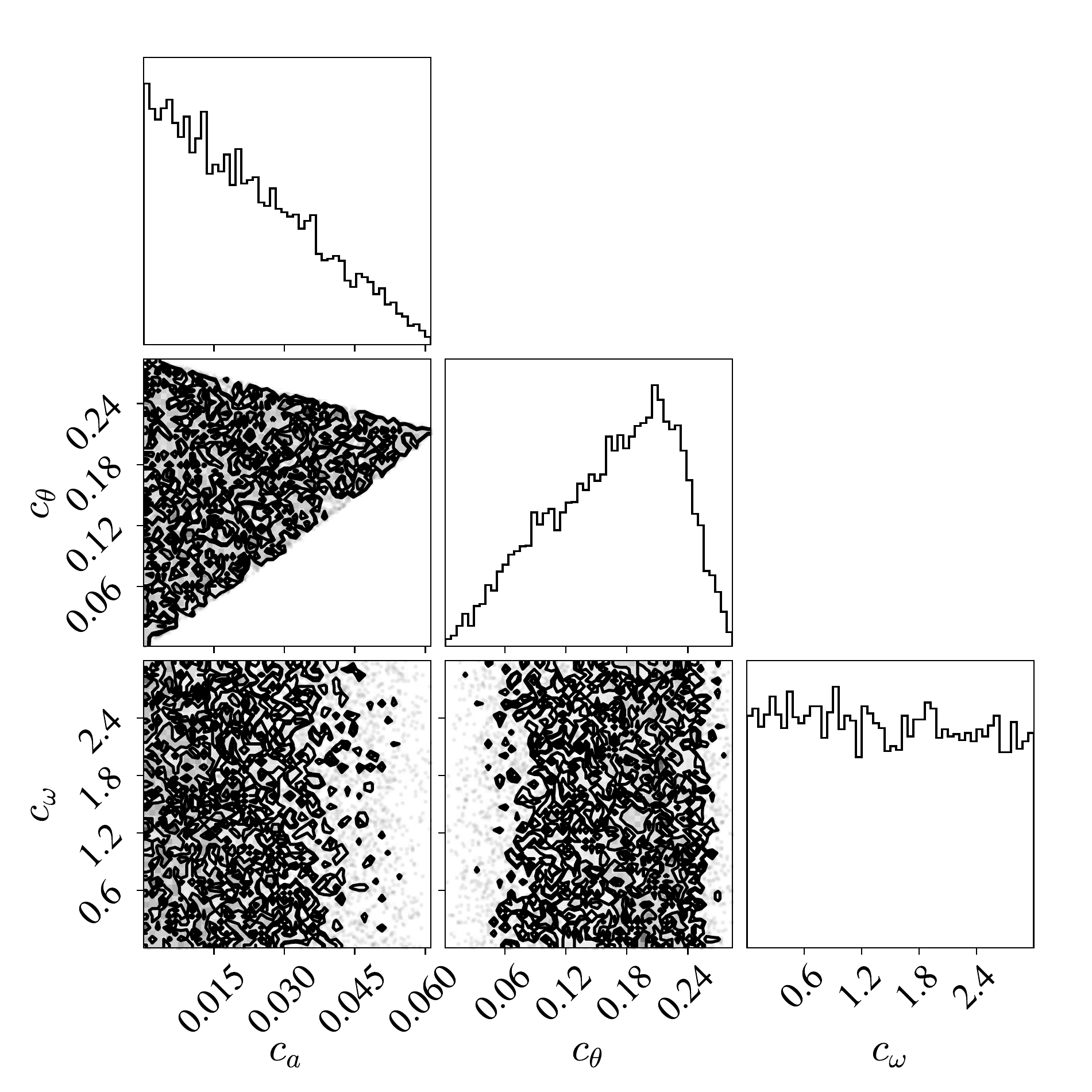}
\caption{Plots demonstrating the effect of successively adding current constraints on Einstein-\ae ther theory to the prior in the c parameterization. Each parameter was sampled uniformly in the region $[-3, 3]$ (the bottom right panel is shown in a smaller range simply so that it is visible). Points that did not obey these constraints were rejected. The constraints were applied in the following order (beginning in the top left corner and ending in the bottom right corner): positive energy conditions, Eq.~\eqref{eqn:positiveEnergyCond}; positive speeds of different GW polarizations, Eq.~\eqref{eqn:positiveSpeeds}; Cherenkov constraint, Eq.~\eqref{eqn:Cherenkov}; BBN constraint, Eq.~\eqref{eqn:BBN}.}
\label{fig:c_beforeSS}
\end{figure*}

Let us first focus on the stability constraints. Equation~\eqref{eqn:positiveEnergyCond} requires that $c_a$ be restricted to the range $[0,2]$ and $c_\omega$ be positive, as shown in the top left panel of Fig.\ \ref{fig:c_beforeSS}, which we generated via rejection sampling. Similarly, Eq.~\eqref{eqn:positiveSpeeds} disallows $c_\theta \in (-2,0)$ because, from Eq.~\eqref{eqn:scalarSpeed} with $c_\sigma = 0$,
\begin{equation}
    c_S^2 = \frac{c_\theta(1 - c_a/2)}{3c_a(1 + c_\theta/2)},
\end{equation}
which by Eq.~\eqref{eqn:positiveSpeeds} must be positive.  Equation~\eqref{eqn:positiveEnergyCond} required already that $c_a \in [0,2]$ and this implies that $(1 - c_a/2)/3c_a \geq 0$ always. Thus, $c_S^2 \geq 0$ requires that 
\begin{equation}
    \frac{c_\theta}{(1 + c_\theta/2)} \geq 0\,,
    \label{eqn:positivectheta}
\end{equation}
which implies $c_\theta \geq 0$ or $c_\theta < -2$, leading to the shape of the top right panel of Fig.\ \ref{fig:c_beforeSS}.

Let us now focus on the Cherenkov constraint, which through rejection sampling leads to the constraints on parameter space shown in the bottom left panel of Fig.\ \ref{fig:c_beforeSS}. To better understand these constraints, consider first the Cherenkov bound $c_S \geq 1$, which leads to 
\begin{align}
      c_S^2 &= \frac{c_\theta(1 - c_a/2)}{3c_a(1 + c_\theta/2)} \geq 1
      \Rightarrow
      \frac{1 - c_a/2}{3c_a} \geq \frac{1 + c_\theta/2}{c_\theta}\,.
      \label{eqn:cSCherenkov}
\end{align}
Using the stability restriction of Eq.~\eqref{eqn:positivectheta}, the above expression becomes
\begin{align}
    \frac{1 - 2c_a}{3c_a} &\geq \frac{1}{c_\theta}.
    \label{eqn:CherenkovInequality}
\end{align}
At this point we must keep careful track of negative signs since both $c_\theta$ and $(1-2c_a)$ can be either positive or negative in the region of parameter space considered. There are four possible combinations with their respective version of the inequality. For example, consider $c_\theta < 0$ and $c_a > 1/2$. Then Eq.~\eqref{eqn:CherenkovInequality} becomes
\begin{equation}
     \frac{1 - 2c_a}{3c_a} c_\theta \leq 1
      \Rightarrow c_\theta \geq \frac{3c_a}{1 - 2c_a} .
\end{equation}
Therefore, in the bottom right corner of the $(c_\theta, c_a)$ correlation of the bottom left panel of Fig.\ \ref{fig:c_beforeSS}, all the points accepted in our rejection sampling must fall above the line $3c_a/(1 - 2c_a)$. The other accepted points in this panel can be explained similarly. 

Let us now focus on the BBN constraints of  Eqs.~\eqref{eqn:BBNgeneralpos} and \eqref{eqn:BBNgeneralneg}. The lower bound on $c_\theta$ [Eq.~\eqref{eqn:BBNgeneralneg}] is minimized when $c_a$ is maximized, and since $c_a \in [0,2]$, this implies that 
\begin{equation}
    c_\theta \gtrsim -2 .
\end{equation}
The stability conditions, however, already required the condition $c_\theta < -2$ or $c_\theta \geq 0$ from Eq.~\eqref{eqn:positiveSpeeds}. Since $c_\theta < -2$ and $c_\theta \geq -2$ cannot simultaneously be true, we must have that $c_\theta \geq 0$. Thus, the BBN constraint becomes 
\begin{equation}
    c_\theta \geq 0, \hspace{5mm}
    c_\theta + \frac{8c_a}{7} \lesssim \frac{2}{7}. \label{eqn:BBN}
\end{equation}
Adding this BBN constraint immediately restricts any sampling to the top left corner of the $c_\theta$-$c_a$ parameter space in the bottom left panel of Fig.\ \ref{fig:c_beforeSS}, 
and adds an upper bound along the line $(2- 8c_a)/7$, resulting in the bottom right panel of Fig.\ \ref{fig:c_beforeSS}. 

Let us now finally discuss solar system constraints, dividing them into two separate cases, as described in previous work~\cite{Sarbach:2019yso, Gupta:2021vdj, Adam:2021vsk}. 
In the first case, $\alpha_1 \lesssim 10^{-4}$ (but \textit{not} $\alpha_1 << 10^{-4}$), which saturates the solar system constraint of Eq.~\eqref{eqn:solarsystem2}. In this region of parameter space, which we will denote \textit{region 1}, $c_a\approx \mathcal{O}(10^{-5})$ and $c_\theta \approx 3c_a (1 + \mathcal{O}(10^{-3}))$ in order to satisfy $\alpha_2 \lesssim 4 \times 10^{-7}$ from Eq.~\eqref{eqn:solarsystem1}. In this limit, the Einstein-\ae ther coupling constants become 
\begin{align}
    \{c_a, c_\theta, c_\omega, c_\sigma \} &= \{c_a, 3c_a(1 + \delta c_\theta), c_\omega, 0\},
    \label{eqn:constantsregion1} \\
    &\edit{\approx \{\mathcal{O}(10^{-5}), \mathcal{O}(10^{-5}), c_\omega, 0\},} \nonumber
\end{align}
where $\delta c_\theta \approx \mathcal{O}(10^{-3})$ \edit{and the only restriction on $c_\omega$ is that it is positive.}. One might wish to assume that $\delta c_\theta <<1$ and thus ignore this term and set $c_\theta = 3c_a$ exactly; however, inserting this expression into Eq.~\eqref{eqn:cSCherenkov} shows that when $c_a \neq 0$, the Cherenkov constraint, $c_S^2 \geq 1$, is no longer satisfied. In this regime, when $c_a \approx \mathcal{O}(10^{-5})$, the BBN constraint [Eq.~\eqref{eqn:BBN}] is automatically satisfied (because when $c_\theta = 3c_a$ the BBN constraint becomes $c_a \lesssim 2/29$), so previous papers did not mention it in association with this region.

Let us now discuss a second way to satisfy the solar system constraints by setting $\alpha_1 \ll 10^{-4}$. In this region of parameter space, which we will denote \textit{region 2}, Eq.~\eqref{eq:alpha1_simped} tells us that $c_a \ll 10^{-4}$ and $c_\theta$ is essentially unconstrained if one forces $c_a \lesssim 10^{-7}$, other than by the BBN constraint. In this case, the BBN constraint simplifies to $0 \leq c_\theta \leq 2/7$, which is consistent with what was reported in \cite{Sarbach:2019yso, Gupta:2021vdj}. \edit{Thus, in this limit, the Einstein-\ae{}ther coupling constants are}
\begin{align}
    \edit{\{c_a, c_\theta, c_\omega, c_\sigma \} }&\edit{= \{c_a, c_\theta, c_\omega, 0\}} 
    \label{eqn:constantsregion2} \\
    &\edit{\approx \{\mathcal{O}(10^{-7}),\mathcal{O}(10^{-1}), c_\omega, 0\} } \nonumber
\end{align} 
\edit{where the only restriction on $c_\omega$ is that it is positive.}
\edit{Notice that this equation defines a region that does not overlap with the region defined in Eq.~\eqref{eqn:constantsregion1}.}

One can show analytically that in \edit{region 2} of parameter space, $Z = 1 + \mathcal{O}(c_a), \kappa_3 = 1 + \mathcal{O}(c_a^{7/2})$, $s = \mathcal{O}(c_a)$ and $\epsilon_x = \mathcal{O}(c_a^{5/2})$, assuming a finite, nonzero $c_\theta$ and $c_\omega$, which were taken to be independent from $c_a$ for the purposes of this expansion. Furthermore, for $c_a \approx 10^{-7}$, $c_a^{5/2} \approx 10^{-18}$ and $c_a^{7/2} \approx 10^{-25}$. Therefore, these quantities barely differ from their values in the GR limit\footnote{Note that if $c_a$ is exaclty zero, the quantities $\{Z, \kappa_3, s, \epsilon_x\}$ are identical to their GR limit, even for a nonzero $c_\theta, c_\omega$. This implies that if $c_a$ were restricted to exactly zero, GW data would not be able to constrain Einstein-\ae{}ther theory.}: $Z = 1, \kappa_3 = 1, s = 0$, and $\epsilon_x = 0$.  On the other hand, in region 1 where we take $c_\theta \approx 3 c_a$, $Z = 4/3 + \mathcal{O}(c_a), \kappa_3 = 1 + \mathcal{O}(c_a), s= \mathcal{O}(c_a)$, and $\epsilon_x = \mathcal{O}(c_a)$. Recall that in region 1, $c_a \approx \mathcal{O}(10^{-5})$. Hence, the Einstein-\ae{}ther modifications to GWs in region 2 of parameter space are negligible compared to those in region 1. Therefore, for the remainder of this work, \textit{we will consider only region 1}.

\begin{figure*}
	\includegraphics[width=0.49\linewidth,clip=true]{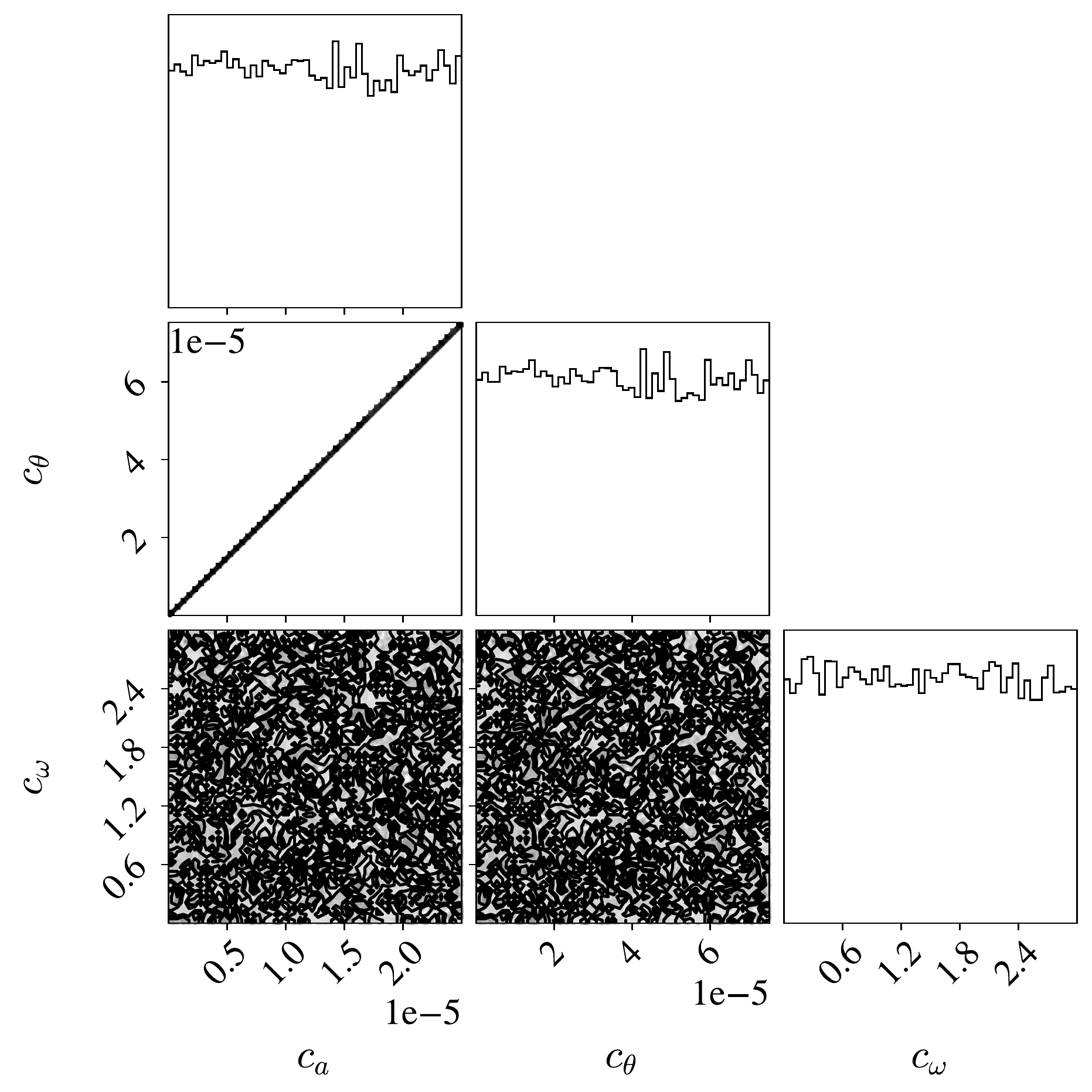}
	\includegraphics[width=0.49\linewidth,clip=true]{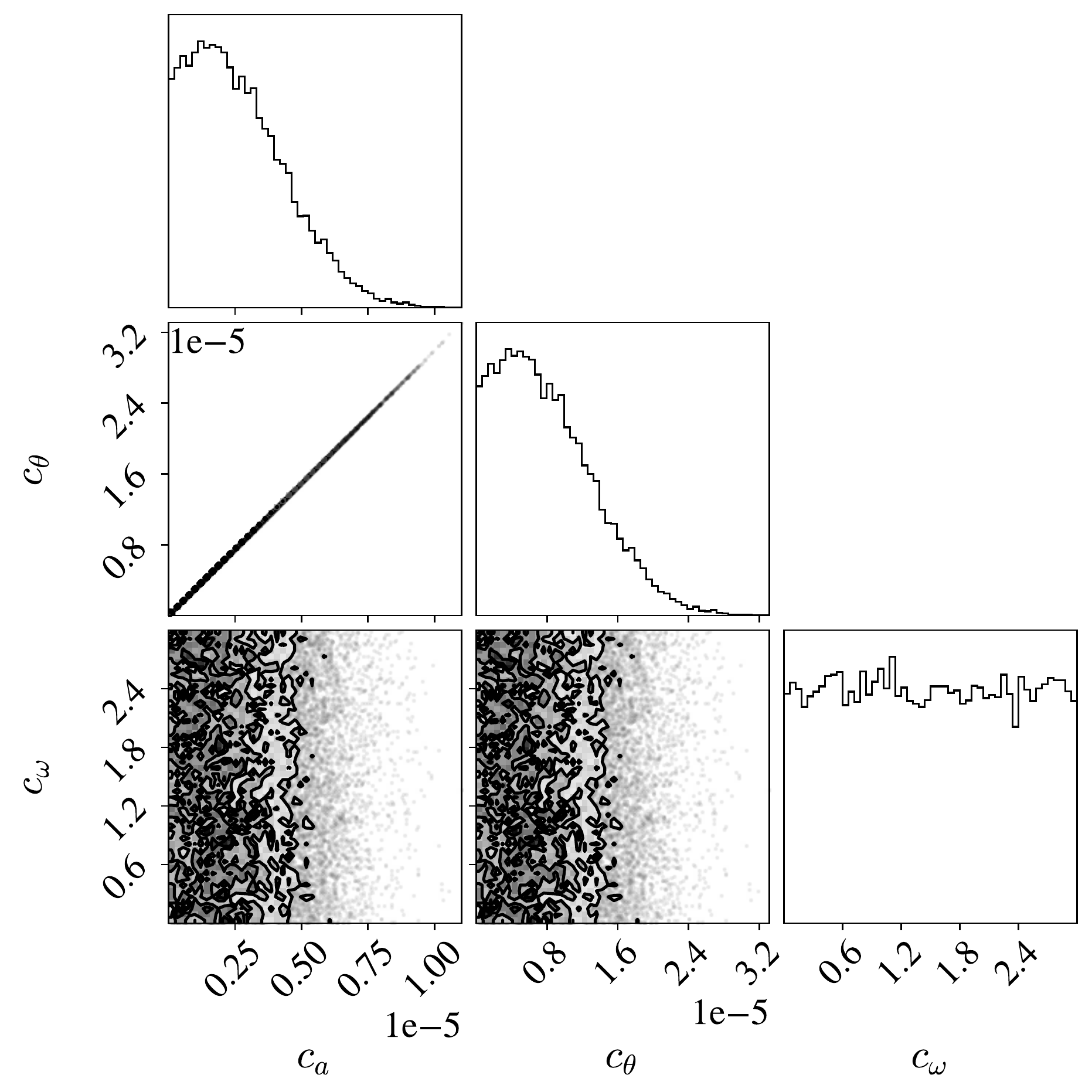}
\caption{Plots showing how the addition of the solar system and binary pulsar constraints affect the prior in region 1 of parameter space. In this region,  we sample uniformly on $c_a, \delta c_\theta,$ and $c_\omega$ as described in Eq.~\eqref{eqn:constantsregion1}. Both plots include all the constraints of Fig.\ \ref{fig:c_beforeSS} as well as the solar system constraints, Eqs.~\eqref{eqn:solarsystem1} and \eqref{eqn:solarsystem2}. The plot on the right further adds the constraint from binary pulsar and triple systems, Eq.~\eqref{eqn:binarypulsarconstraint}.} 
\label{fig:EA_priors_c_region1}
\end{figure*}

Restricting our attention to region 1\footnote{Recall that in region 1, $\alpha_1 \lesssim 10^{-4}$ but it is \textit{not} true that $\alpha_1 << 10^{-4}$.}, we examine the combined constraints. With the addition of the solar system constraints, we arrive at the left panel of Fig.\ \ref{fig:EA_priors_c_region1}.  
We can see that $c_a \approx \mathcal{O}(10^{-5})$ and is uniformly distributed, as expected, and the correlation between $c_a$ and $c_\theta$ gives a clear diagonal line on the parameter space. Furthermore, adding the bound on $\alpha_1$ from binary pulsar and triple systems results in a Gaussian distribution of $c_a$ (and hence $c_\theta$) as in the right panel of Fig.\ \ref{fig:EA_priors_c_region1}. 

\subsection{Priors on ($\mathbf{\alpha_1,\alpha_2,\bar{c}_\omega}$) from existing constraints}
\label{sec:alpha_prior}

In this subsection, we discuss the priors on a simpler reparameterization of the theory in terms of $\{\alpha_1, \alpha_2\}$ instead of $\{c_a, c_\theta\}$ and in terms of a new parameter $\bar{c}_\omega$ instead of $c_\omega$. We will work a lot with this parametrization in the next section because, as you will see here, the priors are simpler and the GW observables depend more cleanly on them. 

Let us first discuss this new parameter $\bar{c}_\omega$. In the previous sections, we saw that $c_\omega$ is unconstrained from $(0, \infty)$ and that both cases $c_\omega\rightarrow 0$ and $c_\omega \rightarrow \infty$ limit to GR. Since we cannot realistically sample across an infinite range, we will define a new variable, 
\begin{equation}
    \bar{c}_\omega = \frac{1}{1 + c_\omega},\label{eqn:cbarw}
\end{equation}
such that as $c_\omega \rightarrow 0$ then $\bar{c}_\omega \rightarrow 1$, and as $c_\omega \rightarrow \infty$ then $\bar{c}_\omega \rightarrow 0$. With this new parameter, the range of the prior becomes $\bar{c}_\omega \in [0,1]$ and one is able to cover the entire $c_\omega$ range.

Let us now discuss the shape of the priors when we impose all existing constraints. To do so, we sample uniformly on $\{\alpha_1, \alpha_2, \bar{c}_\omega \}$, and reject those points that violate the constraints on Einstein-\ae ther theory described in Sec.~\ref{subsec:summary-of-constraints}. We start by sampling each of these parameters in the regions
\begin{subequations}
\begin{align}
    \alpha_1 &\in [-0.25, 0.25], \\
    \alpha_2 &\in [-0.025, 0.025], \\
    \bar{c}_\omega &\in [-1,1],
\end{align}
\label{eqn:samplingRegions} %
\end{subequations}
and show how this parameter space shrinks with the addition of constraints. 

\begin{figure*}
\includegraphics[width=0.49\linewidth,clip=true]{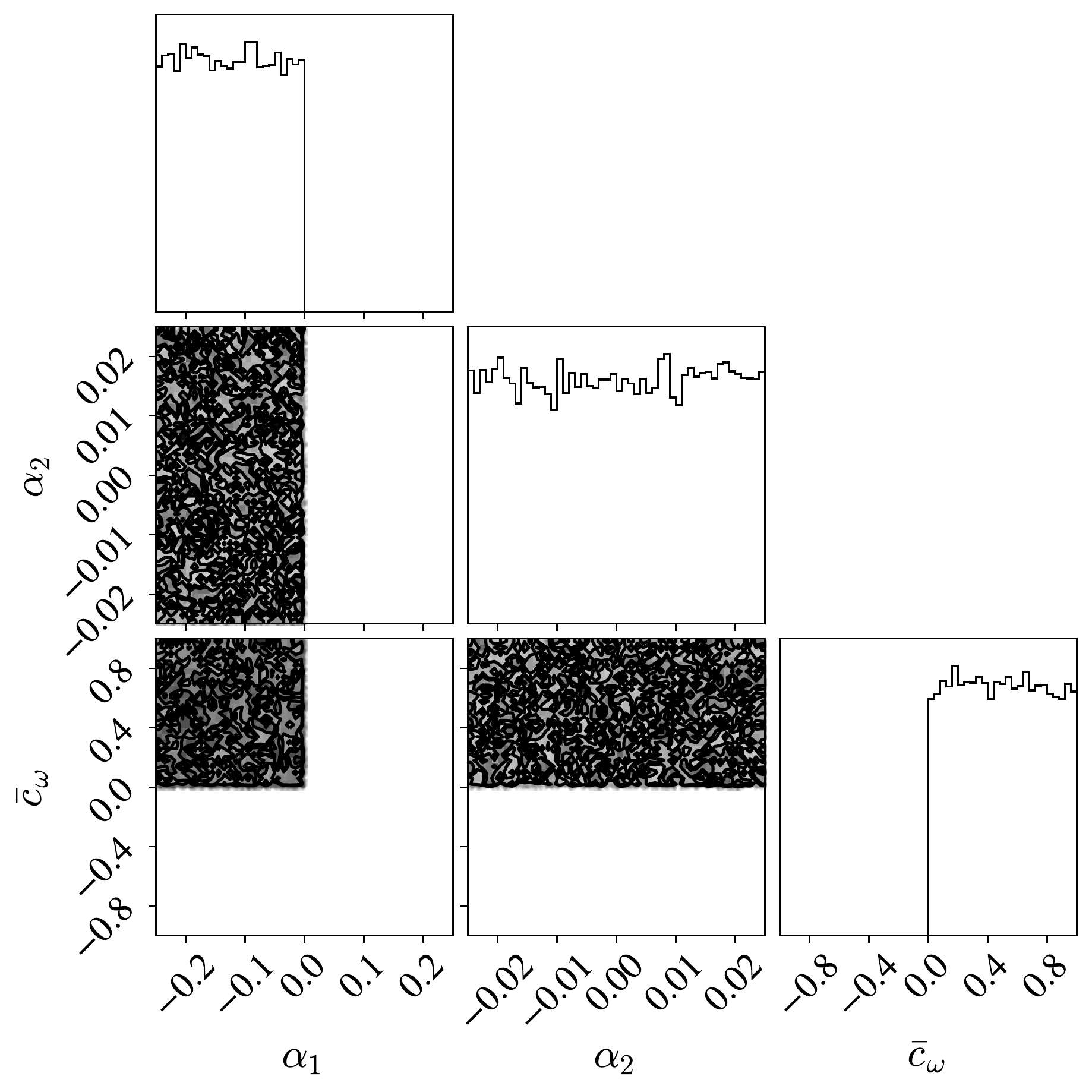}
\includegraphics[width=0.49\linewidth,clip=true]{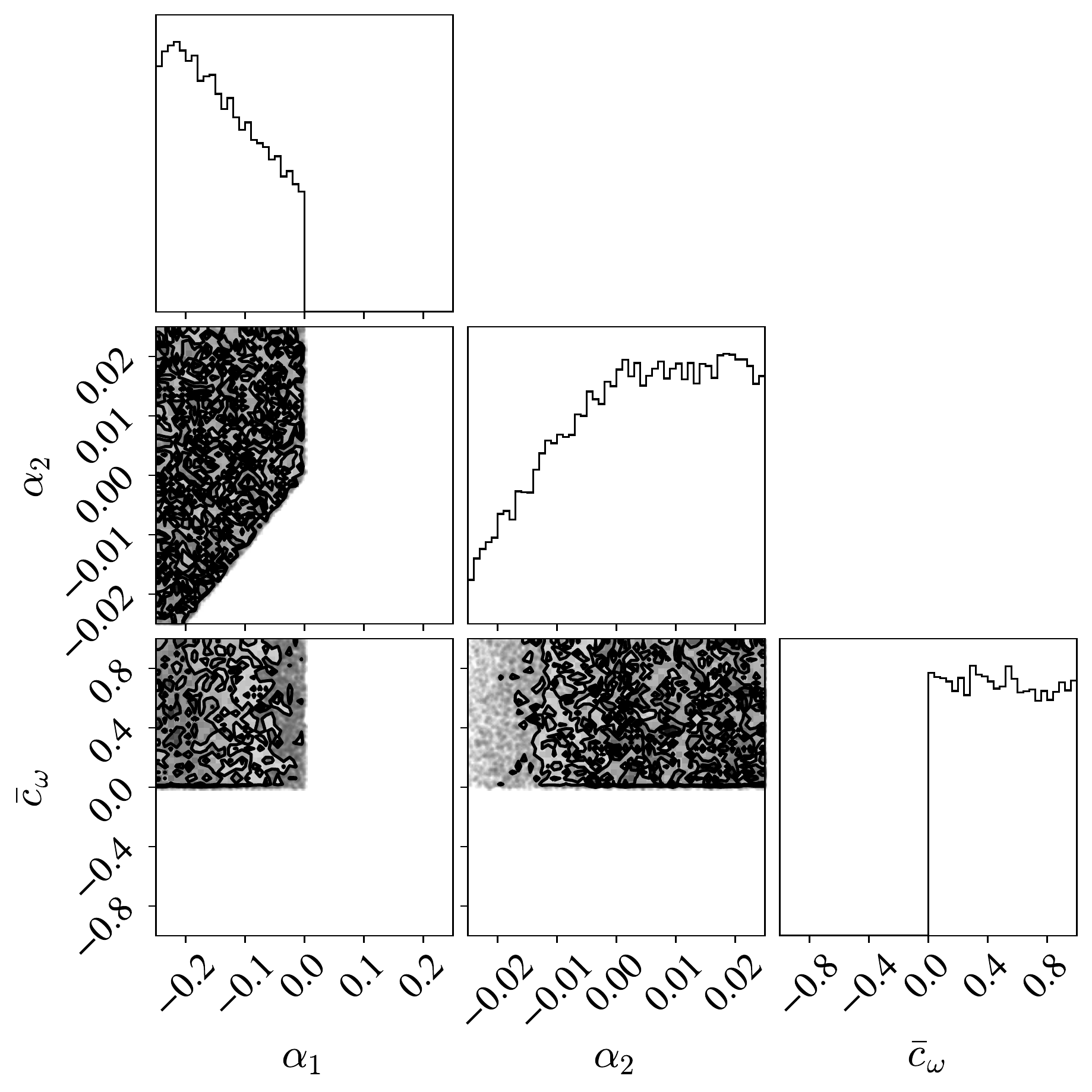} \\
\includegraphics[width=0.49\linewidth,clip=true]{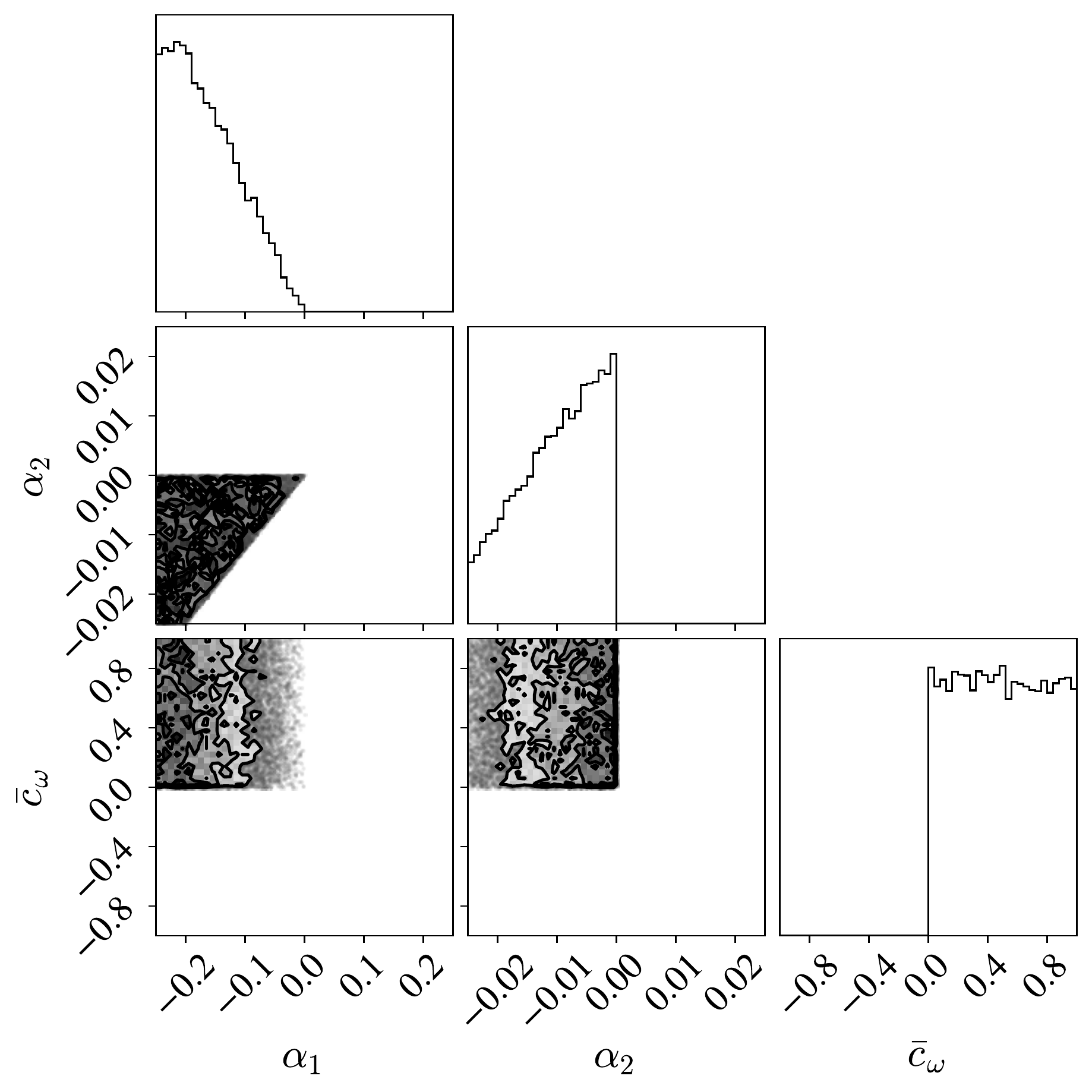}
\includegraphics[width=0.49\linewidth,clip=true]{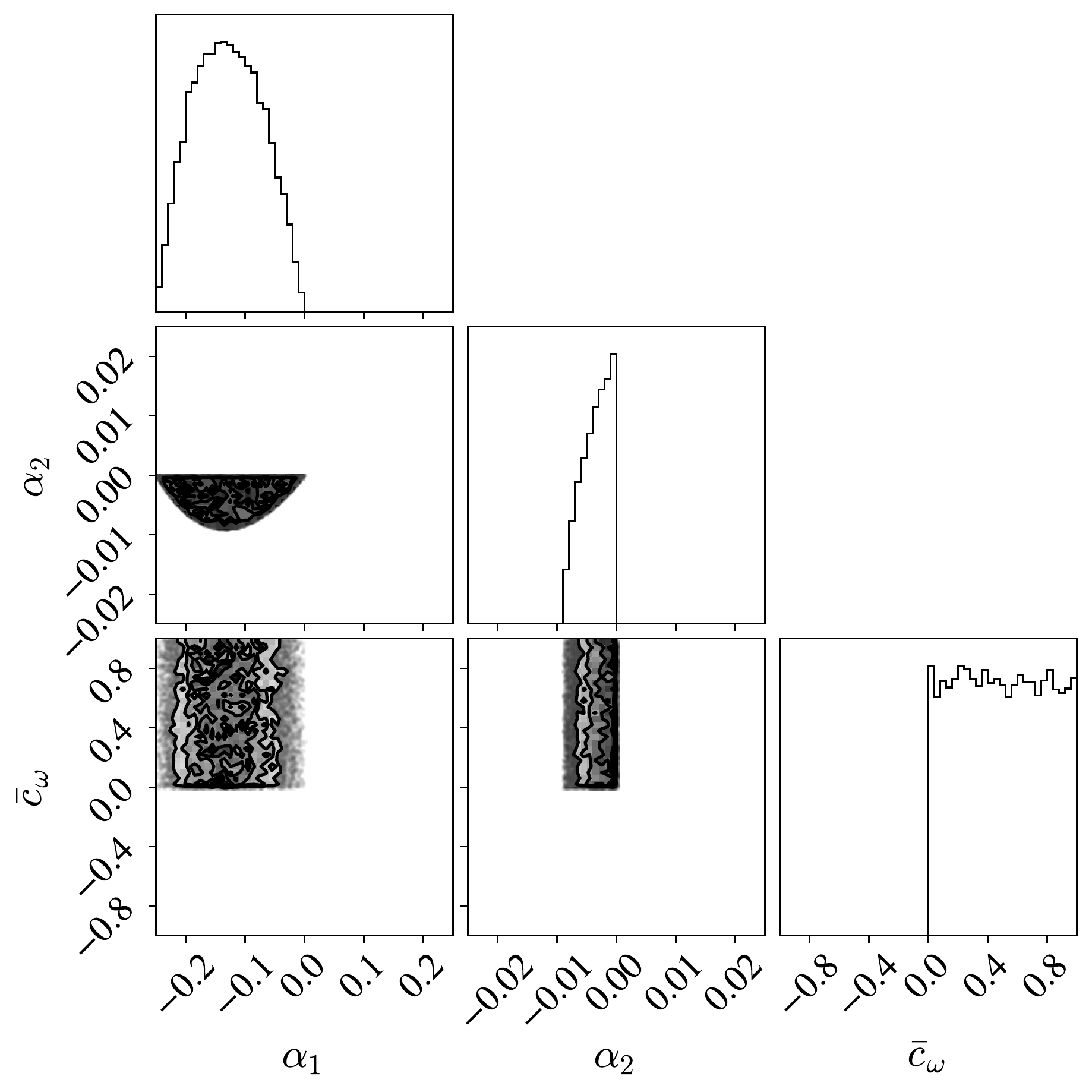}
\caption{Similar to Fig.~\ref{fig:c_beforeSS} but for $(\alpha_1,\alpha_2,\bar c_\omega)$ parameterization uniformly sampled in the region described by Eq.~\eqref{eqn:samplingRegions}. 
Again, the constraints were applied in the following order (beginning in the top left corner and ending in the bottom right corner): positive energy conditions, Eq.~\eqref{eqn:positiveEnergyCond}; positive speeds of different GW polarizations, Eq.~\eqref{eqn:positiveSpeeds}; Cherenkov constraint, Eq.~\eqref{eqn:Cherenkov}; BBN constraint, Eq.~\eqref{eqn:BBN}. 
}
\label{fig:alpha_beforeSS}
\end{figure*}

Let us begin by discussing the stability conditions of Eq.~\eqref{eqn:positiveEnergyCond}. Using the definition of $\alpha_1$ when $c_\sigma =0$, one then finds that $-8\leq \alpha_1 \leq 0$, while $\bar{c}_\omega > 0$ as expected and shown in the top panel of Fig.\ \ref{fig:alpha_beforeSS} through rejection sampling.
As we will see later, this is the only constraint that will have any impact on $\bar{c}_\omega$. Further, requiring that the propagation speeds of the GW polarizations be real [Eq.~\eqref{eqn:positiveSpeeds}] we can derive a constraint on $\alpha_2$. Let us then rewrite $c_S$ in terms of the $\alpha_1$ and $\alpha_2$ to find 
\begin{equation}
    c_S^2 = \frac{\alpha_1}{\alpha_1-8 \alpha_2}\,.
\end{equation}
Since we know that $-8<\alpha_1 < 0$, the numerator of the above equation is negative. Thus, to obtain $c_S^2 \geq 0$, we need the denominator of the above equation to also be negative, which implies that
\begin{subequations}
\begin{align}
    \alpha_2 &\geq  \frac{\alpha_1}{8}.
\end{align}
\end{subequations}
This explains the relationship between $\alpha_1$ and $\alpha_2$ in the top right panel of Fig.\ \ref{fig:alpha_beforeSS}. 

Let us now consider the Cherenkov constraints of Eq.~\eqref{eqn:Cherenkov}. Requiring that the scalar speed be larger than unity now translates to 
\begin{subequations}
\begin{align}
    c_S^2 = \frac{\alpha_1}{\alpha_1-8 \alpha_2} &\geq 1
    \Rightarrow 
    \alpha_1 \leq \alpha_1 - 8\alpha_2,
\end{align}
\end{subequations}
since the denominator of the first expression is negative. This immediately leads to $\alpha_2 \leq 0$. This restriction to negative $\alpha_2$ is the only difference between the top right panel of Fig.\ \ref{fig:alpha_beforeSS} 
and the bottom left panel of Fig.\ \ref{fig:alpha_beforeSS}. 

Let us now study the BBN constraint. Rewriting Eq.~\eqref{eqn:BBN} in terms of $\alpha_1$ and $\alpha_2$, gives two inequalities
\begin{subequations}
\begin{align}
\alpha_2 &\geq \frac{\alpha_1}{2} \left(\frac{\alpha_1 + 2}{\alpha_1 + 8}\right), \\
\alpha_2 &\gtrsim \frac{\alpha_1}{8} \left(\frac{4\alpha_1 + 1}{\alpha_1 +1}\right). 
\end{align}
\end{subequations}
The second constraint is much tighter and results in the curved line visible in the bottom right panel of Fig.\ \ref{fig:alpha_beforeSS}. 
    
\begin{figure*}
\includegraphics[width=0.49\linewidth,clip=true]{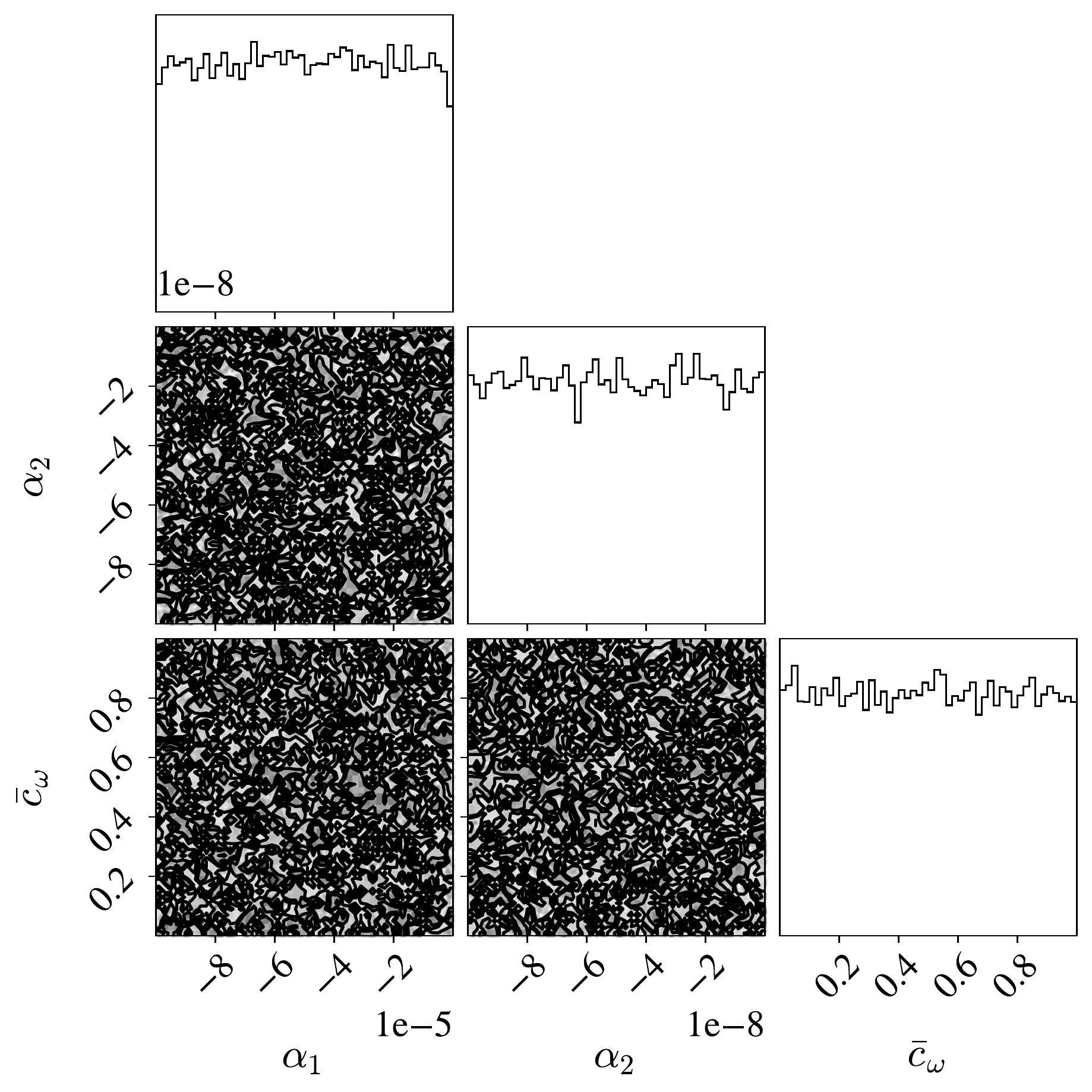}
\includegraphics[width=0.49\linewidth,clip=true]{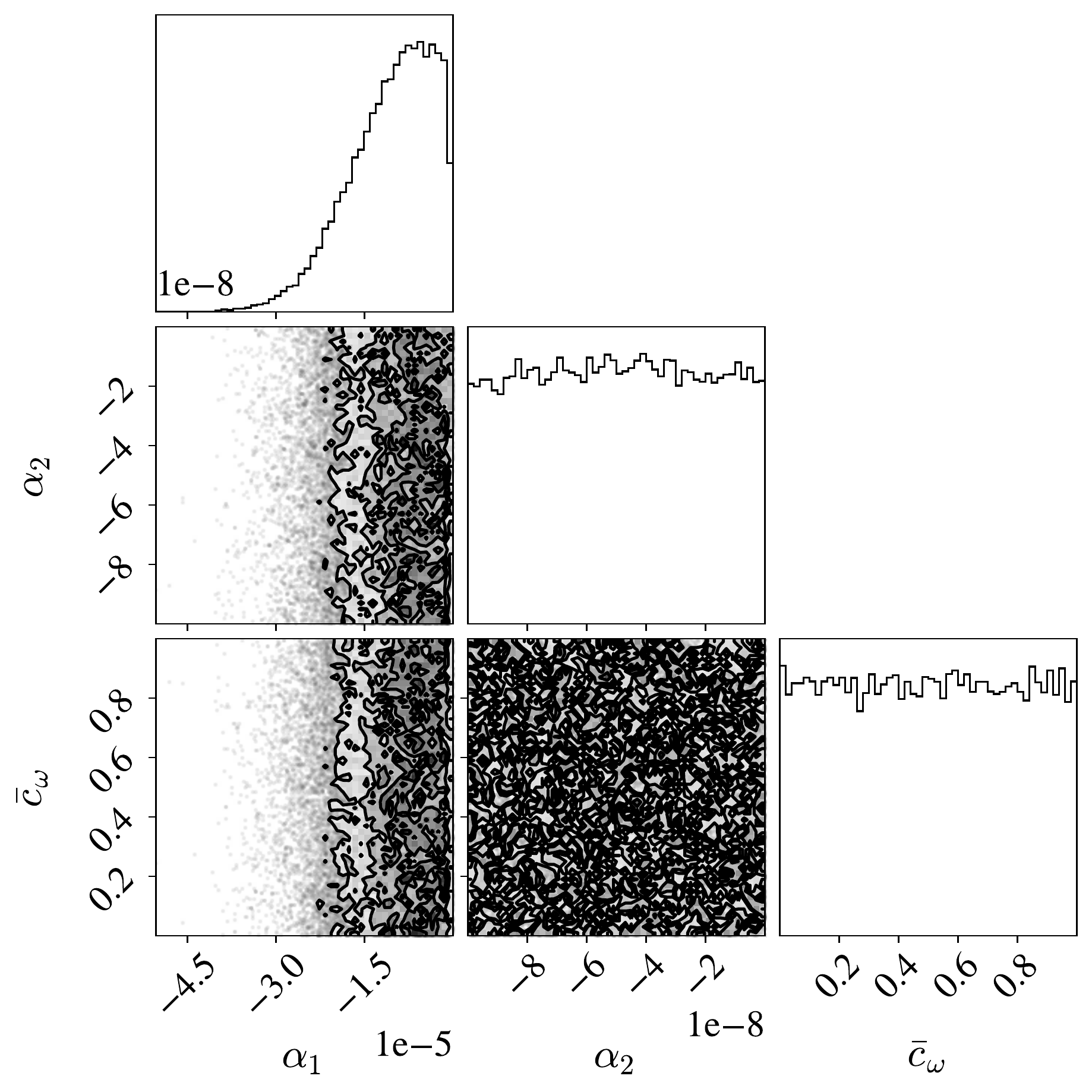}
\caption{Similar to Fig.~\ref{fig:EA_priors_c_region1} but for $(\alpha_1,\alpha_2,\bar c_\omega)$ parameterization uniformly sampled in the region described by Eq.~\eqref{eqn:samplingRegions}.
Both plots include all the constraints of Fig.\ \ref{fig:alpha_beforeSS} as well as the solar system constraints, Eqs.~\eqref{eqn:solarsystem1} and \eqref{eqn:solarsystem2}. The plot on the right further adds the constraint from binary pulsar and triple systems, Eq.~\eqref{eqn:binarypulsarconstraint}.}
\label{fig:EA_priors_alpha}
\end{figure*}

Let us then close by discussing solar system constraints. Since these are bounds on $\alpha_1$ and $\alpha_2$ directly, it is easy to see how they shrink the allowed range for those parameters in the left panel of Fig.\ \ref{fig:EA_priors_alpha}. 
Note that because we are sampling linearly in $\alpha_1$, this is automatically the region 1 of parameter space discussed in the previous section (where $\alpha_1 \approx \mathcal{O}(10^{-4})$). 
We do not have to enforce any extra conditions on $c_\theta$ to be in region 1 when we sample in this parameterization. Finally, we add the binary pulsar and triple system constraint on $\alpha_1$. This takes $\alpha_1$ from a uniform distribution in the allowed region to a Gaussian distribution as seen in the right panel of Fig.\ \ref{fig:EA_priors_alpha}. It has no impact on $\alpha_2$ or $\bar{c}_\omega$. 

Due to the simpler priors in this reparameterization, and the fact that sampling linearly in $\alpha_1$ is equivalent to sampling in region 1 of parameter space, \textit{we will use this parameterization of the theory for the remainder of the paper.} 


\section{Validation of Einstein-\ae{}ther Model through Parameter Estimation Studies with Injections}
\label{sec:injections}
As confirmation that our Einstein-\ae ther waveform template can successfully recover source parameters from GW data, we performed parameter estimation studies on injected data. This section describes those studies, first for data constructed in GR, and then for data constructed in Einstein-\ae ther theory. 

\subsection{GR Injection} 
We begin by constructing a set of injections in GR. We use the {\tt IMRPhenomD\_NRT} waveform template and source parameters similar to the GW170817 event. 
We ``observe'' this data in a three detector network comprised of Hanford, Livingston and Virgo O2-O3 type senstivity for Hanford and Livingston and an optimistic O4 model for Virgo~\cite{ligo_SN_forecast} sensitivities, respectively. The distance to the source was rescaled such that the signal-to-noise ratio (SNR) of the synthetic data as measured by this detector network is 32.4, matching the GW170817 event~\cite{LIGOScientific:2017zic}. 
Explicitly, the parameters used are listed in table \ref{tab:injected_data}. The Einstein-\ae ther parameters, $\{\alpha_1, \alpha_2, \bar{c}_\omega\}$, were not specified, because they are not part of the {\tt IMRPhenomD\_NRT} injection. However, it is useful to note that in the GR limit, $\alpha_1 \rightarrow 0$, $\alpha_2 \rightarrow 0$ and $\bar{c}_\omega \rightarrow 0$ \emph{or}\footnote{$\bar{c}_\omega \rightarrow 0$ or equivalently $c_\omega \rightarrow \infty$ leads to khronometric gravity~\cite{Jacobson:2013xta}, which reduces to GR if the remaining three coupling constants are set to 0 simultaneously.} $\bar{c}_\omega \rightarrow 1$. 

\begin{table}[]
    \centering
    \begin{tabular}{c|c|c|c|c|c|c|c|c|c}
        $\alpha'$ & sin($\delta$) & cos($\iota$) & $t_c$ & $D_L$ & $\bar{\mathcal{M}}$ & $\eta$ & $\chi_1$ & $\chi_2$ & $\Lambda_s$ \\\hline
        3.42 & -.37  & -.82 & 3.0 & 63 & 1.188 & 0.25 & .003 & -.002 & 242
    \end{tabular}
    \caption{Source parameters used for injections. The Einstein-\ae ther parameters were not explicitly set for the GR injection, and were set to nonzero values listed in Eq.~\eqref{eqn:EAnonzero} for the Einstein-\ae ther injection. Note that in the GR case, $\mathcal{M}$ and $\bar{\mathcal{M}}$ are equivalent.
    }
    \label{tab:injected_data}
\end{table}

We then ran an MCMC exploration of the likelihood to perform parameter estimation on this data set, using the {\tt EA\_IMRPhenomD\_NRT} waveform template as our recovery model. The code randomly draws points in the 15-dimensional\footnote{Recall that $c_\sigma$ is set to zero.} parameter space of 
\begin{align*}
    \vec{\theta} &= \{\alpha', \sin\delta, \psi, \cos\iota, \phi_{\text{ref}}, t_c, D_L, \mathcal{M}, \eta, \chi_1, \chi_2, \Lambda_s,\\
    &\hspace{8mm}c_a, c_\theta, c_\omega\},
\end{align*}
using the priors described in Secs.~\ref{subsec:GWATBaseline},~\ref{subsec:NRTmodifications}, and~\ref{sec:alpha_prior}. 
Unfortunately, for the Einstein-\ae{}ther coupling constants, \emph{the posteriors were identical to the priors}. 
This means that the prior was more restrictive than the likelihood and we did not learn any new information from the analysis. However, if the most restrictive of the constraints were removed, the posterior was distinct from the prior. In this way, one can attempt to place constraints on the Einstein-\ae ther parameters from GW data that, even if not competitive with the most restrictive constraints to date, is at least independent of other experimental measurements. Hence, throughout the remainder of this paper, the prior used for the Einstein-\ae ther parameters \textit{include} the stability conditions, the Cherenkov constraint, and the BBN constraint (Eqs.~\eqref{eqn:positiveSpeeds}, \eqref{eqn:positiveEnergyCond}, \eqref{eqn:Cherenkov}, and \eqref{eqn:BBN}), but it \textit{excludes} the solar system constraints and the constraint on $\alpha_1$ from the triple system (Eqs.~\eqref{eqn:solarsystem1}, \eqref{eqn:solarsystem2}, and \eqref{eqn:binarypulsarconstraint}). 

As a test of the code, we performed parameter estimation on the same injected data three different times. In each test, the MCMC began sampling from a different seed point, but all three converged to the same posteriors. The Gelman-Rubin statistic was also used to test convergence~\cite{Gelman:1992zz}. This method takes the square root of the ratio of two estimates of the variance in the MCMC chains to compute a quantity commonly denoted by $\hat{R}$. The numerator of this ratio overestimates the variance and the denominator underestimates it, but both converge to the true value as the number of samples increases. Therefore, $\hat{R}\rightarrow 1$ from above as the number of samples goes to infinity. Reference~\cite{Gelman2013} recommends that $\hat{R}\leq 1.1$ be the condition for convergence. Comparing chains from our three injections, the maximum $\hat{R} = 1.001 <1.1$. Therefore, we are reasonably confident that the MCMC is exploring the parameter space appropriately and converging properly.  

\begin{figure}
    \centering
    \includegraphics[width=\linewidth]{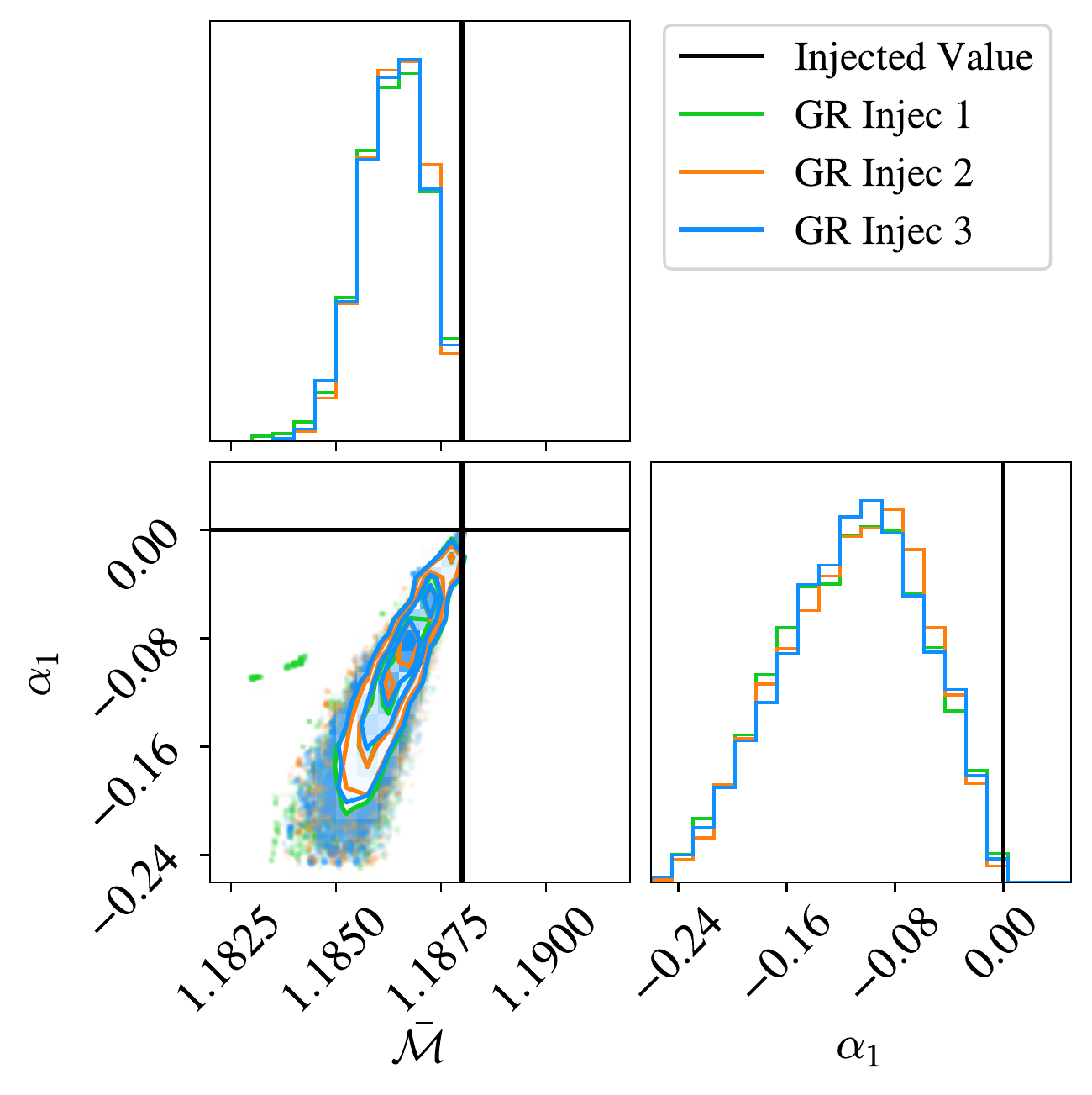}
    \caption{A covariance plot of the posterior of $\bar{\mathcal{M}}$ and $\alpha_1$ recovered with {\tt EA\_IMRPhenomD\_NRT} from three injections of GWs in GR. Note that the $\alpha_1$ prior biases the $\bar{\mathcal{M}}$ posterior to smaller values because it is peaked away from zero. }
    \label{fig:GR_chirpmass}
\end{figure}

Next we compare the posteriors recovered to the injected parameters. 
For everything but the Einstein-\ae ther specific parameters, plots of the posterior distributions recovered from these injections are compared to the injected values in Appendix~\ref{sec:appendixInjecPlots} (labeled as ``GR Injec 1-3''). All were consistent with the injected value, with the chirp mass exhibiting a bias due to correlations with the $\alpha_1$ Einstein-\ae ther parameter. This correlation is better exhibited in Fig.~\ref{fig:GR_chirpmass}, which shows a corner plot in the $\alpha_1$--$\bar{\mathcal{M}}$ plane. Clearly, the injected value is a point in the top-right corner of the covariance panel, which is poorly recovered by the analysis. The reason for this is that the $\alpha_1 = 0$ line in the $\alpha_1$--$\bar{\mathcal{M}}$ plane is strongly disfavoured by the prior (as discussed already in Sec.~\ref{sec:alpha_prior}). This pushes the posterior away from the injected value of $\alpha_1$, which can be compensated for through a different choice of chirp mass. 

\begin{figure}
    \centering
    \includegraphics[width=\linewidth]{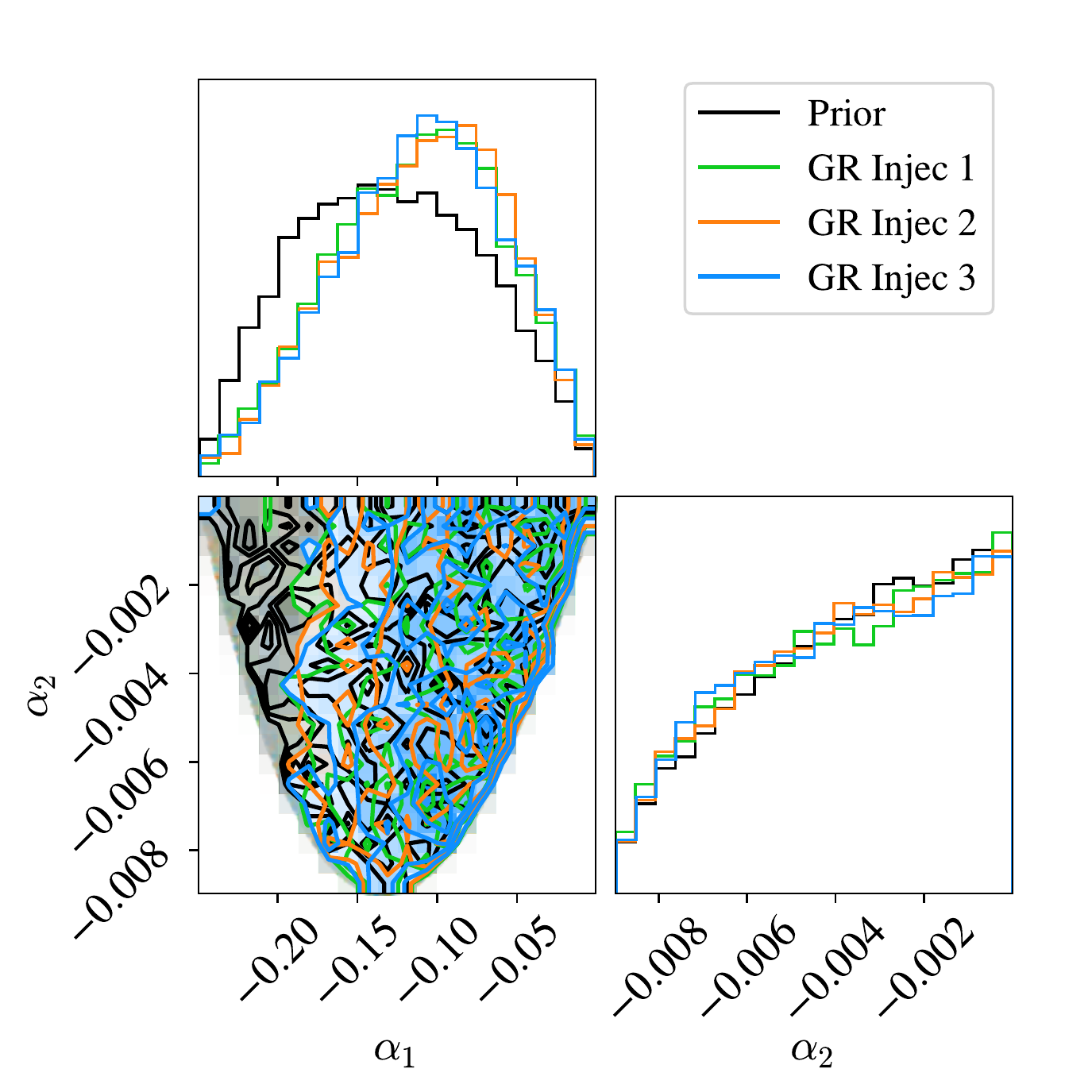}
    \caption{The posterior of $\alpha_1$ and $\alpha_2$ recovered with {\tt EA\_IMRPhenomD\_NRT} from an injection of a GW in GR compared to the prior. Note that both are peaked towards the expected zero-value, though $\alpha_1$ is peaked slightly away from 0 because there are fewer combinations of $\alpha_2$ that lead to $\alpha_1=0$.}
    \label{fig:GR_alphas}
\end{figure}

\begin{figure}
    \centering
    \includegraphics[width=\linewidth]{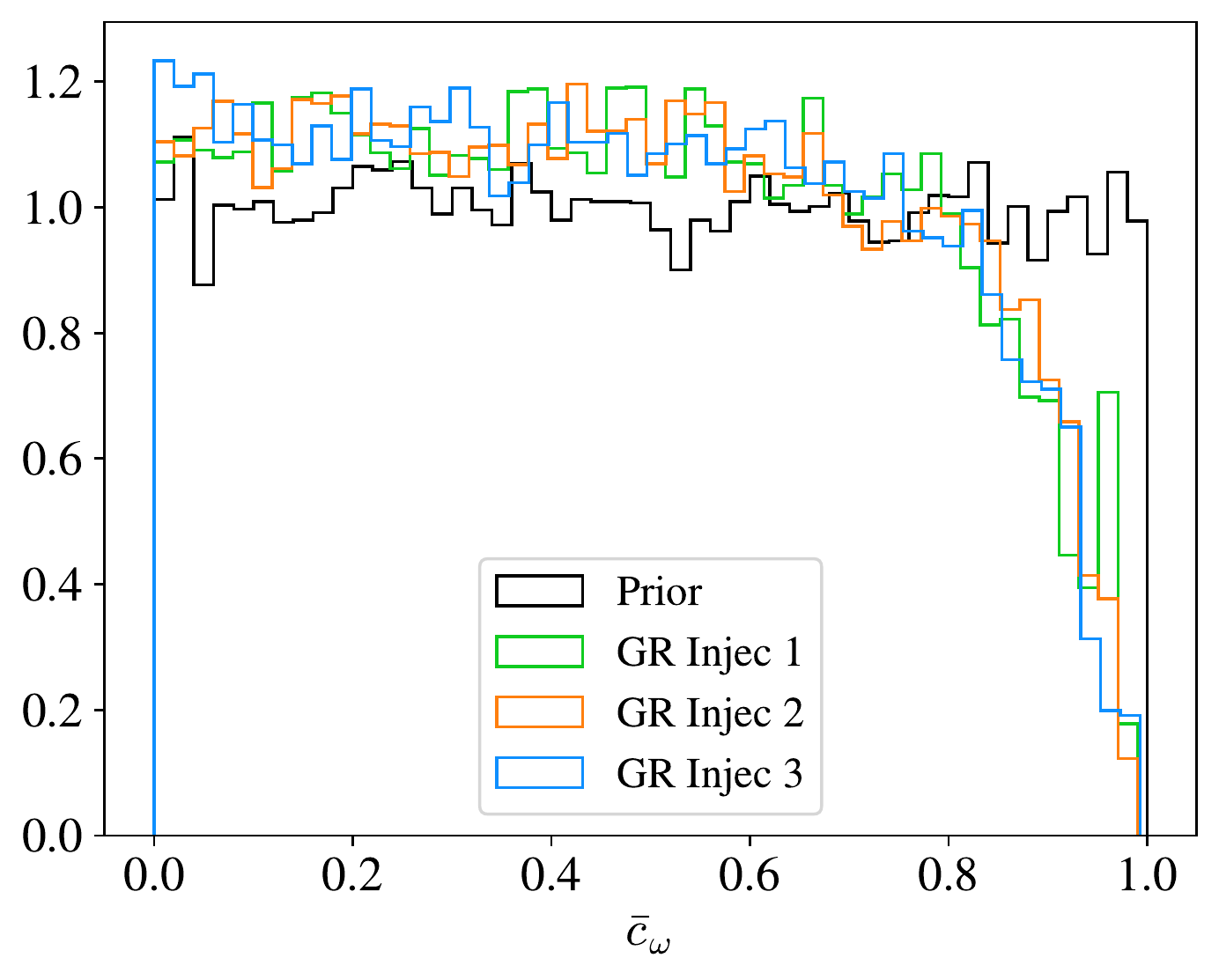}
    \caption{The posterior of $\bar{c}_\omega$ recovered with {\tt EA\_IMRPhenomD\_NRT} from an injection of a GW in GR compared to the prior. Note that both $\bar{c}_\omega = 0$ and $\bar{c}_\omega = 1$ are possible GR values, but $\bar{c}_\omega = 1$ is slightly disfavored by the posterior.}
    \label{fig:GR_cbarw}
\end{figure}

\begin{figure}
    \centering
    \includegraphics[width=\linewidth]{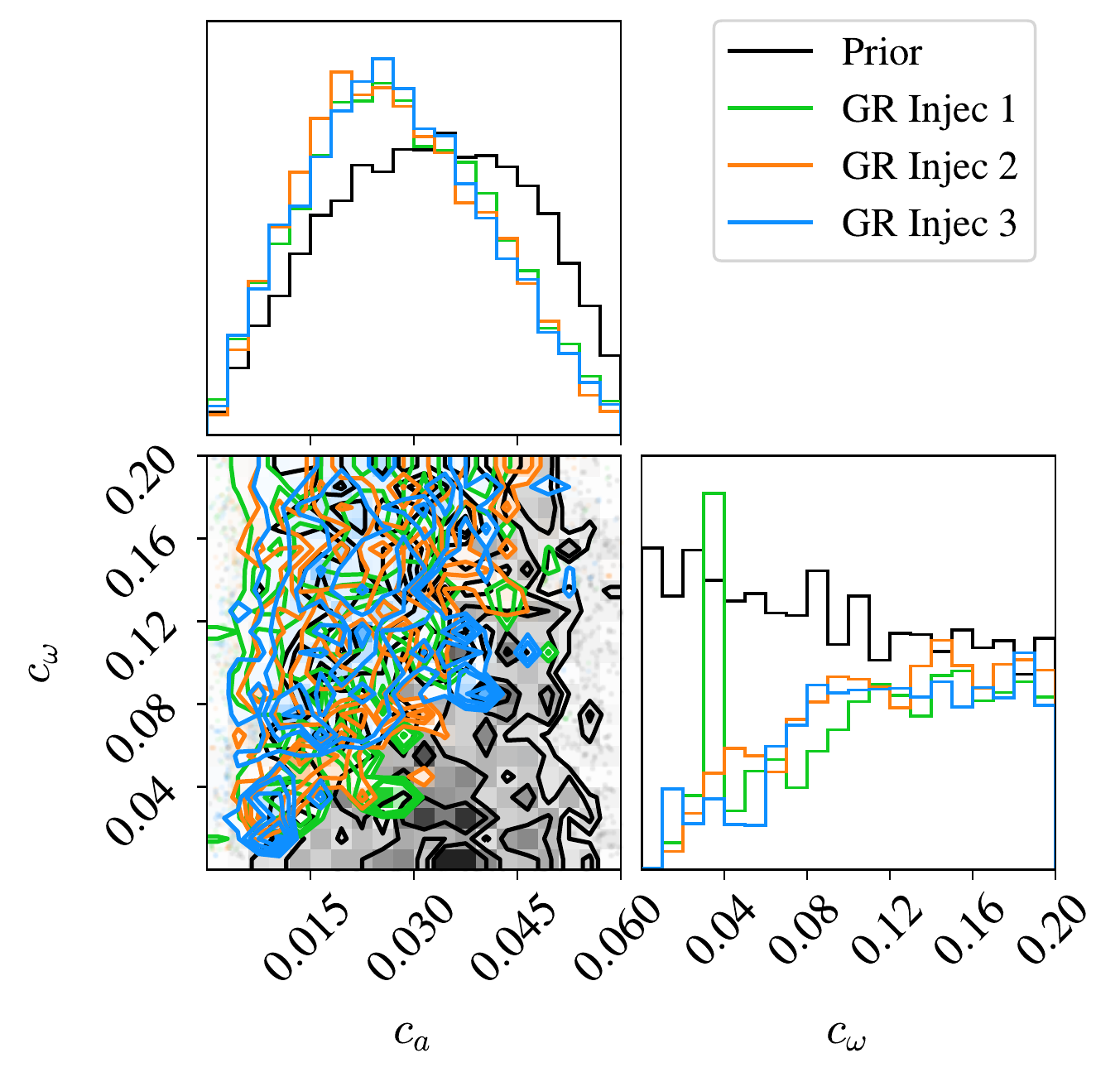}
    \caption{A covariance plot of the posterior of $c_a$ and $c_\omega$ recovered with {\tt EA\_IMRPhenomD\_NRT} from three injections of GWs in GR (in color) as compared to the prior (in black). 
    }
    \label{fig:Injec_cw}
\end{figure}

\begin{figure}
    \centering
    \includegraphics[width=\linewidth]{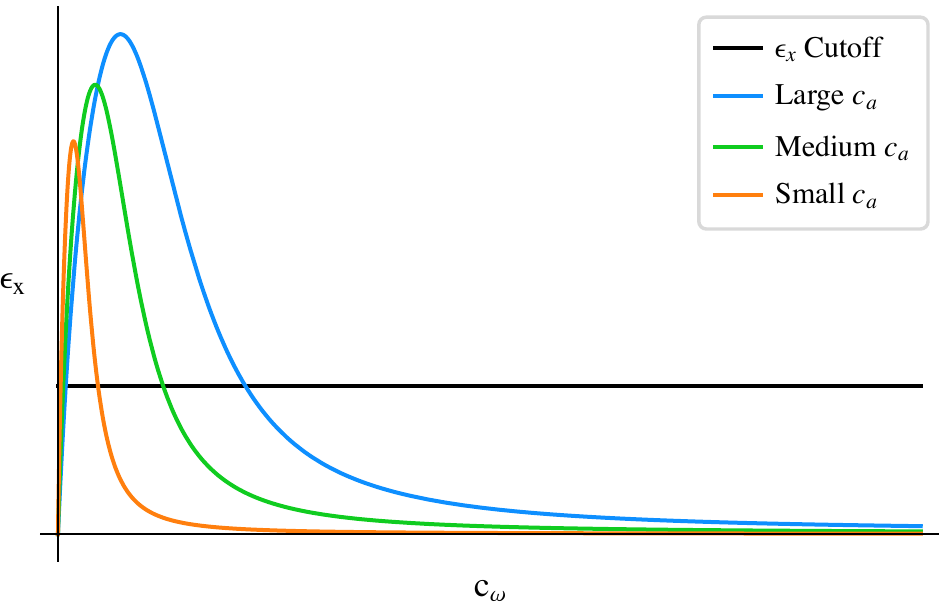}
    \caption{A plot of $\epsilon_x$ (Eq.~\eqref{eqn:epsilon_x}) as a function of $c_\omega$ for three different values of $c_a$. From the shape of this curve, we can see that for small values of $c_\omega$, $\epsilon_x$ is very large. This will make the dipole contribution to the GW very large. If $\epsilon_x$ above some cutoff is disfavored by GW data, then these small values of $c_\omega$ will also be disfavored.}
    \label{fig:epsilon_x}
\end{figure}

Posteriors on the Einstein-\ae ther parameters are presented in Figs.\ \ref{fig:GR_alphas} and \ \ref{fig:GR_cbarw}. The posterior distribution for $\alpha_1$ is distinct from the prior and shifted towards the injected value. 
However, given the shape of the prior, note again that $\alpha_1 = 0$ is possible, but there are fewer combinations of $\alpha_2$ that allow $\alpha_1$ to have this value. This is what pushes the peak of $\alpha_1$ slightly away from the injected value of zero. 
The posterior distribution for $\bar{c}_\omega$ includes both possible GR limits, but seems to favor the limit $\bar{c}_\omega \rightarrow 0$. 
It is easier to understand why if we translate these points into the $\{c_a, c_\theta, c_\omega\}$ parameter space\footnote{Recall that by Eq.~\eqref{eqn:cbarw} $\bar{c}_\omega = 1$ is equivalent to $c_\omega = 0$.}.
Looking at a corner plot of the $c_a$-$c_\omega$ plane for all three injections as compared to the prior (Fig.~\ref{fig:Injec_cw}), we can see that small values of $c_\omega$ are only allowed when $c_a$ is also small.  
Examining the Einstein-\ae{}ther quantities that are important to the likelihood, we find analytically that $\epsilon_x(c_\omega)$ has an interesting shape (Fig.~\ref{fig:epsilon_x}). 
This function is very large for small $c_\omega$, and then quickly drops to very small values as $c_\omega$ increases. 
Plotting this curve for three different values of $c_a$, we see that the larger the $c_a$, the larger the region of $c_\omega$ space in which $\epsilon_x$ is very large. 
Given that the size of $\epsilon_x$ will determine the dipole contribution to the phase and amplitude of the waveform (Eqs.~\eqref{eqn:l2phase} and~\eqref{eqn:l2amp}), it makes sense that large $\epsilon_x$ would be disfavored for a GR injection. 
This seems to explain the disallowed region in the $c_a$-$c_\omega$ covariance plot. 
Translating back to $\bar{c}_\omega$, very small $c_\omega \approx 0$ corresponds to $\bar{c}_\omega \approx 1$. Hence, the lack of support for $\bar{c}_\omega = 1$ in Fig.~\ref{fig:GR_cbarw} is explained.
Note that this dip at $\bar{c}_\omega = 1$ did not happen in the case when all constraints were applied, probably because it was already ruled out by the binary pulsar and the triple system constraints. 

Finally, note that we ran this parameter estimation on injected data with the entire waveform, and separately with just the $\ell =2$ contribution to the waveform. 
The posteriors in both cases were identical. 
This is not surprising, as the $\ell =1$ contribution should be suppressed compared to the $\ell =2$ contribution given how small $\Delta s$ is when $s \approx \mathcal{O}(10^{-3})$ (see Eq.~\eqref{eqn:l1amp} for how this impacts the waveform and Appendix~\ref{sec:appendixSensitivities} for a description of why we expect $s$ to be of this order). However, if we include the $\ell =1$ contribution, the code takes at least twice as long to run, because of all the extra terms in the model that are required to evaluate the likelihood. In the interest of efficiency, and since it makes no difference, for the remainder of the paper, we \textit{do not include} the $\ell =1$ contribution to the waveform. 


\subsection{Non-GR Injection}
The next test of the waveform template involved recovering injected data when the values of the Einstein-\ae ther parameters are distinct from those in GR. To test this we constructed a set of injection data with the {\tt EA\_IMRPhenomD\_NRT} waveform template and Einstein-\ae ther parameter injected values set to 
\begin{subequations}
\begin{align}
    \alpha_1 &= -0.245, \\
    \alpha_2 &= -6.586 \times 10^{-8}, \\
    \bar{c}_\omega &= 0.163453.
\end{align}
\label{eqn:EAnonzero}
\end{subequations}
These values were chosen because they satisfy the complicated Einstein-\ae ther prior and are as distinct as possible from the GR injection (for $\alpha_1$). 
All the other source parameters were the same as in the GR injection and are listed in Table \ref{tab:injected_data}. 

\begin{figure}
    \centering
    \includegraphics[width=\linewidth]{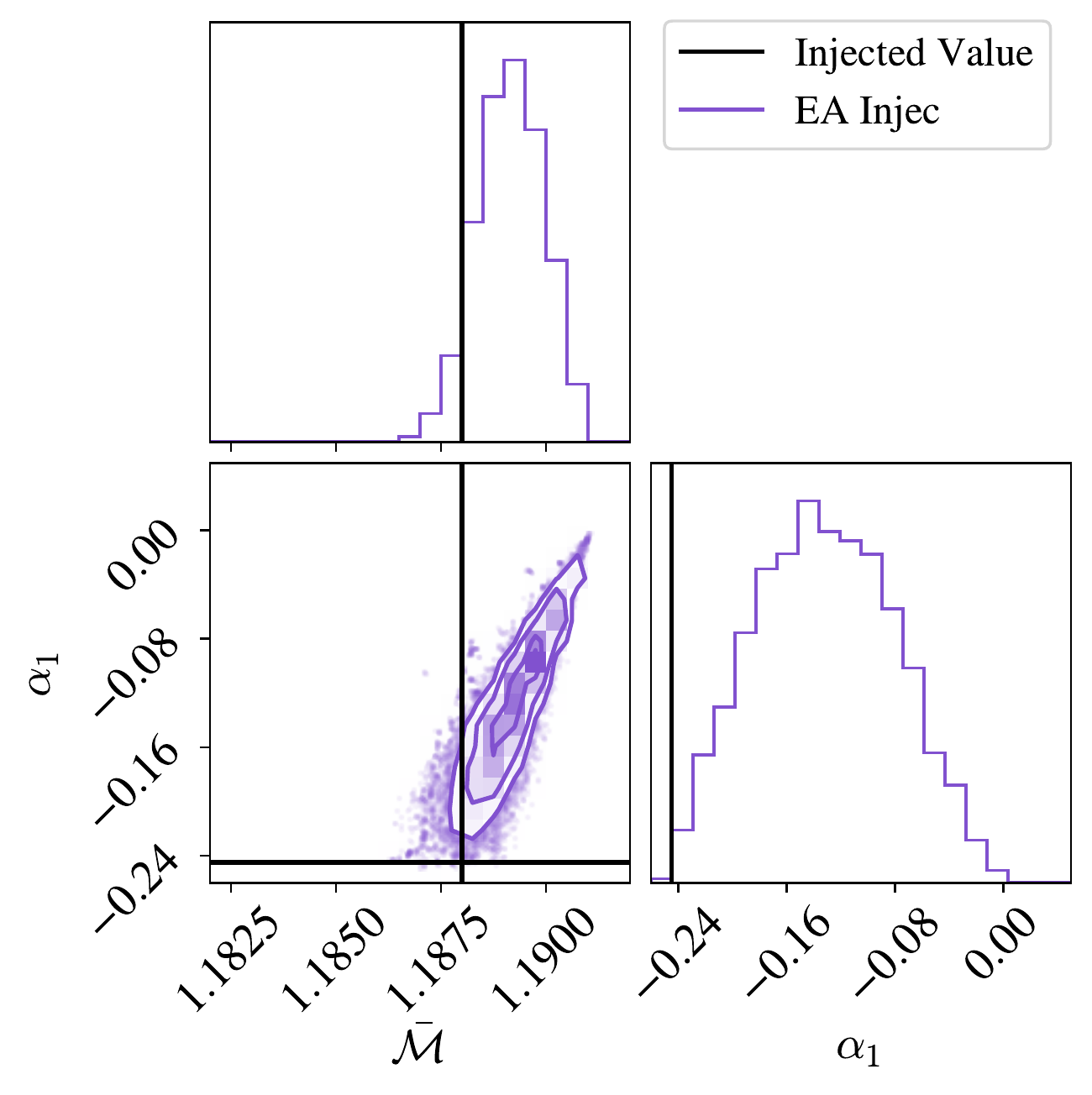}
    \caption{A covariance plot of the posterior of $\bar{\mathcal{M}}$ 
    and $\alpha_1$ recovered from an injection of a GW in Einstein-\ae ther theory. Note that when the injected value of $\alpha_1$ is close to the maximum possible magnitude, the $\bar{\mathcal{M}}$ parameter is biased in the other direction compared to Fig.~\ref{fig:GR_chirpmass}.}
    \label{fig:EA_chirpmass}
\end{figure}

Again, we ran an MCMC to perform parameter estimation on this data set, using the {\tt EA\_IMRPhenomD\_NRT} waveform template as our recovery model. Plots of the posterior distribution recovered from this injection compared to the injected value are in Appendix~\ref{sec:appendixInjecPlots} (this is the ``EA Injec'' data set). All of the posteriors are consistent with the injected parameters. The only posterior that dramatically changes from the recovery of a GR injection, is that of the chirp mass. We can see from Fig.~\ref{fig:EA_chirpmass} that when the value of $\alpha_1$ is at the other edge of the prior, the posterior on the chirp mass is biased in the other direction. 

\begin{figure}
    \centering
    \includegraphics[width=\linewidth]{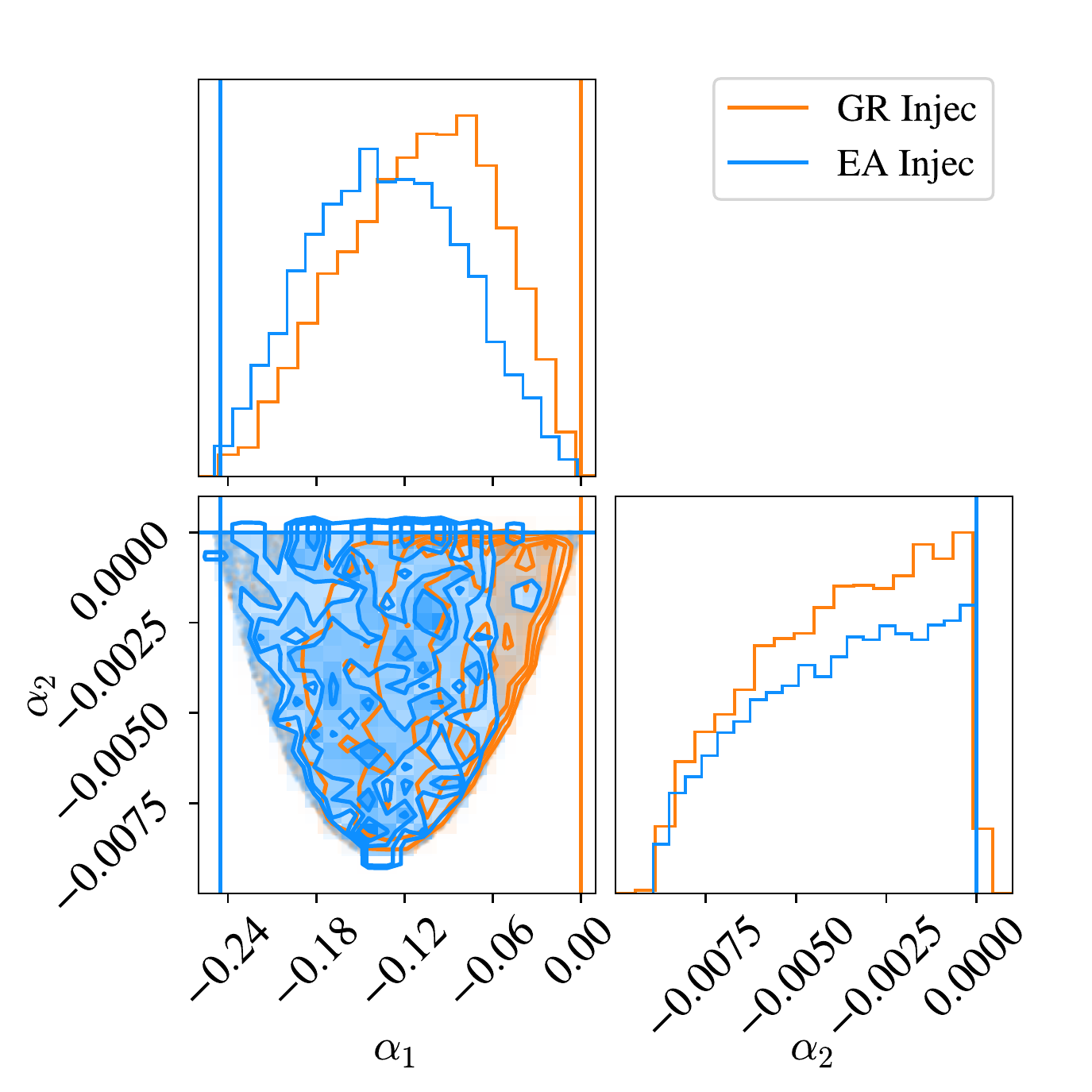}
    \caption{The posterior of $\alpha_1$ and $\alpha_2$ recovered from an injection of a GW in Einstein-\ae ther theory compared to the posterior recovered from an injection in GR. Note that all posteriors are consistent with the injected value, though as in Fig.~\ref{fig:GR_alphas}, $\alpha_1$ is peaked slightly away from the injected value because there are fewer combinations of $\alpha_2$ that lead to $\alpha_1 = -0.245$. }
    \label{fig:EA_alphas}
\end{figure}

\begin{figure}
    \centering
    \includegraphics[width=\linewidth]{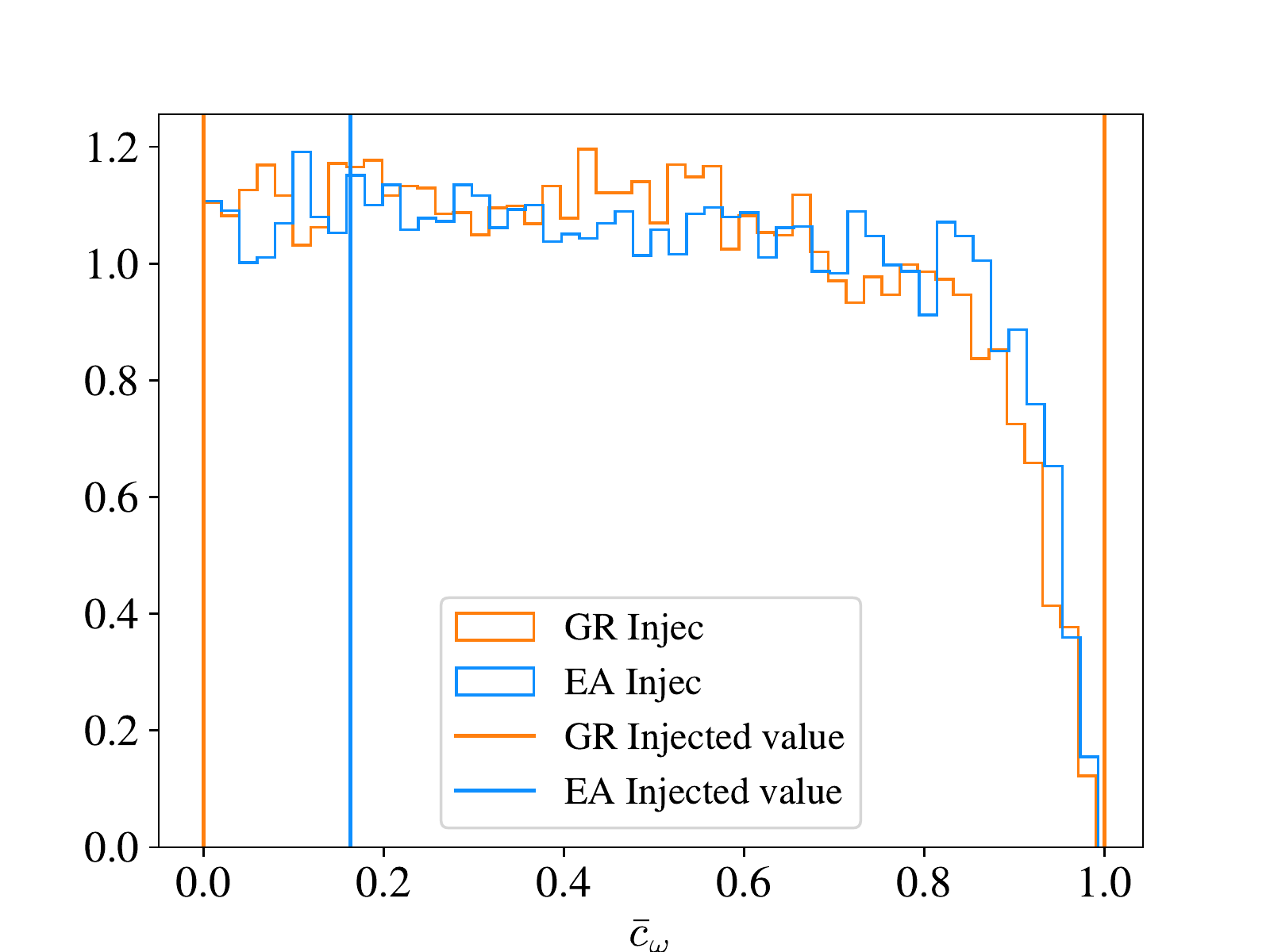}
    \caption{The posterior of $\bar{c}_\omega$ recovered from an injetion of a GW in Einstein-\ae ther theory compared to an injection in GR. These look identical.}
    \label{fig:EA_cbarw}
\end{figure}

As for the Einstein-\ae ther parameters, shown in Figs.\ \ref{fig:EA_alphas} and \ref{fig:EA_cbarw}, the posterior for $\alpha_1$ is \edit{only slightly different when the injection is an EA signal from when it is a GR one, while the posteriors for $\alpha_2$ and $c_\omega$ remain approximately the same.}
\edit{This implies that observations similar to the GW170817 event are not sufficiently informative to distinguish between a GR and an EA model. The small shift in the posterior of $\alpha_1$, however, also implies that future signals at a higher SNR might be able to begin to distinguish EA and GR effects. }
From this, we conclude that \edit{the {\tt EA\_IMRPhenomD\_NRT} model is functioning as expected for both GR and non-GR cases.} 


\section{Constraints on Einstein-\AE ther theory with Gravitational Wave Events from O1--O3}
\label{sec:GW_constraints}
Once the waveform template has undergone testing, we are able to use it to recover the source parameters from GW events. To date, there have been two BNS mergers well above the detection threshold: GW170817 and GW190425 \cite{LIGOScientific:2017zic, LIGOScientific:2020aai}. In this section, we describe the parameter estimation studies we have conducted with these two events. We remind the reader again that we have not considered GW events produced by binaries with one or more BHs, because the Einstein-\ae{}ther sensitivities have not yet been calculated for these objects, and these sensitivities enter the dominant Einstein-\ae{}ther modifications to the GR waveform.  

We performed parameter estimation on both events, using data from the Gravitational Wave Open Science Center \cite{LIGOScientific:2019lzm}. 
The priors used for the {\tt IMRPhenomD} parameters\footnote{$\vec{\theta} = \{\alpha', \sin\delta, \psi, \cos \iota, \phi_{\text{ref}}, t_c, D_L, \mathcal{M}, \eta, \chi_1, \chi_2\}$} 
were those described in Sec.~\ref{subsec:GWATBaseline}.
Note that because of the good sky localization for GW170817, we were further able to narrow the priors on the right ascension and declination for this event to $\alpha' \in [3.4, 3.5]$ and $\sin\delta \in[-.4, -.3]$. 
For both events, the prior on the symmetric tidal deformability, $\Lambda_s$, was the same as that given in Sec.~\ref{subsec:NRTmodifications}. 
Finally, the prior on the Einstein-\ae{}ther parameters was the less restrictive prior described in Sec.~\ref{sec:injections}, which included the stability conditions (Eqs.~\eqref{eqn:positiveEnergyCond}, \eqref{eqn:positiveSpeeds}), Cherenkov constraints (Eq.~\eqref{eqn:Cherenkov}), and the BBN constraint(Eq.~\eqref{eqn:BBN}). The complicated shape of this prior is shown in the bottom right panel of Fig.~\ref{fig:alpha_beforeSS}.

We will start by examining the results we obtain when we analyze the GW170817 event. We perform three different parameter estimation studies on this data, starting the MCMC from three different seed points. The posteriors from each run are identical, giving us good reason to believe that the MCMC explored the space adequately and converged. 
Visual inspection of the MCMC chains suggests the analysis has converged to a stable distribution.
Furthermore, the Gelman-Rubin statistic for these runs gave an $\hat{R} = 1.0009< 1.1$, which also indicates convergence.

\begin{figure*}
\includegraphics[width=0.43\linewidth,clip=true]{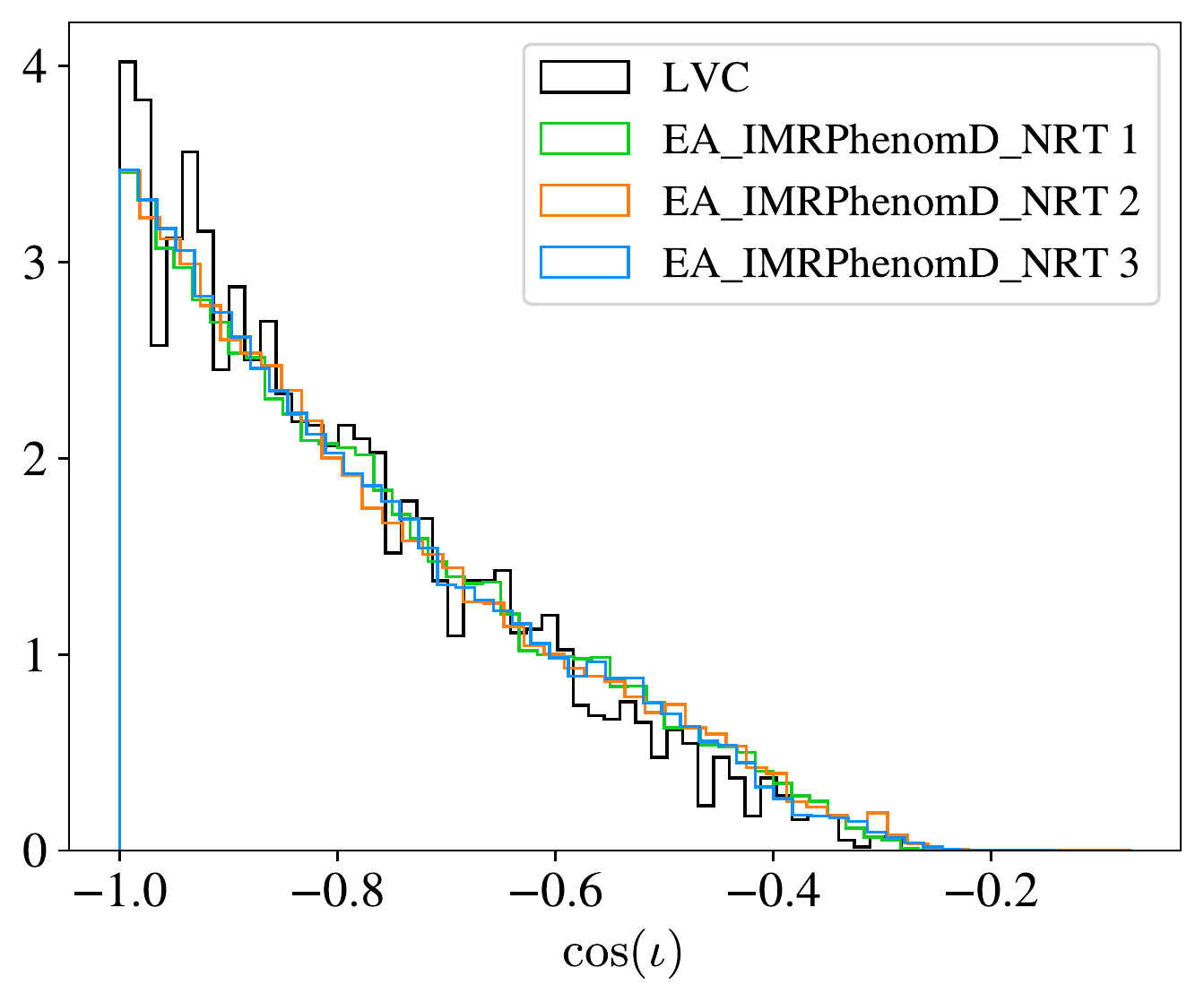}
\includegraphics[width=0.45\linewidth,clip=true]{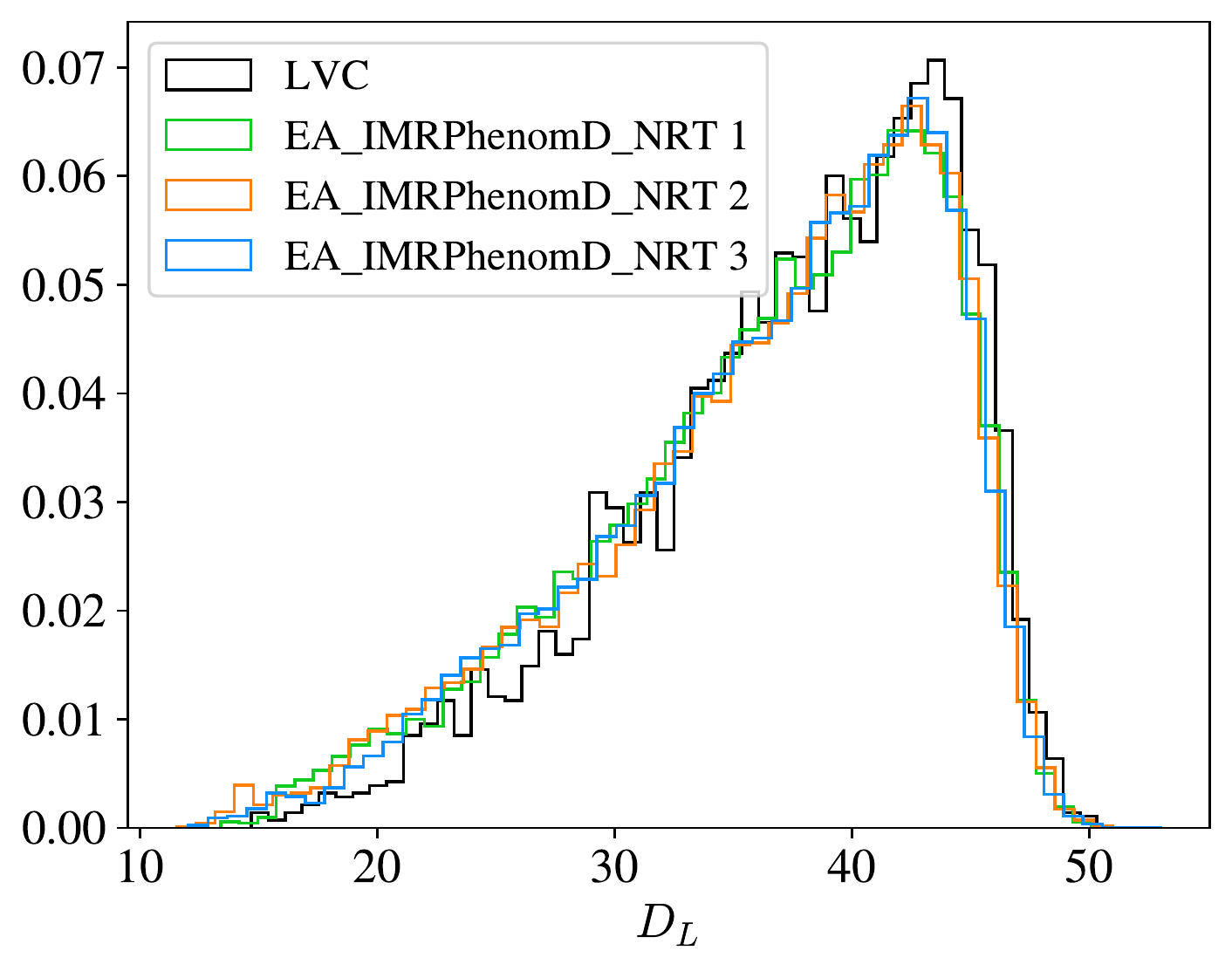} \\
\includegraphics[width=0.45\linewidth,clip=true]{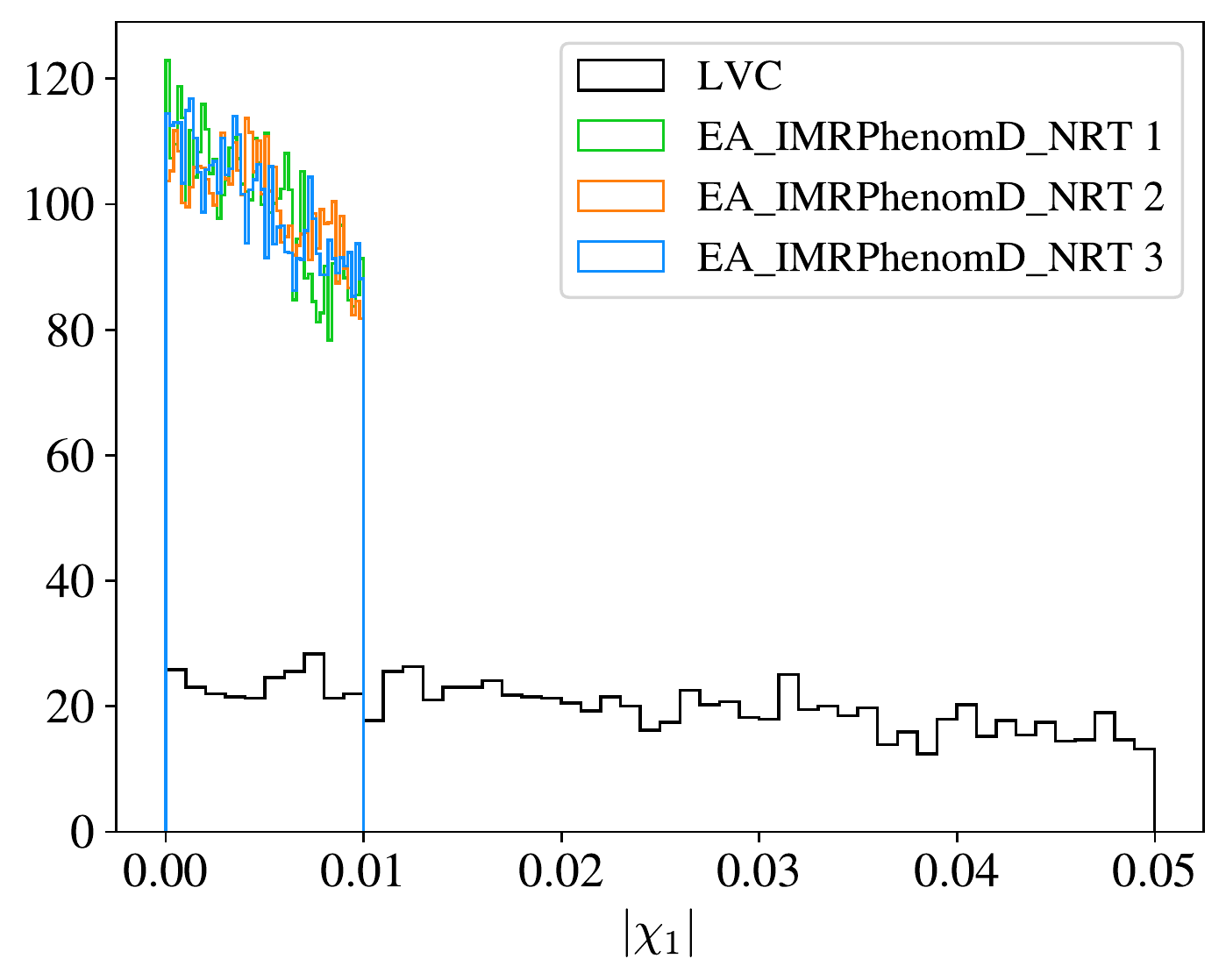}
\includegraphics[width=0.45\linewidth,clip=true]{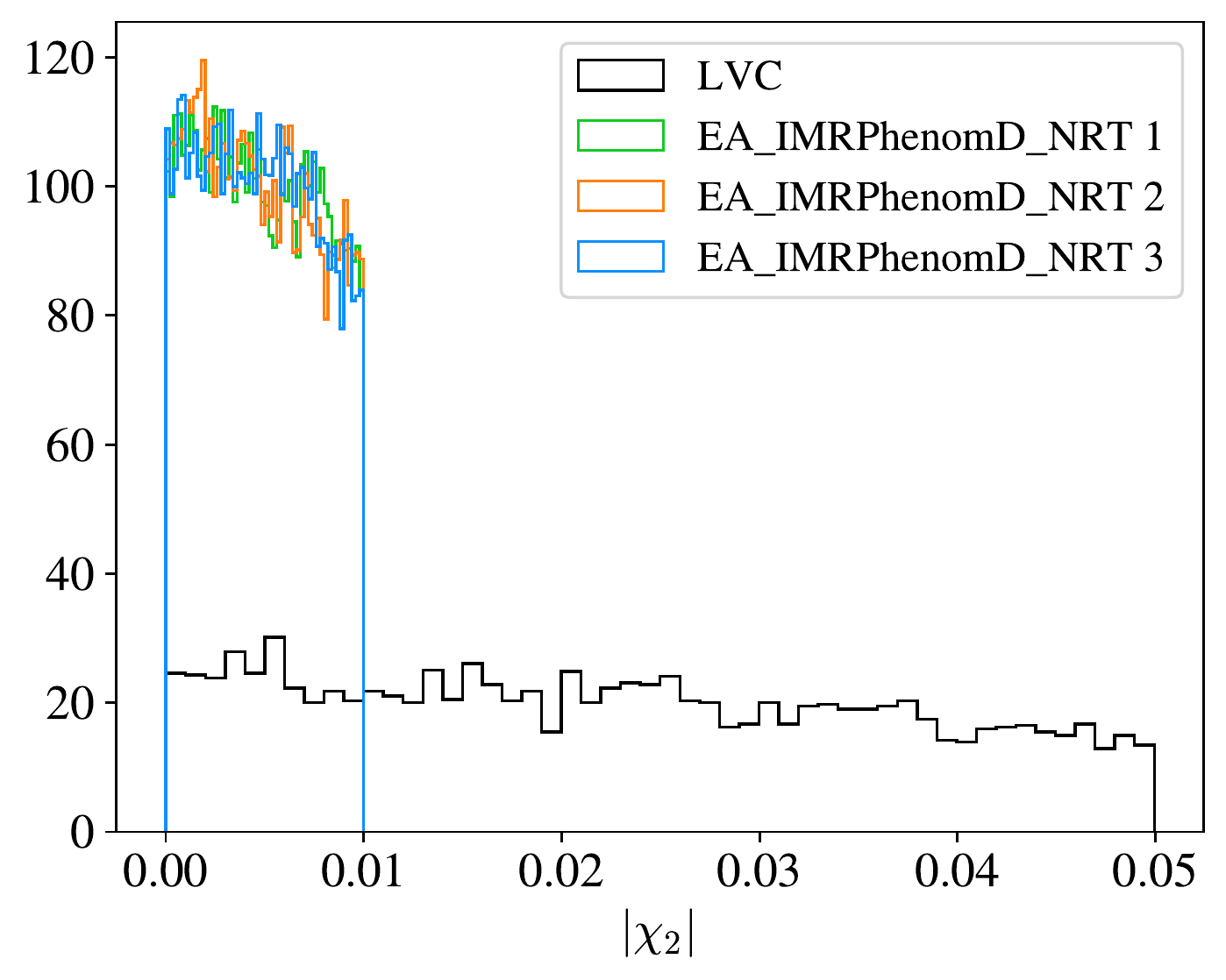} \\
\includegraphics[width=0.45\linewidth,clip=true]{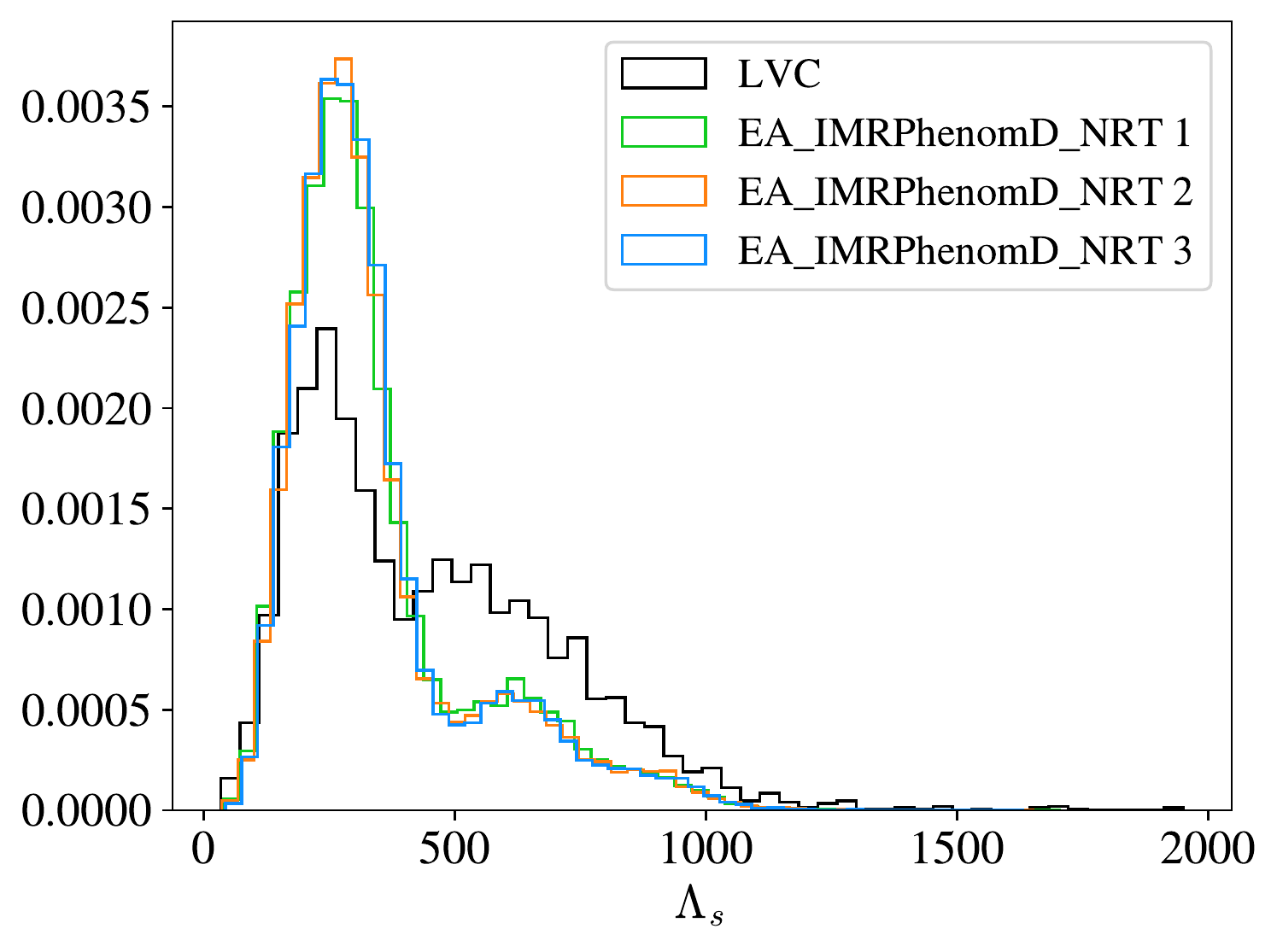}
\includegraphics[width=0.45\linewidth,clip=true]{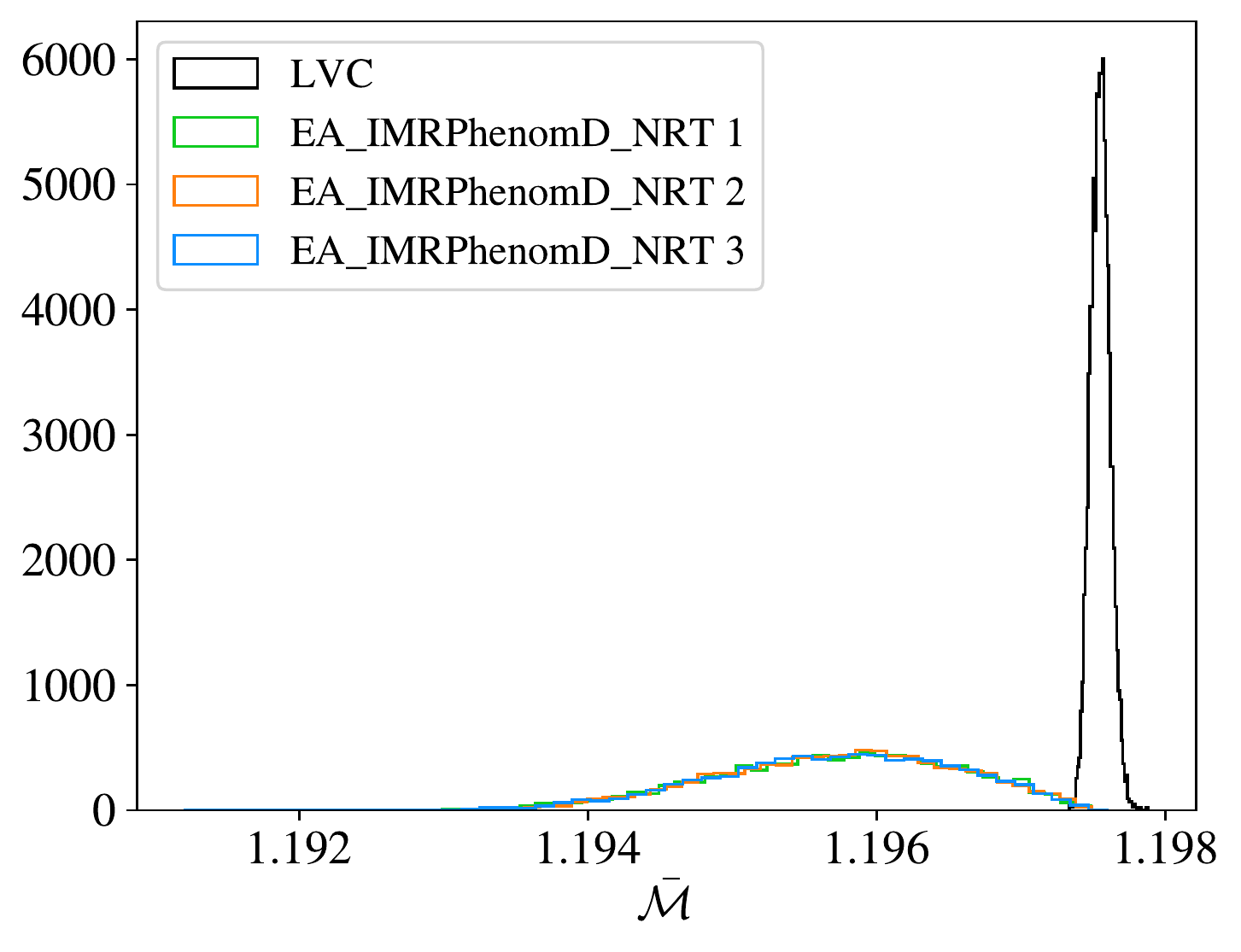} \\
\caption{Comparison of our posteriors with those published by the LIGO/Virgo (LVC) collaboration for six of the source parameters of GW170817. All are consistent except for the chirp mass, which, as discussed in the text, is shifted due to Einstein-\ae{}ther correlations. Our spin posteriors are also different from LVC's because of our use of a small spin prior.} 
\label{fig:GW170817_compare_LIGO}
\end{figure*}

\begin{figure}    
\includegraphics[width=\linewidth,clip=true]{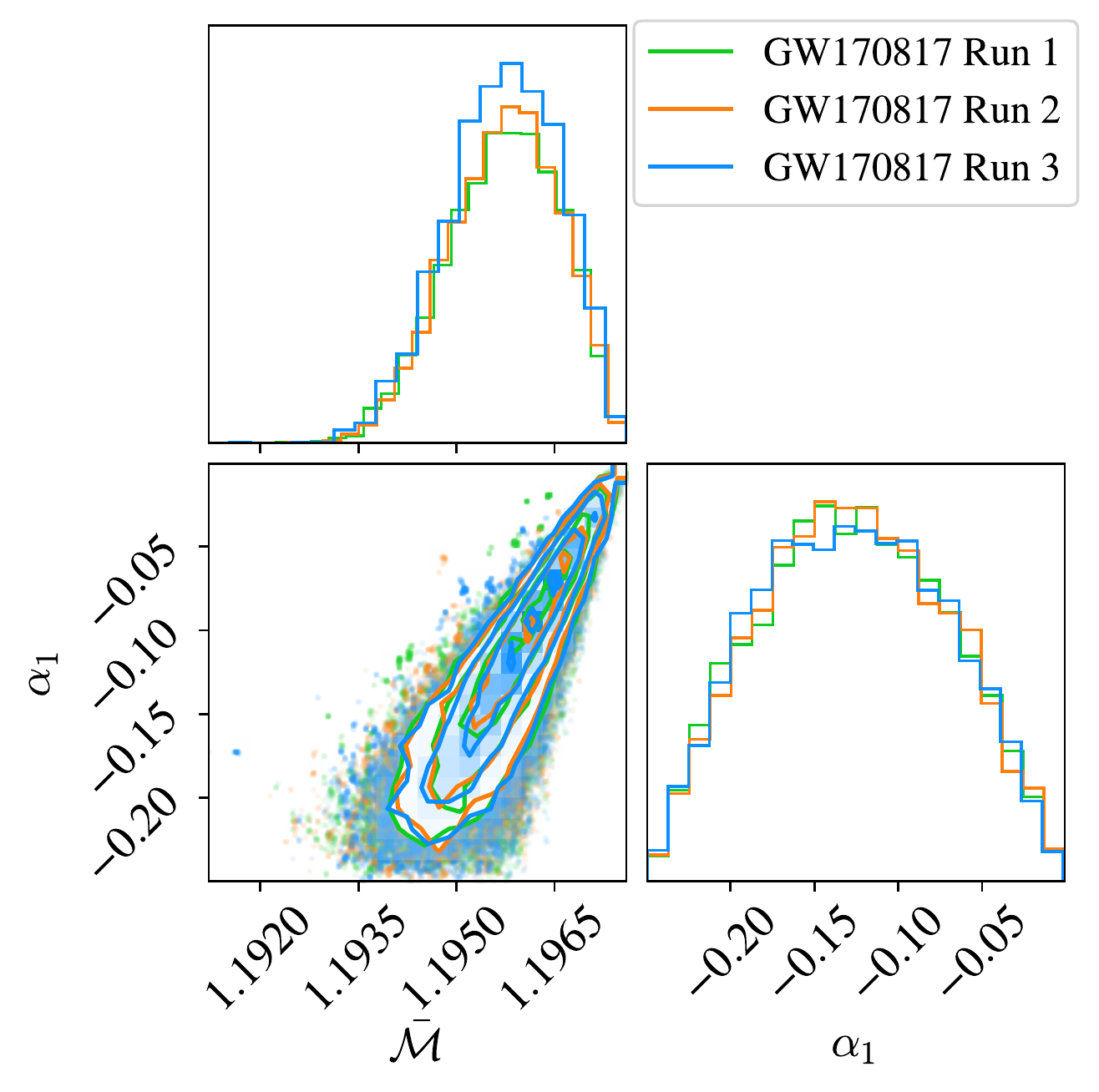}   
\caption{Correlation between the $\bar{\mathcal{M}}$ and $\alpha_1$ parameters for GW170817. Just as in the injections, this correlation tends to widen the chirp mass posterior.
}
\label{fig:GWevents_chirpmass}
\end{figure}

We plot the posteriors we obtain when we analyze the GW170817 event directly on top of LIGO's for convenient comparison (Fig.~\ref{fig:GW170817_compare_LIGO})\cite{LIGOScientific:2018hze}. Note that the prior we use for the $\chi_1$ and $\chi_2$ parameters is narrower than that used by LIGO. If we use the same prior as LIGO's for $\chi_1$ and $\chi_2$, our posteriors for these parameters match LIGO's and the results for all the other parameters are statistically consistent with our previous posteriors. Comparing the plots in Fig.~\ref{fig:GW170817_compare_LIGO}, we find that all the posteriors for the GR parameters are consistent with LIGO's except for the chirp mass. Given what we saw with this parameter in the injection studies \edit{(Fig.~\ref{fig:GR_chirpmass} and Fig.~\ref{fig:EA_chirpmass})}, this is not surprising. 
Correlations between the Einstein-\ae ther parameter $\alpha_1$ and the chirp mass tend to dramatically increase the width of the posterior on the latter parameter and expand it asymmetrically. \edit{Furthermore, if the injected value of $\alpha_1$ is on the edge of the prior, the recovery of chirp mass will be skewed by the correlation}.
This \edit{widened posterior on chirp mass is} explicitly demonstrated \edit{for the GW170817 event} in Fig.\ \ref{fig:GWevents_chirpmass}.

\begin{figure}
    \centering
    \includegraphics[width=\linewidth]{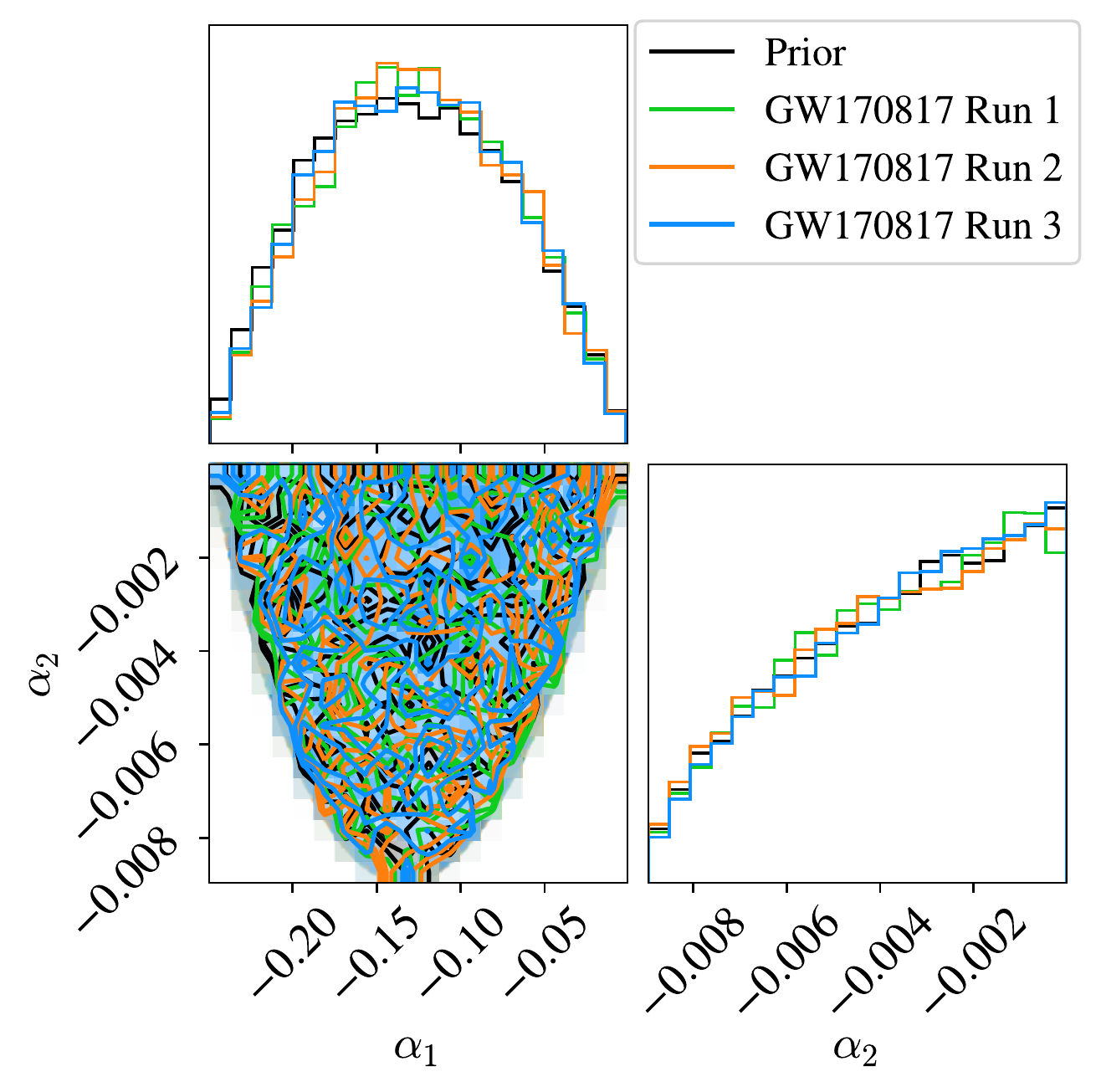}
    \caption{The posteriors for $\alpha_1$ and $\alpha_2$ from GW170817 plotted over the prior. Three separate runs are shown here and they all converge to the same answer, which is indistinguishable from the prior.}
    \label{fig:alphas_combined}
\end{figure}

\begin{figure}
    \centering
    \includegraphics[width=\linewidth]{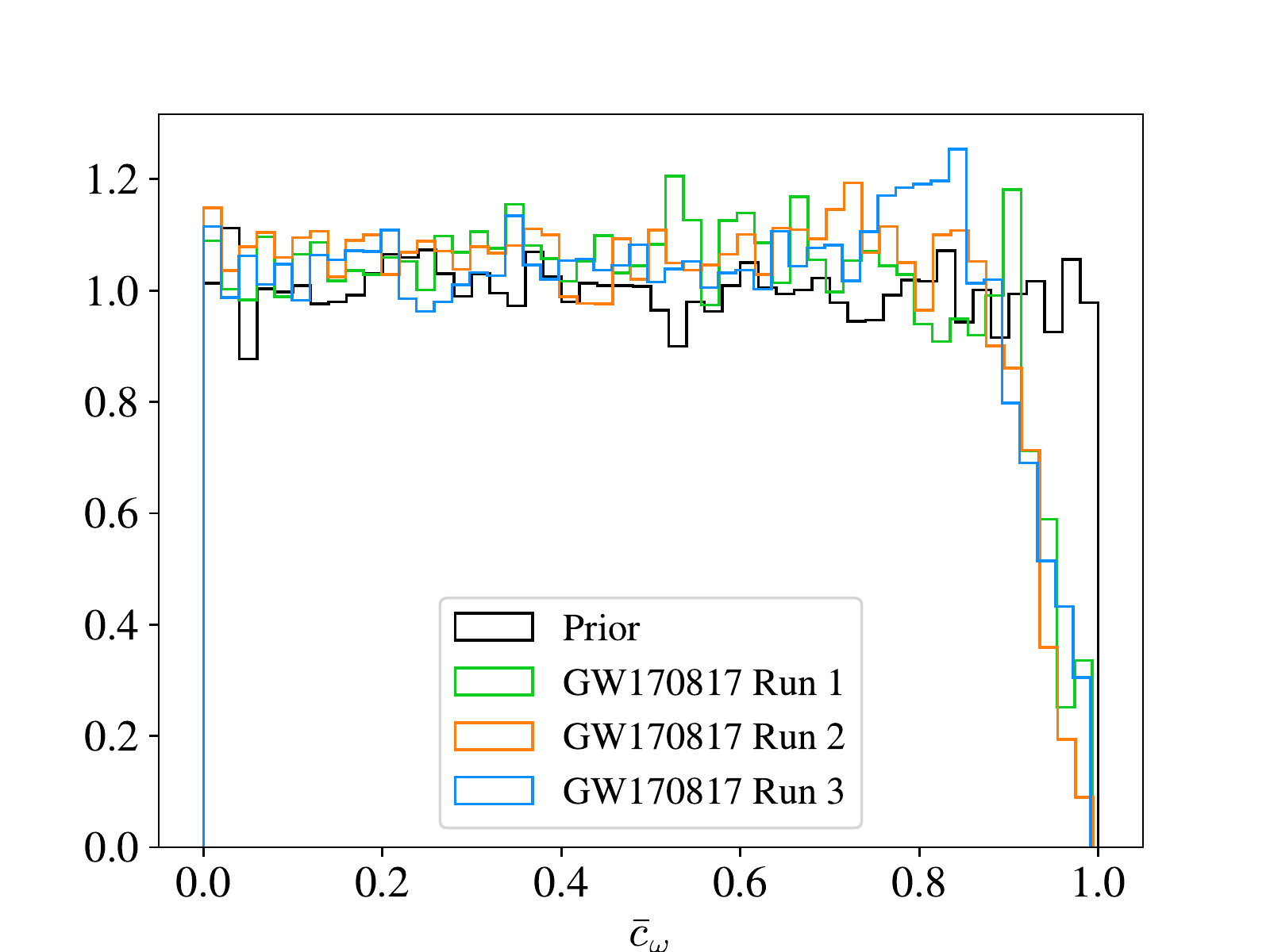}
    \caption{The posteriors for $\bar{c}_\omega$ from GW170817 plotted over the prior. Again three separate runs are shown that are all consistent with each other and indistinguishable from the prior aside from a slight disfavoring of $\bar{c}_\omega = 1$ (reason explained in Sec.~\ref{sec:injections}). }
    \label{fig:cbarw_combined}
\end{figure}

The posteriors for the Einstein-\ae ther parameters are shown in Figs.\ \ref{fig:alphas_combined} and \ref{fig:cbarw_combined}.
There is no improvement over the prior aside from a slight disfavoring of $\bar{c}_\omega = 1$ (equivalent to $c_\omega = 0$).
The reason for this was explained in Sec.~\ref{sec:injections}.
We did not expand the prior on the Einstein-\ae{}ther parameters further to explore wider regions of parameter space because numerical instabilities and floating-point errors in the waveform calculation prevented us from performing the inference analysis. 
Furthermore, the sensitivity model breaks down for certain combinations of the Einstein-\ae{}ther coupling constants outside the priors we have chosen (see Appendix~\ref{sec:appendixSensitivities} for more detail).

For the GW190425 event, the Einstein-\ae{}ther posteriors were no more informative (they were identical to those obtained from the analysis of the GW170817 event). This is not surprising given the lower SNR of this signal. The combined SNR of GW170817 was estimated to be 32.4 (accounting for the SNR in each of the three detectors, LIGO Hanford, LIGO Livingston and Virgo), while the SNR of GW190425 was just 12.9 (in the LIGO Livingston detector) \cite{LIGOScientific:2017zic, LIGOScientific:2020aai}. The SNR of the GW170817 detection was about 2.5 times larger than that of GW190425. We expect statistical error to be inversely proportional to SNR. Therefore, as the SNR increases, the statistical error decreases. Thus, it makes sense that posteriors from GW190425 do not contain more information than those from GW170817. 

\section{Conclusions}
\label{sec:conclusions}
The posteriors shown in the previous section represent the first direct search for Einstein-\ae{}ther modifications in GW data.
Our study is also one of the first tests to compare LVC data to a waveform with the GR transverse-traceless polarization and with additional non-GR polarizations simultaneously, as predicted from a specific modified theory. 
While this study was unable to place tight constraints on the Einstein-\ae{}ther parameters, 
there is still a lot to learn from it. 
Our analysis reveals the complications that may arise in modified theories with multiple coupling constants to constrain, especially if any of those constants is degenerate with astrophysical parameters. 
Our analysis further demonstrates that constraints from the absence of a dipole term in GW radiation may continue to dominate other constraints from GW observations.
Finally, this work summarizes all of the current constraints on Einstein-\ae{}ther theory, giving a careful description of each region of parameter space and how sensitivities in this theory are affected in those regions. From this study, it is clear that region 1 of parameter space (as described in Sec.~\ref{subsec:c_prior}) will be accessible to GW studies before region 2 is. 

The results of this study prompt the question: what might improve the constraints that GW data can place on Einstein-\ae{}ther theory? There are several possible avenues to approach this question. Firstly we can consider the types of events that are being studied. It is possible that there are certain combinations of astrophysical source parameters that are better for constraining this theory than others. We only considered source parameters similar to those detected with BNS mergers to date. Perhaps there is some type of ``golden event'', that if we were fortunate enough to observe it, would greatly constrain the theory further. A good candidate for such a golden event is a mixed compact binary consisting of a low-mass BH and a neutron star. 
The analysis of such a system would require first the calculation of Einstein-\ae{}ther sensitivities for BHs.  
We can also consider what might be achieved with future events and future GW detectors. As detectors continue to improve and higher SNR events are detected, how will constraints on Einstein-\ae{}ther theory change? 
It seems reasonable to expect some improvement that scales as $1/$SNR, but it is unclear exactly how much the posteriors will change, because of the strong correlations between the Einstein-\ae{}ther parameters and other system parameters (like the chirp mass). 
Furthermore, as more BNS events are detected, constraints from each event can be combined, since the value of the Einstein-\ae{}ther coupling constants must be consistent across all events. 
On the order of 10 BNS events are predicted for the LVC fourth observing run, O4, starting later this year~\cite{Abbott2020}. 

Another possible consideration is improvement of the waveform template itself. This waveform template was built off of {\tt IMRPhenomD\_NRT}, which was fit to numerical relativity simulations in GR. There have so far been no numerical relativity simulations of binary NS mergers in Einstein-\ae ther theory. It is possible that fitting a waveform template to NR simulations in this theory would make it more accurate and better able to constrain the theory. However, developing such a simulation comes with its own set of challenges, and we doubt that the modifications would be so large to improve constraints beyond what has already been achieved with binary pulsar and solar system observations. 

Another large avenue of possible future work would be to extend this analysis to BHNS mergers or BBH mergers, if Einstein-\ae{}ther theory sensitivities were known for BHs, as mentioned before. If that were accomplished, the number of events that could be used for this study would increase dramatically, even before the next observing runs begin. At the very least, one could make assumptions about what the sensitivity for BHs in this theory is likely to be, and then examine the BHNS merger events. This would not place true constraints on Einstein-\ae{}ther theory parameters, because simplifying assumptions would have been made, but it may give some idea of what we might hope to learn from these events in the future. Ultimately, there is still much that could be investigated about GWs in Einstein-\ae{}ther theory. It would be especially useful to determine if there is any point at which GW constraints on Einstein-\ae{}ther theory will surpass those from current experiments. 

\acknowledgements
The authors would like to thank Toral Gupta, Enrico Barausse, Anzhong Wang, Chao Zhang, Tim Dietrich, and Nathan Johnson-McDaniel for helpful discussions. K.S. would like to acknowledge that this material is based upon work supported by the National Science Foundation Graduate Research Fellowship Program under Grant No. DGE – 1746047.
S. P. acknowledges partial support by the Center for AstroPhysical Surveys (CAPS) at the National Center for Supercomputing Applications (NCSA), University of Illinois Urbana-Champaign. 
K.Y. acknowledges support from NSF Grant PHY-1806776, PHY-2207349, a Sloan Foundation Research Fellowship and the Owens Family Foundation.
N. Y. acknowledges support from NSF Grant No. PHY 2207650.

This work made use of the Illinois Campus Cluster, a computing resource that is operated by the Illinois Campus Cluster Program (ICCP) in conjunction with the National Center for Supercomputing Applications (NCSA) and which is supported by funds from the University of Illinois at Urbana-Champaign.

Additionally, this research has made use of data or software obtained from the Gravitational Wave Open Science Center (gw-openscience.org), a service of LIGO Laboratory, the LIGO Scientific Collaboration, the Virgo Collaboration, and KAGRA. LIGO Laboratory and Advanced LIGO are funded by the United States National Science Foundation (NSF) as well as the Science and Technology Facilities Council (STFC) of the United Kingdom, the Max-Planck-Society (MPS), and the State of Niedersachsen/Germany for support of the construction of Advanced LIGO and construction and operation of the GEO600 detector. Additional support for Advanced LIGO was provided by the Australian Research Council. Virgo is funded, through the European Gravitational Observatory (EGO), by the French Centre National de Recherche Scientifique (CNRS), the Italian Istituto Nazionale di Fisica Nucleare (INFN) and the Dutch Nikhef, with contributions by institutions from Belgium, Germany, Greece, Hungary, Ireland, Japan, Monaco, Poland, Portugal, Spain. The construction and operation of KAGRA are funded by Ministry of Education, Culture, Sports, Science and Technology (MEXT), and Japan Society for the Promotion of Science (JSPS), National Research Foundation (NRF) and Ministry of Science and ICT (MSIT) in Korea, Academia Sinica (AS) and the Ministry of Science and Technology (MoST) in Taiwan.

\appendix
\renewcommand{\theequation}{\thesection.\arabic{equation}}
\section{{\tt IMRPhenomD\_NRTidal} Modifications}
\label{sec:appendixNRTidal}
To minimize confusion for anyone attempting to reproduce our code, we will describe here in detail the modifications that were made to the {\tt IMRPhenomD} waveform template to make it consistent with {\tt IMRPhenomD\_NRTidalv2}.

Eq.\ 17 of Dietrich \emph{et. al} gives the tidal phase correction in the frequency domain \cite{Dietrich:2019kaq}:
\begin{equation} 
\psi_T (x) = -\kappa_{\text{eff}}^T \frac{39}{16\eta} x^{5/2} \tilde{P}_{\text{NRTidalv2}}(x),
\label{eqn:NRT_phase}
\end{equation}
with 
\begin{equation} 
\kappa_{\text{eff}}^T = \frac{3}{16} \tilde{\Lambda},
\end{equation}
where $\tilde{\Lambda}$ is the commonly used mass-weighted tidal deformability (Eq.~\eqref{eqn:mass_weighted_tidal}), $\eta$ is the symmetric mass ratio (Eq.~\eqref{eqn:eta}), and
\begin{equation}
x = \left(\frac{\hat{\omega}}{2} \right)^{2/3} = \left(\pi m f_{GW} \right)^{2/3}, 
\end{equation} 
since $\hat{\omega} = 2\pi m f_{\text{GW}}$, with $m = m_1 + m_2$, is the dimensionless GW frequency. The last expression in Eq.~\eqref{eqn:NRT_phase} is the Pad\'e approximant (Eq.\ 18 of Dietrich \emph{et. al}) which is a function of $x$ with eight numerical coefficients, four of which were determined by fitting to data \cite{Dietrich:2018uni}: 
\begin{equation}
\tilde{P}_{\text{NRTidalv2}} (x) = \frac{1 + \sum_{i=0}^4 \tilde{n}_{1 + i/2} x^{1 + i/2}}{1 + \sum_{j=0}^{2} \tilde{d}_{1+j/2} x^{1 + j/2} }.
\end{equation}
The coefficients are given in Eqs.\ 19-21 of the NRTidal paper \cite{Dietrich:2019kaq}. However, in order for our waveform to match LALSuite as well as it does, we needed to use the same number of significant digits.\textit{ Hence, we took the values of these coefficients directly from LALSuite's code}. We copy them here in table \ref{tab:NRTcoeff} for convenience. 

\begin{table}[]
    \centering
    \begin{tabular}{c|c|c}
        i & $\tilde{n}_{1 + i/2}$ & $\tilde{d}_{1 + i/2}$  \\\hline 
        0 & $-12.615214237993088$ & $-15.111207827736678$ \\
        1 & $19.0537346970349$ & $22.195327350624694$ \\
        2 & $-21.166863146081035$ & $8.064109635305156$\\
        3 & $90.55082156324926$ & 0 \\
        4 & $-60.25357801943598$ & 0
    \end{tabular}
    \caption{The coefficients of the Pad\'{e} approximant used in the tidal correction to the phase. To make our code consistent with LALSuite it was necessary to use these exact numbers.}
    \label{tab:NRTcoeff}
\end{table}

The tidal amplitude correction in the frequency domain is given by Eq.\ 24 of Dietrich \emph{et. al} \cite{Dietrich:2019kaq}: 
\begin{align} \tilde{A}_T^{\text{NRTidalv2}} &= -\sqrt{\frac{5\pi \nu}{24}} \frac{9m^2}{R} \kappa^T_{\text{eff}} x^{13/4} \frac{1 + \frac{449}{108} x + \frac{22672}{9} x^{2.89}}{1 + 13477.8 x^4}. \end{align}
A Planck taper is used to end the inspiral waveform, beginning at the merger frequency \cite{Dietrich:2018uni}, 
\begin{equation}
    f_{\text{merger}} = \frac{0.3586}{2m\pi} \sqrt{\frac{m_2}{m_1}} \frac{1 + n_1 \kappa_{\text{eff}}^T + n_2 \left(\kappa_{\text{eff}}^T\right)^2}{1 + d_1 \kappa_{\text{eff}}^T + d_2 \left(\kappa_{\text{eff}}^T\right)^2},
\end{equation}
with $n_1 = 3.354\times 10^{-2}, n_2 = 4.315 \times 10^{-5},d_1 = 7.542\times 10^{-2}, d_2 = 2.236 \times 10^{-4}$,
and reducing the amplitude to zero by the time $f = 1.2 f_{\text{merger}}$ \cite{McKechan:2010kp, Dietrich:2018uni}. 
The exact form of this taper\edit{, $\tilde{A}_{\text{Planck}}$,} can be found in Eq. 7 of~\cite{McKechan:2010kp}\edit{. We repeat it here for convenience,}
\begin{equation}
    \edit{\tilde{A}_{\text{Planck}} = \begin{cases}
        1, & f \leq f_{\text{merger}}, \\
        \frac{1}{\exp{(z(f))} + 1}, & f_{\text{merger}} \leq f \leq 1.2 f_{\text{merger}}, \\
        0, &  f \geq 1.2f_{\text{merger}},
    \end{cases}}
\end{equation}
\edit{where}
\begin{equation}
    \edit{z(f) = \frac{f_{\text{merger}} - 1.2f_{\text{merger}}}{f - f_{\text{merger}}} + \frac{f_{\text{merger}} - 1.2f_{\text{merger}}}{f - 1.2f_{\text{merger}}}}.
\end{equation}
Putting it all together,  the final amplitude in the frequency domain is \cite{Dietrich:2019kaq}, 
\begin{equation}
\tilde{A} = (\tilde{A}_{\text{BBH}} + \tilde{A}_T^{\text{NRTidalv2}} ) \times \tilde{A}_{\text{Planck}}.
\end{equation} 

The {\tt IMRPhenomD\_NRTidalv2} waveform template also accounts for spin-spin effects in the phase. The terms added to the BBH baseline phase are \cite{Dietrich:2019kaq}, 
\begin{align}
    \Psi_{SS} &= \frac{3x^{-5/2}}{128\eta} \left(\psi_{SS,2PN}^{(1)} x^2 + \psi_{SS,3PN}^{(1)} x^3 \right. \nonumber \\
    &\2\2 \left.+ \psi_{SS,3.5PN}^{(1)} x^{7/2} \right) + [(1) \leftrightarrow (2)] 
\end{align}
where (1) and (2) represent the two bodies in the binary system (with $m_1 \geq m_2$ as before). The 2PN and 3PN terms were already implemented in LALSuite \cite{Bohe:2015ana, Mishra:2016whh, Krishnendu:2017shb}: 
\begin{align}
    \psi_{SS,2PN}^{(1)} &= -50 (C_Q^{(1)} - 1) \mu_1^2 \chi_1^2, \\
\psi_{SS,3PN}^{(1)} &= \frac{5}{84} \left(9407 + 8218 \mu_1 - 2016 \mu_1^2 \right) \nonumber \\
&\2\2 \times (C_Q^{(1)} - 1) \mu_1^2 \chi_1^2, 
\end{align}
and the 3.5PN term was added by \cite{Dietrich:2019kaq}:
\begin{align}
\psi_{SS,3.5PN}^{(1)} &= 10 \left[ \left(\mu_1^2 + \frac{308}{3} \mu_1 \right) \chi_1 + \left(\mu_2^2 - \frac{89}{3} \mu_2 \right) \chi_2 \nonumber \right. \\
&\2\2  - 40\pi \Big] (C_Q^{(1)} - 1) \mu_1^2 \chi_1^2 \nonumber \\
&\2\2 - 440 (C_{Oc}^{(1)} - 1) \mu_1^3 \chi_1^3,
\end{align}
where $\mu_{1,2} = m_{1,2}/m$ as before, $\chi_{1,2}$ are the spins of each body, and $C_Q^{(1,2)}$ and $C_{Oc}^{(1,2)}$ are the spin-induced deformabilities for the individual stars which can be related to the tidal deformability with the universal relations \cite{Yagi:2016bkt}, 
\begin{align}
     C_Q^{(1,2)}  = \exp\left( \sum_{i=0}^{4} q_i \ln(\Lambda_{1,2})^i\right) , \\
     C_{Oc}^{(1,2)} = \exp\left( \sum_{i=0}^{4} o_i \ln(C_Q^{(1,2)})^i\right) ,
\end{align}
with coefficients in table \ref{tab:quadoctcoeff}. We computed $C_Q$ and $C_{Oc}$ for the specific case $\Lambda_1 = \Lambda_2 = 350$ to compare against the values used for Fig.\ 7 of \cite{Dietrich:2019kaq}, and caught a small typo in the caption of that image. The correct values, which were used to create the plot, are $C_Q = 5.29$ and $C_{Oc} = 10.5$.

\textit{Note that because the 2PN and 3PN spin-spin terms were added to the code earlier, they are implemented in a different way from the 3.5PN spin-spin term and the tidal effects.} To make our code consistent with LALSuite, we had to follow their convention. Thus, the 2PN and 3PN spin-spin terms were added to the PN terms in the inspiral only. This carries through to higher frequencies via boundary conditions when the different parts of the waveform are stitched together. Meanwhile, the 3.5PN spin-spin term and the tidal modifications to the phase and amplitude are added to the entire waveform so that the underlying BBH model did not need to be recalibrated. 

\begin{table}[]
    \centering
    \begin{tabular}{c|c|c|c|c|c}
        $i$ & 0 & 1 & 2 & 3 & 4 \\\hline
        $q_i$ & $0.1940$ & $0.09163$ & $0.04812$ & $-0.004283$ &  $0.00012450$ \\
        $o_i$ & $0.003131$ & $2.071$ &  $-0.7152$ &  $0.2458$ & $-0.03309$ \\
    \end{tabular}
    \caption{The coefficients for the quadrupolar and octupolar spin-induced deformabilities as a function of tidal deformability.}
    \label{tab:quadoctcoeff}
\end{table}

\section{Order of Magnitude of the Sensitivity Parameter}
\label{sec:appendixSensitivities}
The implementation of the sensitivity model in {\tt GWAT} was tested in Sec.~\ref{subsec:EA_in_code} and compared against previous work~\cite{Gupta:2021vdj}. 
However, this was done for the most restrictive prior on the Einstein-\ae{}ther parameters (described in detail in Sec.~\ref{sec:alpha_prior}) and as discussed in Sec.~\ref{sec:injections}, in this region of parameter space the prior is more informative than the likelihood. 
Thus, we also considered a slightly less restrictive prior as outlined in Sec.~\ref{sec:injections}. 
In this appendix we demonstrate how this new prior affects the calculation of sensivities in Einstein-\ae{}ther theory. 

\begin{figure}
    \centering
    \includegraphics[width=\linewidth]{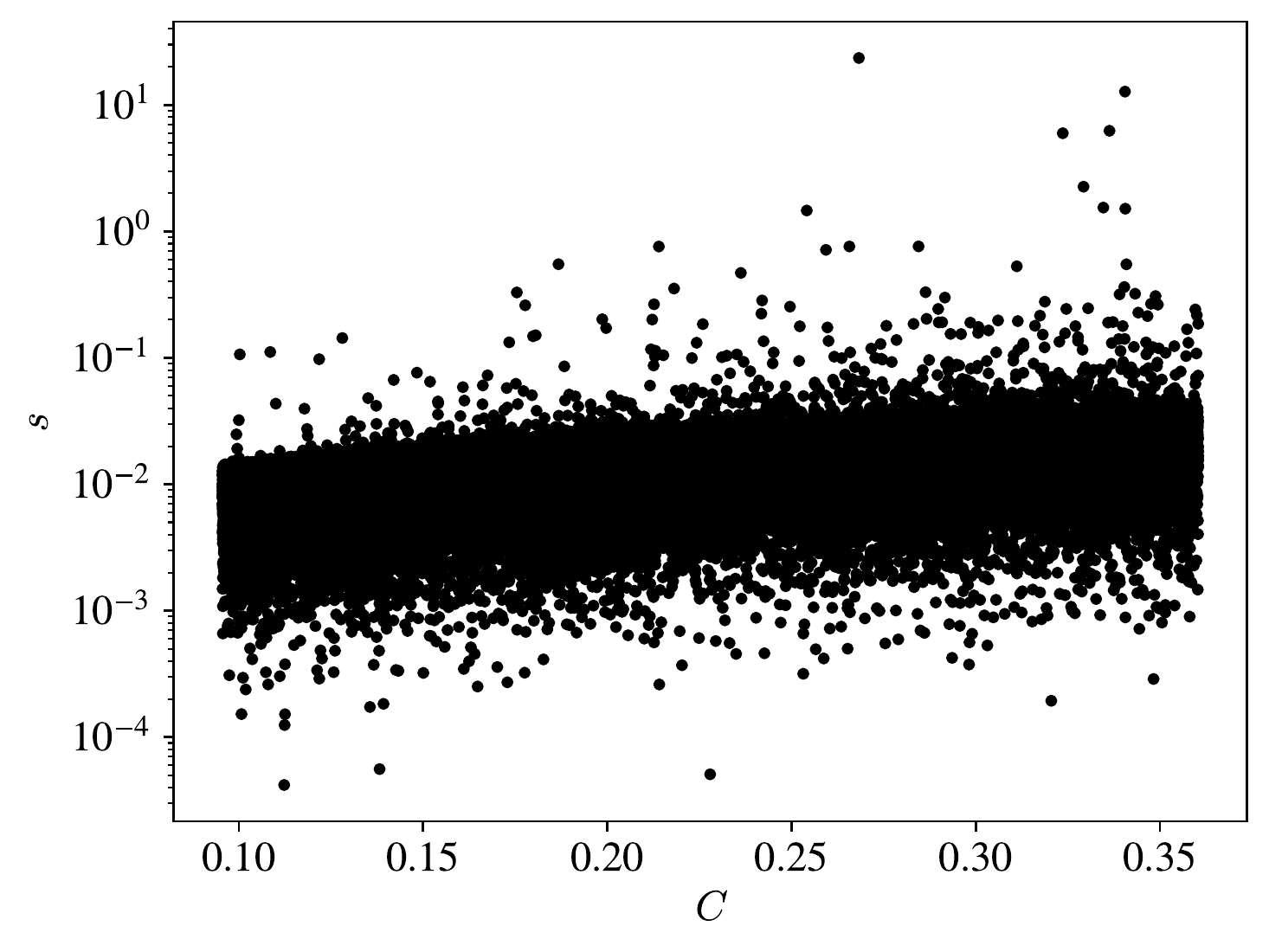}
    \caption{Sensitivity as a function of compactness varying the Einstein-\ae{}ther parameters in the region of parameter space used for this study. A comparison with Fig.~\ref{fig:CLove_sC} reveals that in this region, the sensitivities are approximately three orders of magnitude larger than in the most restrictive region of parameter space.}
    \label{fig:sensitivity_expanded}
\end{figure}

We begin by plotting sensitivity as a function of compactness for 50,000 random values of compactness when the Einstein-\ae{}ther parameters are varied in the region of parameter space relevant to this work (Fig.~\ref{fig:sensitivity_expanded}). 
Recall from Sec.~\ref{sec:injections}, the prior includes the stability conditions, the Cherenkov constraint, and the BBN constraint, while it excludes the solar system constraints and the constraint on $\alpha_1$ from binary pulsars and the triple system. 
In this region, \textit{the sensitivities calculated are approximately three orders of magnitude larger than in the region considered in previous work}. 
This increase is consistent with the increase in magnitude of $\alpha_1$ from one region to the other since the dominant contribution to sensitivity from the Einstein-\ae{}ther coupling constants is linear in $\alpha_1$ (recall Eq.~\eqref{eqn:sensitivity} and the fact that $\alpha_2$ is much smaller than $\alpha_1$).

One important consequence of working in a less restrictive region of parameter space is that it is possible to select a combination of coupling constants with $s \geq 1$. 
Given the definition of $s$ in terms of $\sigma$ (Eq.~\eqref{eqn:s_def}), $s \geq 1$ is unphysical.
Furthermore, when $s > 1$, there are quantities in the waveform (namely $A_{(2)}(f)$, Eq.~\eqref{eqn:l2amp}) that depend on $\sqrt{(1-s)}$ that the code will fail to calculate. 
Therefore, points with $s \geq 1$ should also be rejected. 

Note that in the region of parameter space we use in this study, only $10$ out of $50,000$ points had $s \geq 1$. So the problematic points are occurring with a frequency of $0.02\%$ and can be safely removed from our data without affecting our result. 
However, this issue only gets worse as one moves to larger regions of parameter space and the magnitude of $\alpha_1$ increases. We recommend that anyone wishing to examine a less restrictive region of the parameter space thoroughly test the sensitivity model in that region to ensure it does not break down. 

\begin{figure}
    \centering
    \includegraphics[width=\linewidth]{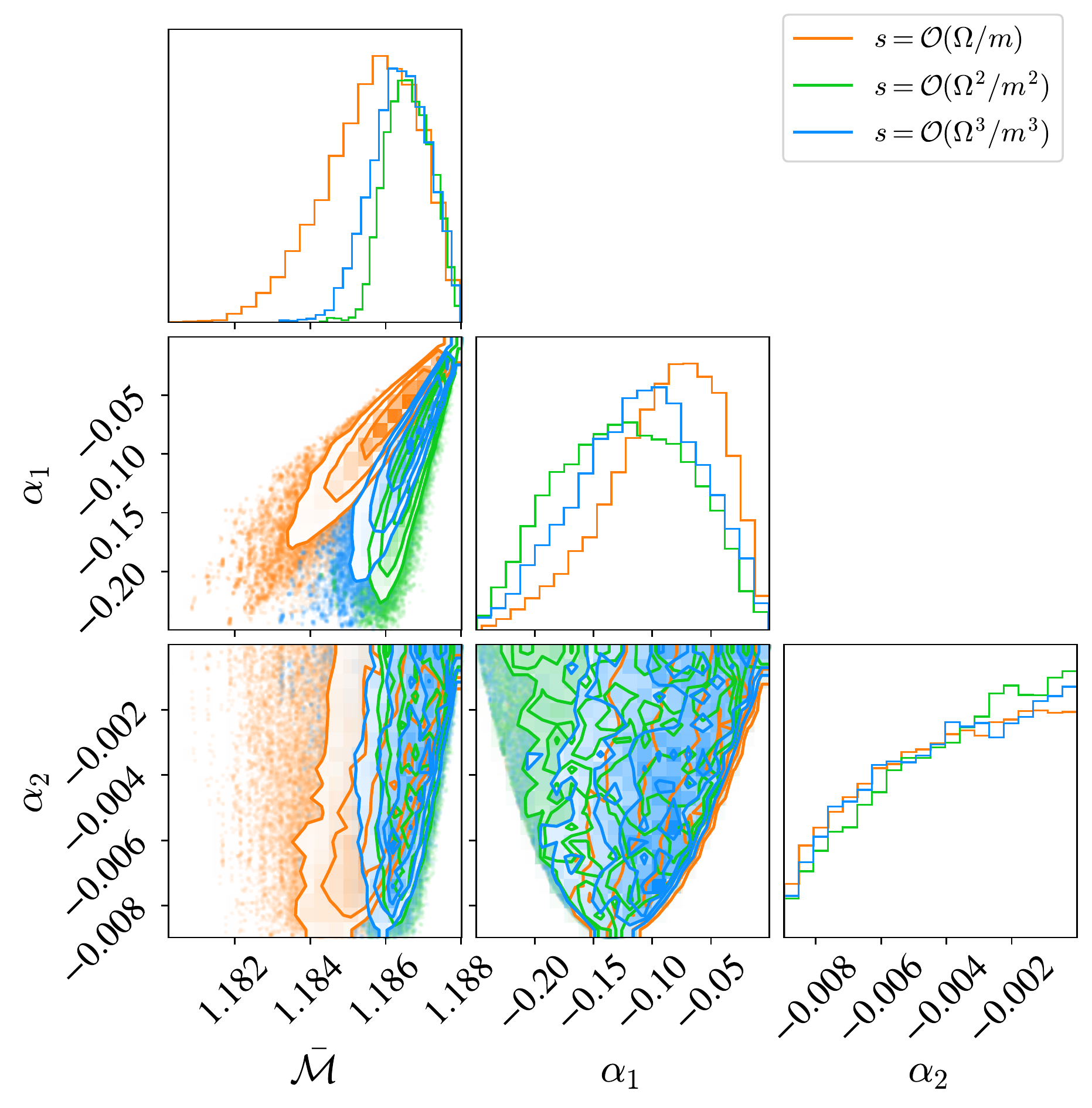}
    \caption{The posteriors for $\bar{\mathcal{M}}$, $\alpha_1$ and $\alpha_2$ for injected data when sensitivity is computed to different orders. This demonstrates how much our results might depend on the sensitivity model.}
    \label{fig:sensitivity_orders}
\end{figure}

To explicitly illustrate how much our result depends on the sensitivity model, we performed parameter estimation on the same injected data\footnote{The injected data was generated with the {\tt IMRPhenomD\_NRT} waveform template and used the source parameters listed in Table~\ref{tab:injected_data}. } while computing the sensitivity to different orders in the binding energy to mass ratio. 
In Fig.~\ref{fig:sensitivity_orders}, we compare three different runs with $s$ computed to $\{\mathcal{O}(\Omega/m), \mathcal{O}(\Omega^2/m^2), \mathcal{O}(\Omega^3/m^3)\}$ respectively. 
The difference in shape for the correlation between $\bar{\mathcal{M}}$ and $\alpha_1$ can be explained with Eqs.~\eqref{eqn:sensitivity} and~\eqref{eqn:sensitivity_chirpmass}.
To explain this shape analytically, we will treat $\alpha_2$ as negligible compared to $\alpha_1$ (a good approximation in the region we sample in) and keep $\Omega/m$ constant. 
Then as $\alpha_1$ is varied from $[-.25, 0]$ the first term in Eq.~\eqref{eqn:sensitivity} is the largest and is positive, the second term is smaller and negative, and the third term provides a very small positive contribution. 
All three terms tend to zero as $\alpha_1 \rightarrow 0$. Adding these terms together order by order, we get three different expressions for $s$ and we can see how they depend on $\alpha_1$. 
This same dependence appears in the correlation plot between $\bar{\mathcal{M}}$ and $\alpha_1$ because of the dependence of $\bar{\mathcal{M}}$ on $s$ (Eq.~\eqref{eqn:sensitivity_chirpmass}). 
Given how much our posteriors depend on how many terms are included in the sensitivity calculation, we recommend that the sensitivity model be further investigated (for instance, computed to higher orders) before constraints are placed with GW data. 

\section{Cherenkov Constraints}
\label{sec:appendixCherenkov}
To summarize the constraints of \cite{Elliott:2005va}, when $c_T < 1$, 
\begin{equation}
\frac{-c_\sigma}{2} < 5 \times10^{-16}, \label{eqn:tensorCherenkov}
\end{equation}
when $c_V < 1$, 
\begin{equation}
\Big|\frac{2c_\sigma^2 [c_\sigma^2 - 2c_a - (c_\sigma + c_\omega)]}{\left(c_\omega + c_\sigma\right)^2}\Big| < 7 \times10^{-32}, \label{eqn:vectorCherenkov}
\end{equation}
and when $c_S < 1$, 
\begin{equation}
\Big|\frac{(c_\sigma  - c_a)^2}{c_a} \Big|  < 1 \times 10^{-30}. \label{eqn:scalarCherenkov}
\end{equation}
This last constraint only holds when
\begin{equation}
\Big|\frac{2 \left[ (c_\theta + 2c_\sigma)/3 - c_a \right]}{c_\omega + c_\sigma}\Big| > 10^{-22}
\end{equation}
is also satisfied. 

Note that all of the emission processes which would place the constraints of Eqs.~\eqref{eqn:tensorCherenkov} - \eqref{eqn:scalarCherenkov} vanish as the $c_i$'s tend to zero. However, the emission of two scalar \ae ther field excitations via an off-shell graviton propagator does not vanish in this limit and provides a bound on the ratios of the $c_i$ for $c_S < 1$, namely 
\begin{equation}
\Big| \frac{2\left[c_a - \left(2c_\sigma + c_\theta \right)/3 \right]}{c_\omega + c_\sigma} \Big| < 3 \times 10^{-19}. \label{eqn:scalarCherenkov2}
\end{equation}

Together, Eqs.~\eqref{eqn:tensorCherenkov}-\eqref{eqn:scalarCherenkov2} are the conditions explicitly checked by {\tt GWAT} as part of the prior. Any points that meet the conditions for the constraint to be imposed (e.g. $c_V < 1$), but do not satisfy these equations (in this example, Eq.~\eqref{eqn:vectorCherenkov}) are rejected. It is important to note that because we are setting $c_\sigma = 0$ identically, the constraint of Eq.~\eqref{eqn:vectorCherenkov} will be satisfied for every combination of the Einstein-\ae{}ther parameters. Thus, for the prior, note that $c_V < 1$ is actually allowed. However, given that $c_V = c_\omega/2c_a$ when $c_\sigma = 0$, there are conditions in the likelihood that disfavor $c_\omega < 2c_a $ (or equivalently $c_V < 1$) in the posterior, as discussed in Sec.~\ref{sec:injections}.

\section{Recovery of Injected Parameters}
\label{sec:appendixInjecPlots}
\begin{figure*}
\includegraphics[width=0.45\linewidth,clip=true]{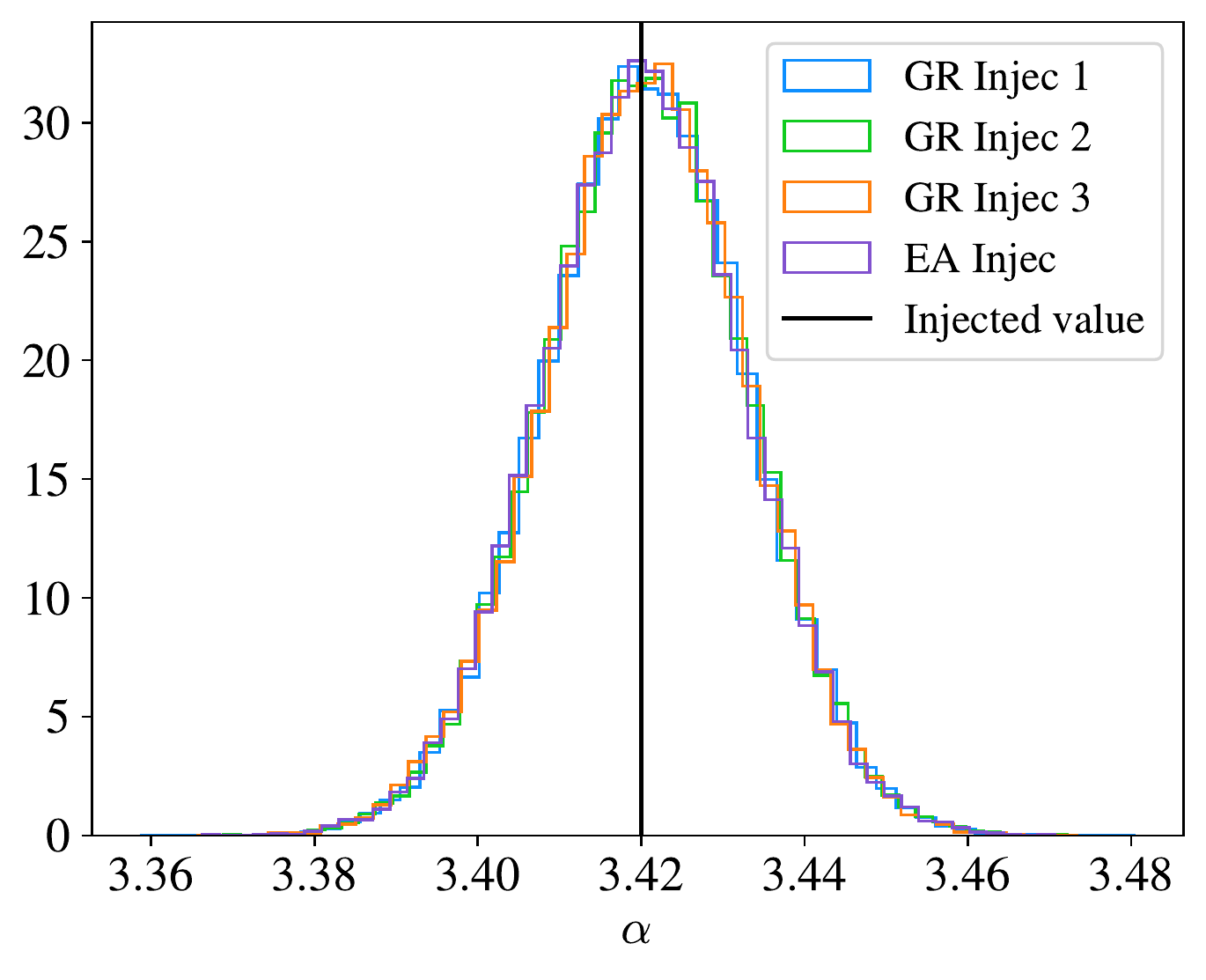}
\includegraphics[width=0.45\linewidth,clip=true]{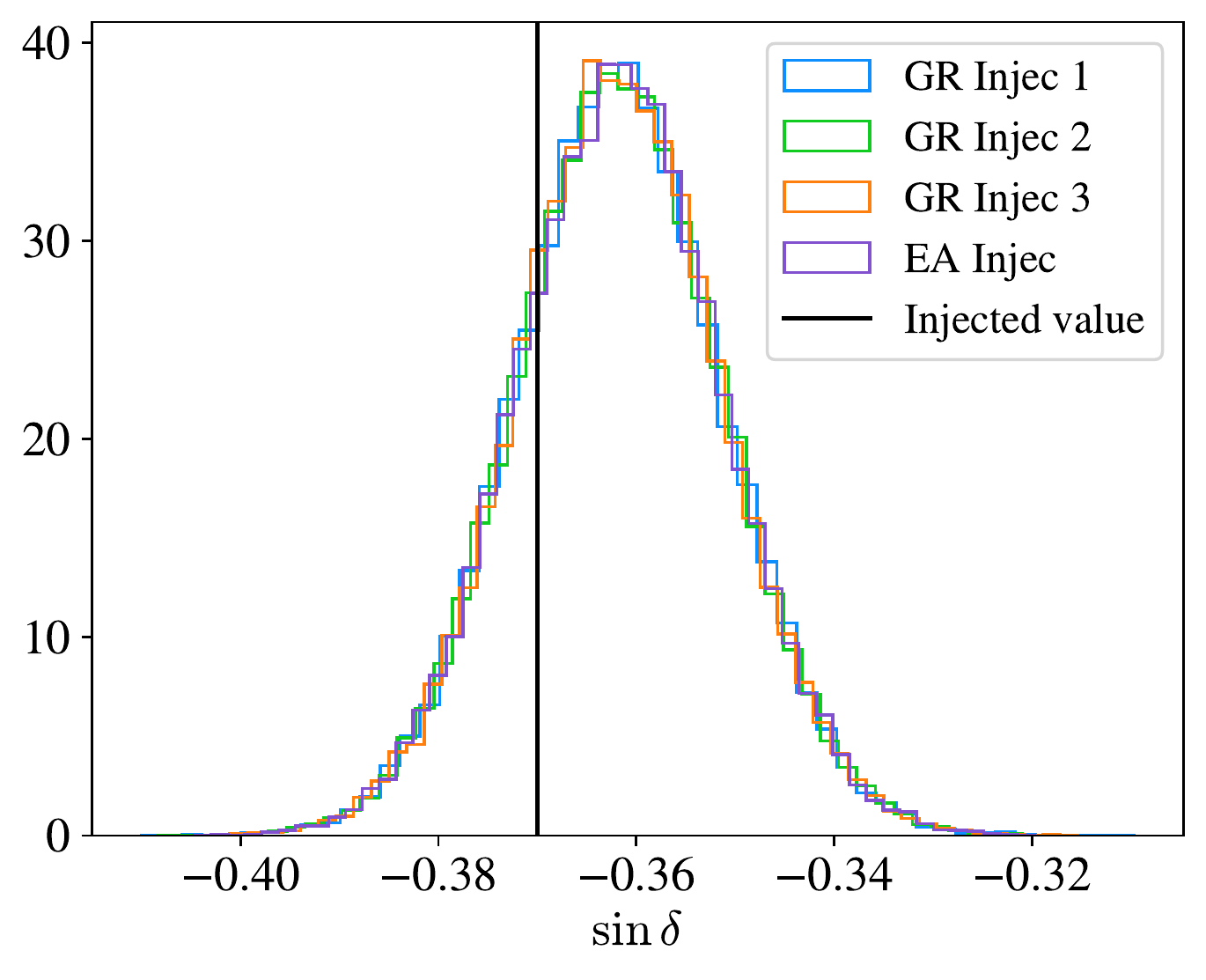} \\
\includegraphics[width=0.43\linewidth,clip=true]{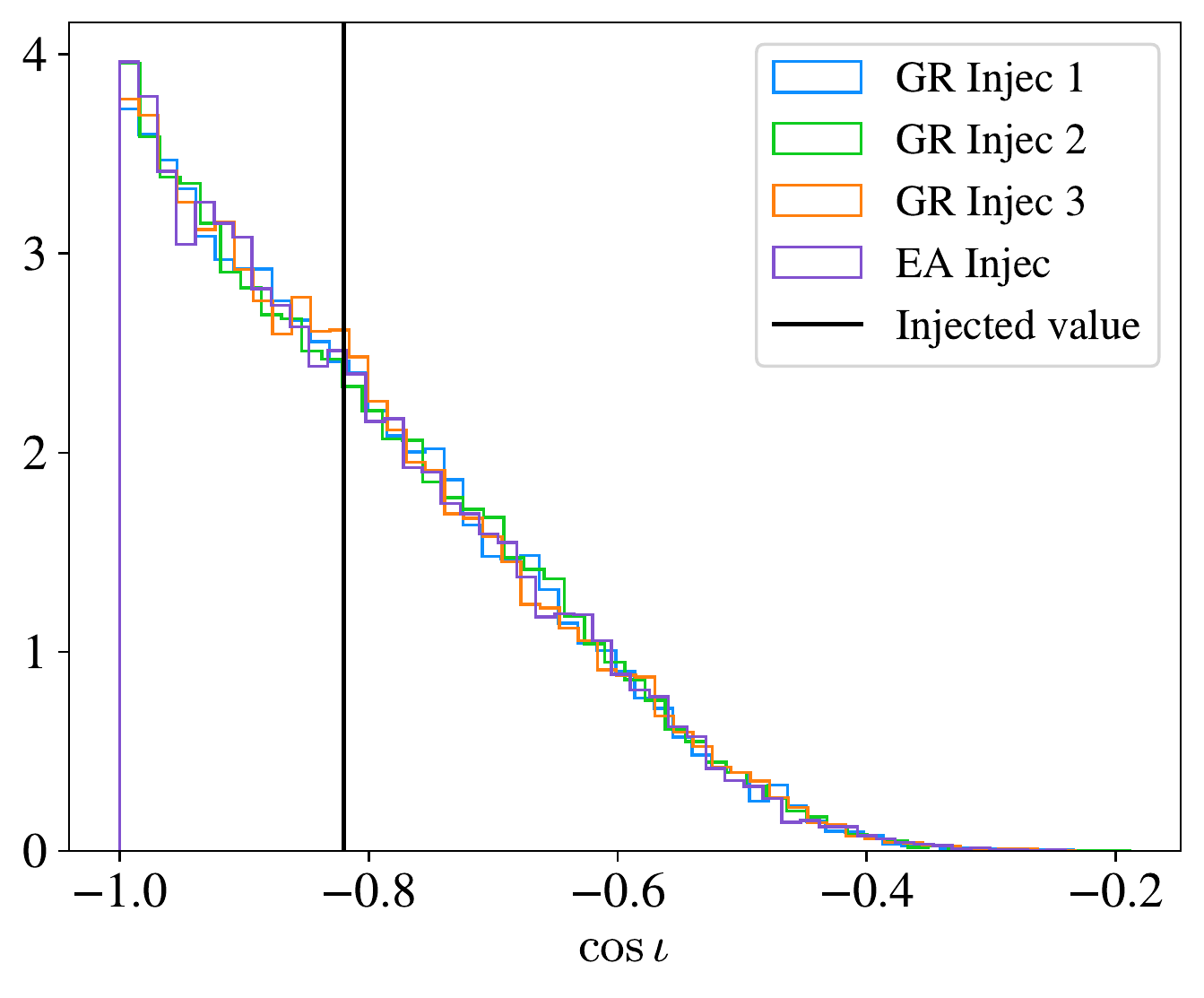}
\includegraphics[width=0.45\linewidth,clip=true]{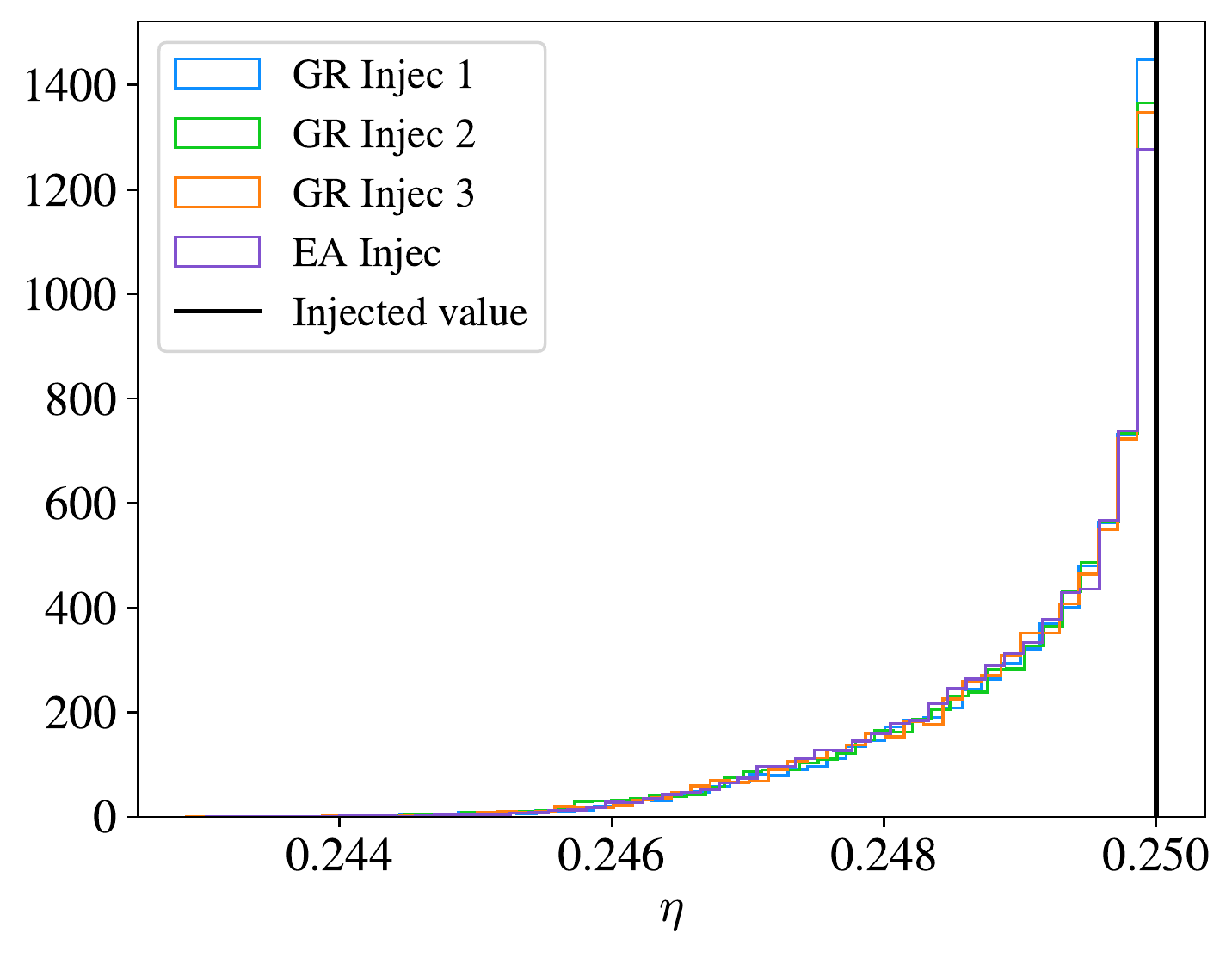} \\
\includegraphics[width=0.45\linewidth,clip=true]{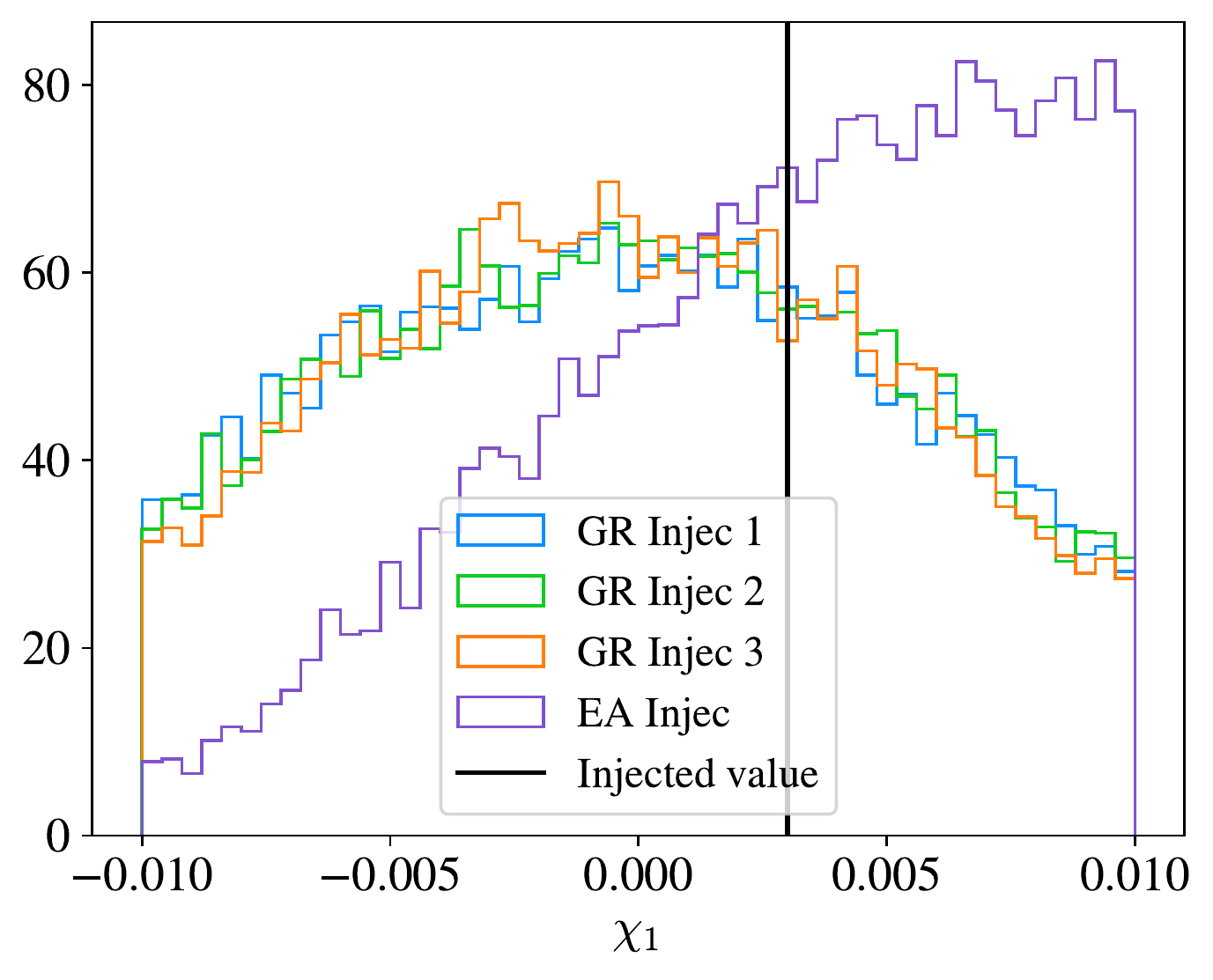}
\includegraphics[width=0.45\linewidth,clip=true]{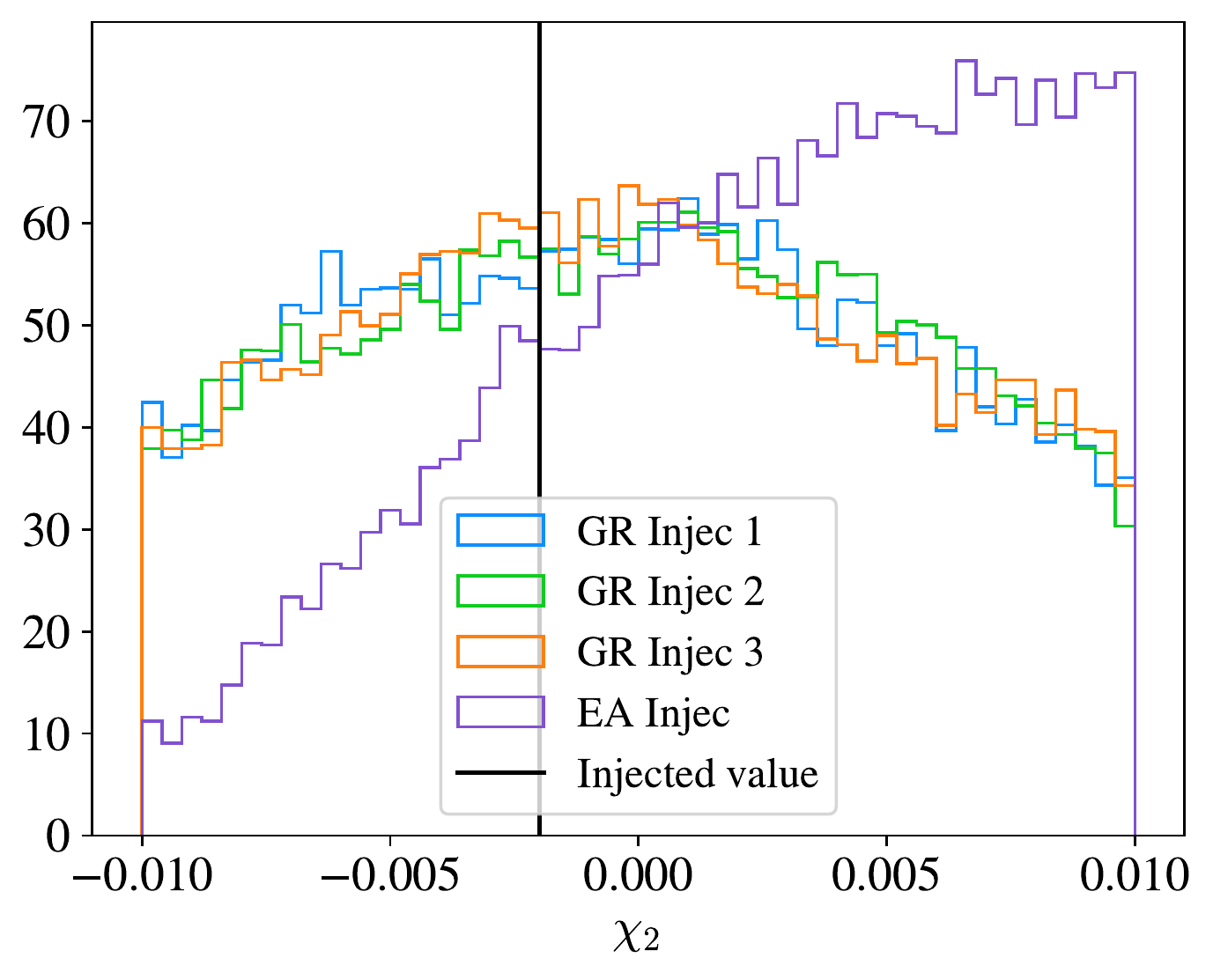}
\caption{Posteriors recovered from injections (GWs in GR and then GWs in Einstein-\ae ther, labeled EA on the plots) with the {\tt EA\_IMRPhenomD\_NRT} waveform template. All injected values lie within the 90\% credible region. }
\label{fig:general_param_recovery_1}
\end{figure*}

\begin{figure*}
\includegraphics[width=0.47\linewidth,clip=true]{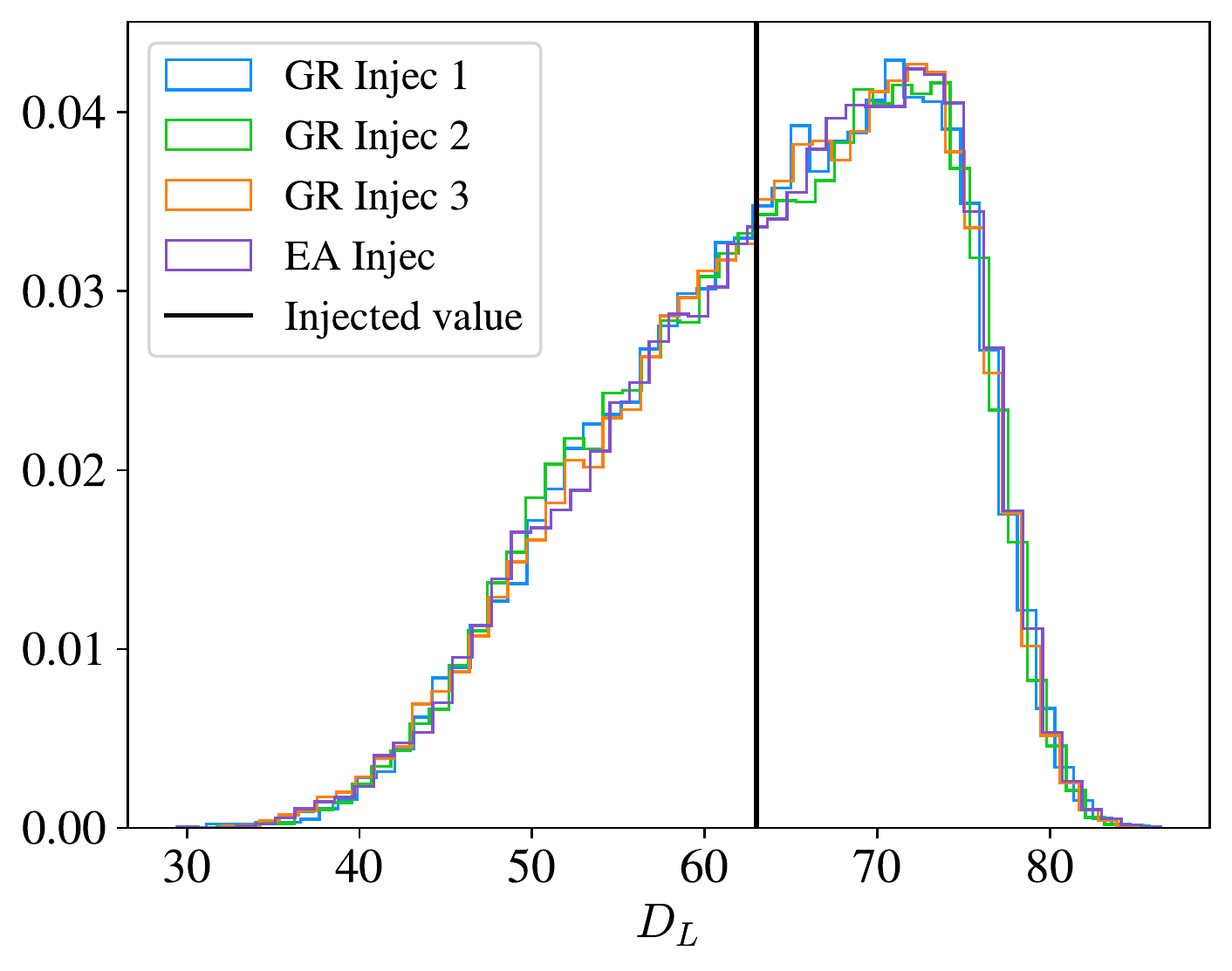}
\includegraphics[width=0.49\linewidth,clip=true]{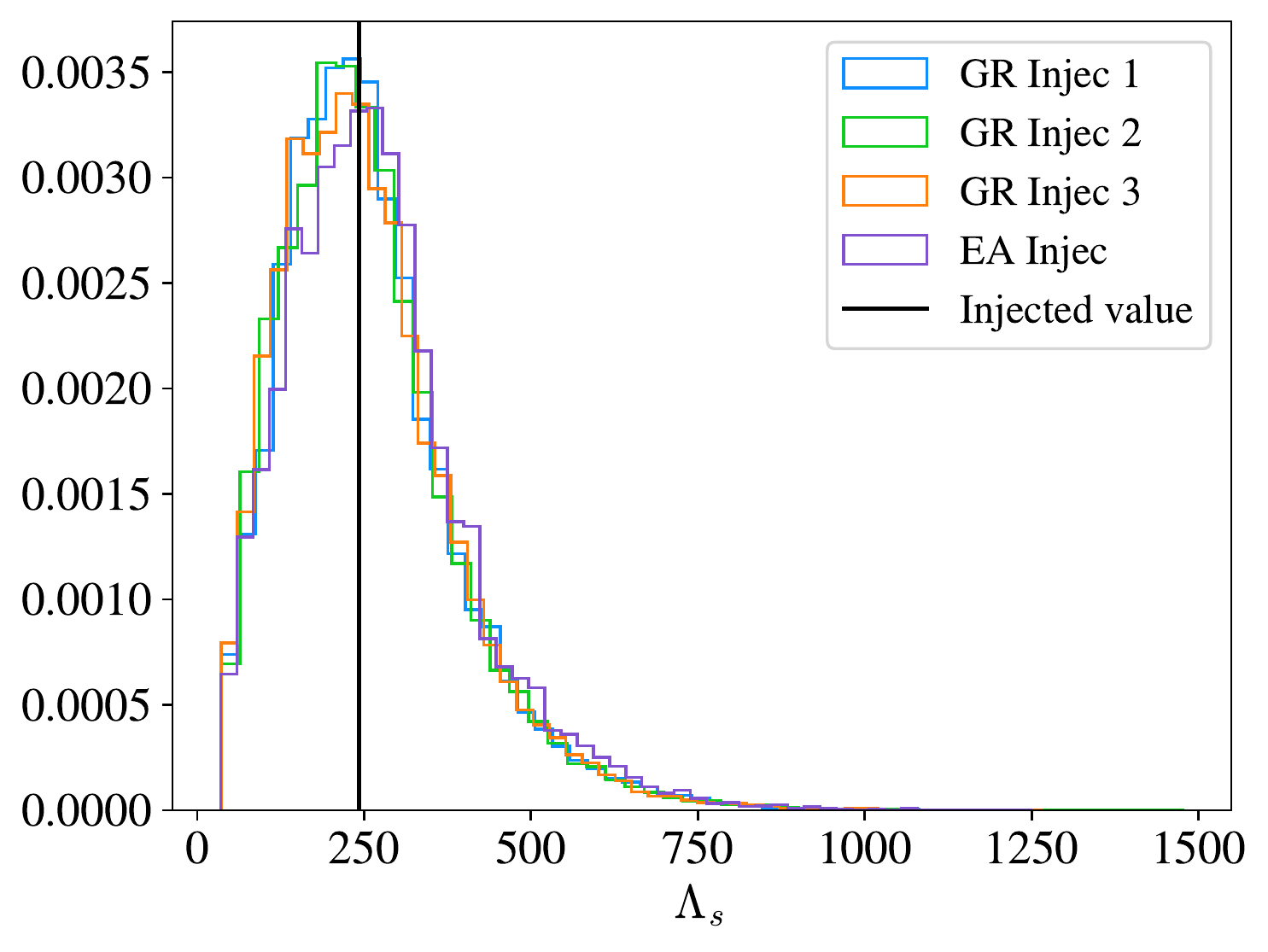} \\
\includegraphics[width=0.49\linewidth,clip=true]{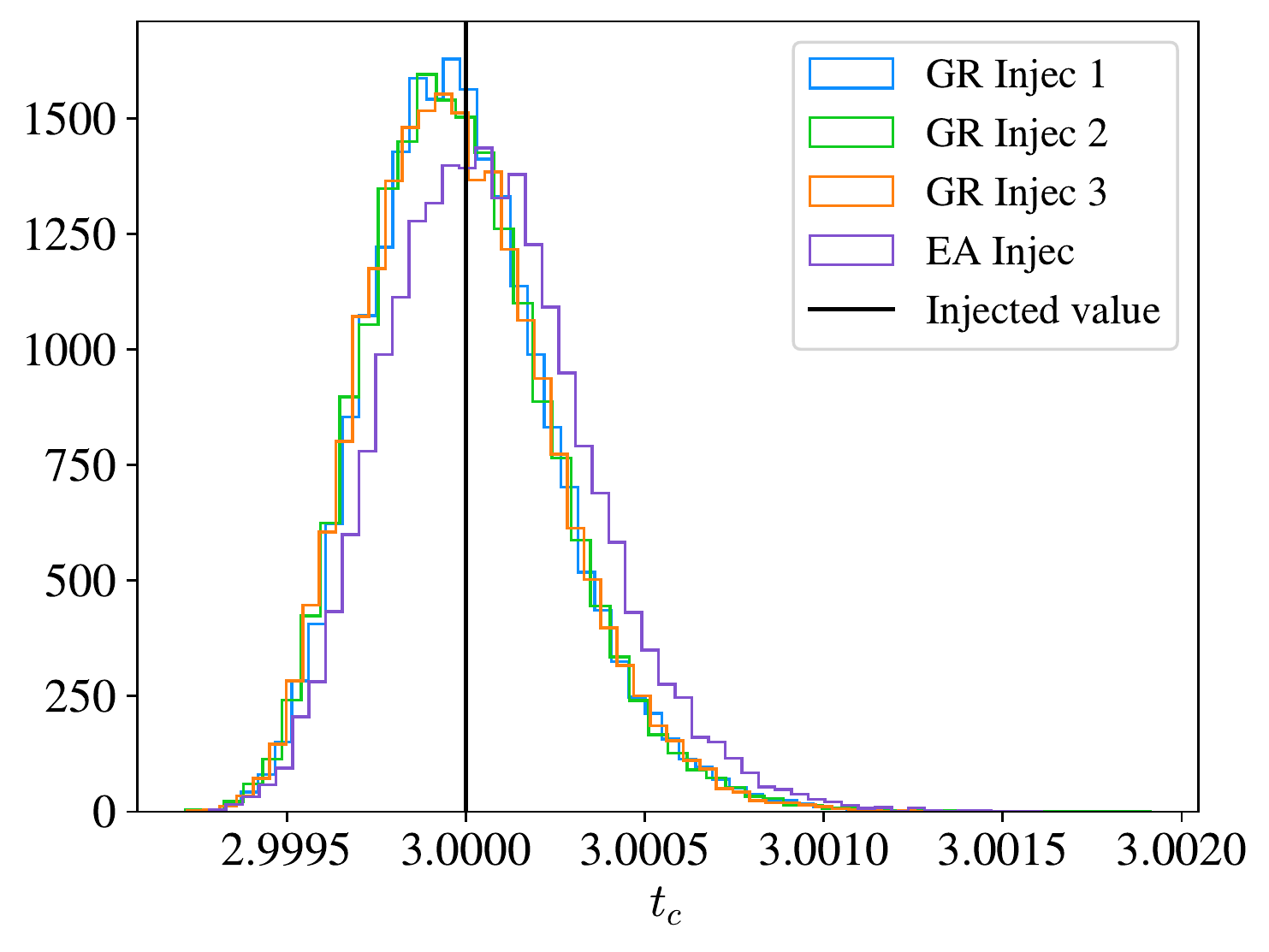}
\includegraphics[width=0.47\linewidth,clip=true]{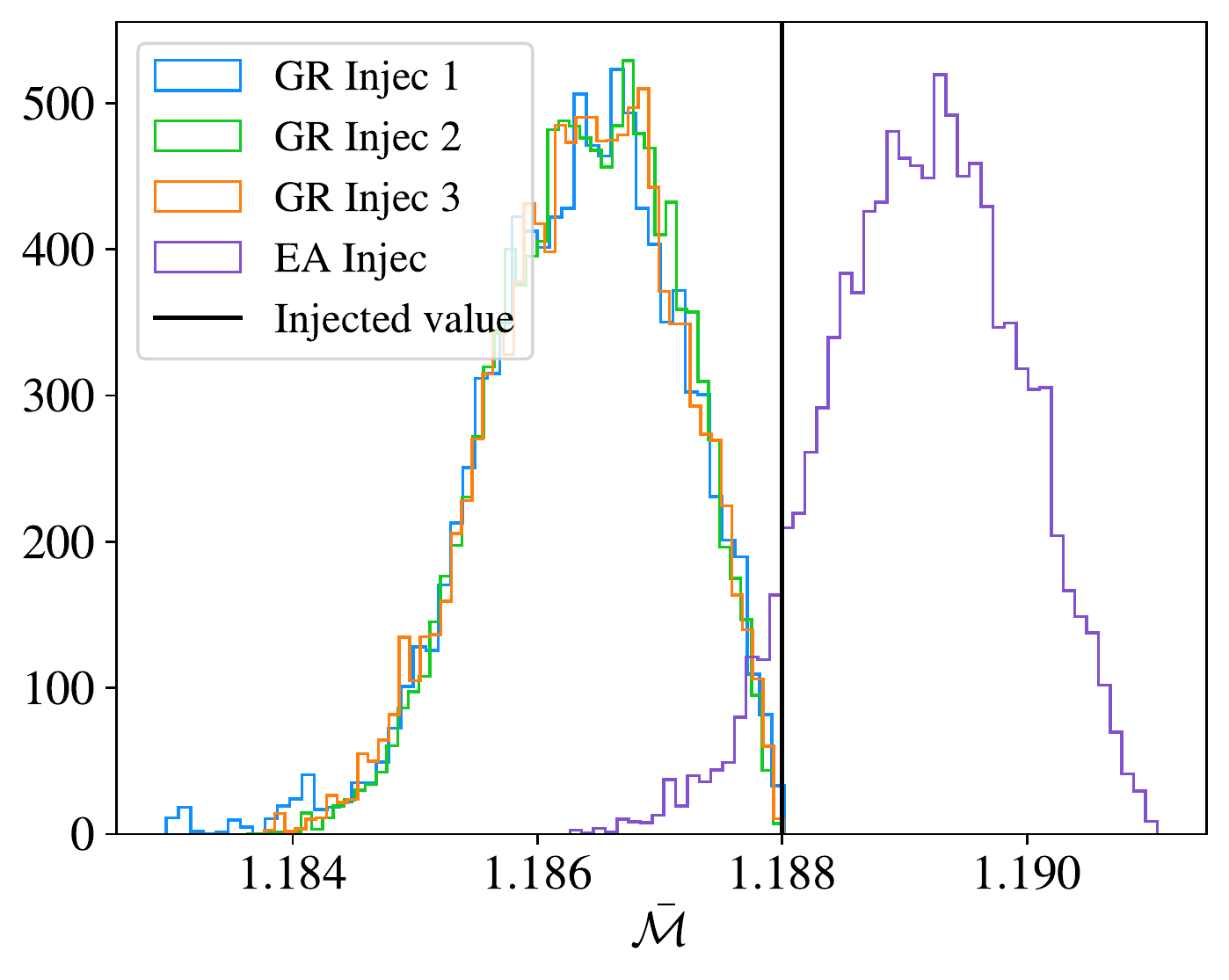}
\caption{Posteriors recovered from injections (GWs in GR and then GWs in Einstein-\ae ther, labeled EA on the plots) with the {\tt EA\_IMRPhenomD\_NRT} waveform template. All injected values were successfully recovered within the 90\% credible region \textit{except} for the chirp mass posterior which is extremely asymmetric in the case of the GR injection. This is because of the correlation between $\bar{\mathcal{M}}$ and $\alpha_1$ and is discussed in Sec.\ \ref{sec:injections}.}
\label{fig:general_param_recovery_2}
\end{figure*}

In this section we present comparisons between  posteriors recovered with the {\tt EA\_IMRPhenomD\_NRT} waveform template and injected values (Figs.~\ref{fig:general_param_recovery_1} and~\ref{fig:general_param_recovery_2}). As described in section \ref{sec:injections}, this was done for two different cases: a GR case and a non-GR case. In the GR case, the input data was constructed with the {\tt IMRPhenomD\_NRT} waveform template which does not specify the Einstein-\ae ther parameters. In the non-GR case, the EA injection, the input data was constructed with the {\tt EA\_IMRPhenomD\_NRT} waveform template and the Einstein-\ae ther parameters were given values distinct from those in the GR case (no longer zero or 1). 

\bibliography{introduction, background, GWs, waveform_template, current_constraints, injections, GW_constraints, conclusions, appendixNRTidal}

\begin{thebibliography}{61}%
\makeatletter
\providecommand \@ifxundefined [1]{%
 \@ifx{#1\undefined}
}%
\providecommand \@ifnum [1]{%
 \ifnum #1\expandafter \@firstoftwo
 \else \expandafter \@secondoftwo
 \fi
}%
\providecommand \@ifx [1]{%
 \ifx #1\expandafter \@firstoftwo
 \else \expandafter \@secondoftwo
 \fi
}%
\providecommand \natexlab [1]{#1}%
\providecommand \enquote  [1]{``#1''}%
\providecommand \bibnamefont  [1]{#1}%
\providecommand \bibfnamefont [1]{#1}%
\providecommand \citenamefont [1]{#1}%
\providecommand \href@noop [0]{\@secondoftwo}%
\providecommand \href [0]{\begingroup \@sanitize@url \@href}%
\providecommand \@href[1]{\@@startlink{#1}\@@href}%
\providecommand \@@href[1]{\endgroup#1\@@endlink}%
\providecommand \@sanitize@url [0]{\catcode `\\12\catcode `\$12\catcode `\&12\catcode `\#12\catcode `\^12\catcode `\_12\catcode `\%12\relax}%
\providecommand \@@startlink[1]{}%
\providecommand \@@endlink[0]{}%
\providecommand \url  [0]{\begingroup\@sanitize@url \@url }%
\providecommand \@url [1]{\endgroup\@href {#1}{\urlprefix }}%
\providecommand \urlprefix  [0]{URL }%
\providecommand \Eprint [0]{\href }%
\providecommand \doibase [0]{https://doi.org/}%
\providecommand \selectlanguage [0]{\@gobble}%
\providecommand \bibinfo  [0]{\@secondoftwo}%
\providecommand \bibfield  [0]{\@secondoftwo}%
\providecommand \translation [1]{[#1]}%
\providecommand \BibitemOpen [0]{}%
\providecommand \bibitemStop [0]{}%
\providecommand \bibitemNoStop [0]{.\EOS\space}%
\providecommand \EOS [0]{\spacefactor3000\relax}%
\providecommand \BibitemShut  [1]{\csname bibitem#1\endcsname}%
\let\auto@bib@innerbib\@empty
\bibitem [{\citenamefont {Abbott}\ \emph {et~al.}(2021{\natexlab{a}})\citenamefont {Abbott} \emph {et~al.}}]{LIGOScientific:2021djp}%
  \BibitemOpen
  \bibfield  {author} {\bibinfo {author} {\bibfnamefont {R.}~\bibnamefont {Abbott}} \emph {et~al.} (\bibinfo {collaboration} {LIGO Scientific, VIRGO, KAGRA}),\ }\href@noop {} {\bibinfo {title} {{GWTC-3: Compact Binary Coalescences Observed by LIGO and Virgo During the Second Part of the Third Observing Run}}} (\bibinfo {year} {2021}{\natexlab{a}}),\ \Eprint {https://arxiv.org/abs/2111.03606} {arXiv:2111.03606 [gr-qc]} \BibitemShut {NoStop}%
\bibitem [{\citenamefont {Abbott}\ \emph {et~al.}(2021{\natexlab{b}})\citenamefont {Abbott} \emph {et~al.}}]{LIGOScientific:2020tif}%
  \BibitemOpen
  \bibfield  {author} {\bibinfo {author} {\bibfnamefont {R.}~\bibnamefont {Abbott}} \emph {et~al.} (\bibinfo {collaboration} {LIGO Scientific, Virgo}),\ }\bibfield  {title} {\bibinfo {title} {{Tests of general relativity with binary black holes from the second LIGO-Virgo gravitational-wave transient catalog}},\ }\href {https://doi.org/10.1103/PhysRevD.103.122002} {\bibfield  {journal} {\bibinfo  {journal} {Phys. Rev. D}\ }\textbf {\bibinfo {volume} {103}},\ \bibinfo {pages} {122002} (\bibinfo {year} {2021}{\natexlab{b}})},\ \Eprint {https://arxiv.org/abs/2010.14529} {arXiv:2010.14529 [gr-qc]} \BibitemShut {NoStop}%
\bibitem [{\citenamefont {Will}(2014)}]{Will:2014kxa}%
  \BibitemOpen
  \bibfield  {author} {\bibinfo {author} {\bibfnamefont {C.~M.}\ \bibnamefont {Will}},\ }\bibfield  {title} {\bibinfo {title} {{The Confrontation between General Relativity and Experiment}},\ }\href {https://doi.org/10.12942/lrr-2014-4} {\bibfield  {journal} {\bibinfo  {journal} {Living Rev. Rel.}\ }\textbf {\bibinfo {volume} {17}},\ \bibinfo {pages} {4} (\bibinfo {year} {2014})},\ \Eprint {https://arxiv.org/abs/1403.7377} {arXiv:1403.7377 [gr-qc]} \BibitemShut {NoStop}%
\bibitem [{\citenamefont {Clifton}\ \emph {et~al.}(2012)\citenamefont {Clifton}, \citenamefont {Ferreira}, \citenamefont {Padilla},\ and\ \citenamefont {Skordis}}]{Clifton:2011jh}%
  \BibitemOpen
  \bibfield  {author} {\bibinfo {author} {\bibfnamefont {T.}~\bibnamefont {Clifton}}, \bibinfo {author} {\bibfnamefont {P.~G.}\ \bibnamefont {Ferreira}}, \bibinfo {author} {\bibfnamefont {A.}~\bibnamefont {Padilla}},\ and\ \bibinfo {author} {\bibfnamefont {C.}~\bibnamefont {Skordis}},\ }\bibfield  {title} {\bibinfo {title} {{Modified Gravity and Cosmology}},\ }\href {https://doi.org/10.1016/j.physrep.2012.01.001} {\bibfield  {journal} {\bibinfo  {journal} {Phys. Rept.}\ }\textbf {\bibinfo {volume} {513}},\ \bibinfo {pages} {1} (\bibinfo {year} {2012})},\ \Eprint {https://arxiv.org/abs/1106.2476} {arXiv:1106.2476 [astro-ph.CO]} \BibitemShut {NoStop}%
\bibitem [{\citenamefont {Mattingly}(2005)}]{Mattingly:2005re}%
  \BibitemOpen
  \bibfield  {author} {\bibinfo {author} {\bibfnamefont {D.}~\bibnamefont {Mattingly}},\ }\bibfield  {title} {\bibinfo {title} {{Modern tests of Lorentz invariance}},\ }\href {https://doi.org/10.12942/lrr-2005-5} {\bibfield  {journal} {\bibinfo  {journal} {Living Rev. Rel.}\ }\textbf {\bibinfo {volume} {8}},\ \bibinfo {pages} {5} (\bibinfo {year} {2005})},\ \Eprint {https://arxiv.org/abs/gr-qc/0502097} {arXiv:gr-qc/0502097} \BibitemShut {NoStop}%
\bibitem [{\citenamefont {Jacobson}(2007)}]{Jacobson:2007veq}%
  \BibitemOpen
  \bibfield  {author} {\bibinfo {author} {\bibfnamefont {T.}~\bibnamefont {Jacobson}},\ }\bibfield  {title} {\bibinfo {title} {{Einstein-aether gravity: A Status report}},\ }\href {https://doi.org/10.22323/1.043.0020} {\bibfield  {journal} {\bibinfo  {journal} {PoS}\ }\textbf {\bibinfo {volume} {QG-PH}},\ \bibinfo {pages} {020} (\bibinfo {year} {2007})},\ \Eprint {https://arxiv.org/abs/0801.1547} {arXiv:0801.1547 [gr-qc]} \BibitemShut {NoStop}%
\bibitem [{\citenamefont {Eling}\ \emph {et~al.}(2004)\citenamefont {Eling}, \citenamefont {Jacobson},\ and\ \citenamefont {Mattingly}}]{Eling:2004dk}%
  \BibitemOpen
  \bibfield  {author} {\bibinfo {author} {\bibfnamefont {C.}~\bibnamefont {Eling}}, \bibinfo {author} {\bibfnamefont {T.}~\bibnamefont {Jacobson}},\ and\ \bibinfo {author} {\bibfnamefont {D.}~\bibnamefont {Mattingly}},\ }\bibfield  {title} {\bibinfo {title} {{Einstein-Aether theory}},\ }in\ \href@noop {} {\emph {\bibinfo {booktitle} {{Deserfest: A Celebration of the Life and Works of Stanley Deser}}}}\ (\bibinfo {year} {2004})\ pp.\ \bibinfo {pages} {163--179},\ \Eprint {https://arxiv.org/abs/gr-qc/0410001} {arXiv:gr-qc/0410001} \BibitemShut {NoStop}%
\bibitem [{\citenamefont {Abbott}\ \emph {et~al.}(2017)\citenamefont {Abbott} \emph {et~al.}}]{LIGOScientific:2017zic}%
  \BibitemOpen
  \bibfield  {author} {\bibinfo {author} {\bibfnamefont {B.~P.}\ \bibnamefont {Abbott}} \emph {et~al.} (\bibinfo {collaboration} {LIGO Scientific, Virgo, Fermi-GBM, INTEGRAL}),\ }\bibfield  {title} {\bibinfo {title} {{Gravitational Waves and Gamma-rays from a Binary Neutron Star Merger: GW170817 and GRB 170817A}},\ }\href {https://doi.org/10.3847/2041-8213/aa920c} {\bibfield  {journal} {\bibinfo  {journal} {Astrophys. J. Lett.}\ }\textbf {\bibinfo {volume} {848}},\ \bibinfo {pages} {L13} (\bibinfo {year} {2017})},\ \Eprint {https://arxiv.org/abs/1710.05834} {arXiv:1710.05834 [astro-ph.HE]} \BibitemShut {NoStop}%
\bibitem [{\citenamefont {Elliott}\ \emph {et~al.}(2005)\citenamefont {Elliott}, \citenamefont {Moore},\ and\ \citenamefont {Stoica}}]{Elliott:2005va}%
  \BibitemOpen
  \bibfield  {author} {\bibinfo {author} {\bibfnamefont {J.~W.}\ \bibnamefont {Elliott}}, \bibinfo {author} {\bibfnamefont {G.~D.}\ \bibnamefont {Moore}},\ and\ \bibinfo {author} {\bibfnamefont {H.}~\bibnamefont {Stoica}},\ }\bibfield  {title} {\bibinfo {title} {{Constraining the new Aether: Gravitational Cerenkov radiation}},\ }\href {https://doi.org/10.1088/1126-6708/2005/08/066} {\bibfield  {journal} {\bibinfo  {journal} {JHEP}\ }\textbf {\bibinfo {volume} {08}},\ \bibinfo {pages} {066}},\ \Eprint {https://arxiv.org/abs/hep-ph/0505211} {arXiv:hep-ph/0505211} \BibitemShut {NoStop}%
\bibitem [{\citenamefont {Carroll}\ and\ \citenamefont {Lim}(2004)}]{Carroll:2004ai}%
  \BibitemOpen
  \bibfield  {author} {\bibinfo {author} {\bibfnamefont {S.~M.}\ \bibnamefont {Carroll}}\ and\ \bibinfo {author} {\bibfnamefont {E.~A.}\ \bibnamefont {Lim}},\ }\bibfield  {title} {\bibinfo {title} {{Lorentz-violating vector fields slow the universe down}},\ }\href {https://doi.org/10.1103/PhysRevD.70.123525} {\bibfield  {journal} {\bibinfo  {journal} {Phys. Rev. D}\ }\textbf {\bibinfo {volume} {70}},\ \bibinfo {pages} {123525} (\bibinfo {year} {2004})},\ \Eprint {https://arxiv.org/abs/hep-th/0407149} {arXiv:hep-th/0407149} \BibitemShut {NoStop}%
\bibitem [{\citenamefont {Muller}\ \emph {et~al.}(2008)\citenamefont {Muller}, \citenamefont {Williams},\ and\ \citenamefont {Turyshev}}]{Muller:2005sr}%
  \BibitemOpen
  \bibfield  {author} {\bibinfo {author} {\bibfnamefont {J.}~\bibnamefont {Muller}}, \bibinfo {author} {\bibfnamefont {J.~G.}\ \bibnamefont {Williams}},\ and\ \bibinfo {author} {\bibfnamefont {S.~G.}\ \bibnamefont {Turyshev}},\ }\bibfield  {title} {\bibinfo {title} {{Lunar laser ranging contributions to relativity and geodesy}},\ }\href {https://doi.org/10.1007/978-3-540-34377-6_21} {\bibfield  {journal} {\bibinfo  {journal} {Astrophys. Space Sci. Libr.}\ }\textbf {\bibinfo {volume} {349}},\ \bibinfo {pages} {457} (\bibinfo {year} {2008})},\ \Eprint {https://arxiv.org/abs/gr-qc/0509114} {arXiv:gr-qc/0509114} \BibitemShut {NoStop}%
\bibitem [{\citenamefont {Nordtvedt}(1987)}]{solarSystem2}%
  \BibitemOpen
  \bibfield  {author} {\bibinfo {author} {\bibfnamefont {K.}~\bibnamefont {Nordtvedt}},\ }\bibfield  {title} {\bibinfo {title} {Probing gravity to the second post-newtonian order and to one part in $10^{7}$ using the spin axis of the sun.},\ }\href {https://doi.org/10.1086/165603} {\bibfield  {journal} {\bibinfo  {journal} {Astrophys. Journal}\ }\textbf {\bibinfo {volume} {320}},\ \bibinfo {pages} {871} (\bibinfo {year} {1987})}\BibitemShut {NoStop}%
\bibitem [{\citenamefont {Gupta}\ \emph {et~al.}(2021)\citenamefont {Gupta}, \citenamefont {Herrero-Valea}, \citenamefont {Blas}, \citenamefont {Barausse}, \citenamefont {Cornish}, \citenamefont {Yagi},\ and\ \citenamefont {Yunes}}]{Gupta:2021vdj}%
  \BibitemOpen
  \bibfield  {author} {\bibinfo {author} {\bibfnamefont {T.}~\bibnamefont {Gupta}}, \bibinfo {author} {\bibfnamefont {M.}~\bibnamefont {Herrero-Valea}}, \bibinfo {author} {\bibfnamefont {D.}~\bibnamefont {Blas}}, \bibinfo {author} {\bibfnamefont {E.}~\bibnamefont {Barausse}}, \bibinfo {author} {\bibfnamefont {N.}~\bibnamefont {Cornish}}, \bibinfo {author} {\bibfnamefont {K.}~\bibnamefont {Yagi}},\ and\ \bibinfo {author} {\bibfnamefont {N.}~\bibnamefont {Yunes}},\ }\bibfield  {title} {\bibinfo {title} {{New binary pulsar constraints on Einstein-\ae{}ther theory after GW170817}},\ }\href {https://doi.org/10.1088/1361-6382/ac1a69} {\bibfield  {journal} {\bibinfo  {journal} {Class. Quant. Grav.}\ }\textbf {\bibinfo {volume} {38}},\ \bibinfo {pages} {195003} (\bibinfo {year} {2021})},\ \Eprint {https://arxiv.org/abs/2104.04596} {arXiv:2104.04596 [gr-qc]} \BibitemShut {NoStop}%
\bibitem [{\citenamefont {Hansen}\ \emph {et~al.}(2015)\citenamefont {Hansen}, \citenamefont {Yunes},\ and\ \citenamefont {Yagi}}]{Hansen:2014ewa}%
  \BibitemOpen
  \bibfield  {author} {\bibinfo {author} {\bibfnamefont {D.}~\bibnamefont {Hansen}}, \bibinfo {author} {\bibfnamefont {N.}~\bibnamefont {Yunes}},\ and\ \bibinfo {author} {\bibfnamefont {K.}~\bibnamefont {Yagi}},\ }\bibfield  {title} {\bibinfo {title} {{Projected Constraints on Lorentz-Violating Gravity with Gravitational Waves}},\ }\href {https://doi.org/10.1103/PhysRevD.91.082003} {\bibfield  {journal} {\bibinfo  {journal} {Phys. Rev. D}\ }\textbf {\bibinfo {volume} {91}},\ \bibinfo {pages} {082003} (\bibinfo {year} {2015})},\ \Eprint {https://arxiv.org/abs/1412.4132} {arXiv:1412.4132 [gr-qc]} \BibitemShut {NoStop}%
\bibitem [{\citenamefont {Zhang}\ \emph {et~al.}(2020)\citenamefont {Zhang}, \citenamefont {Zhao}, \citenamefont {Wang}, \citenamefont {Wang}, \citenamefont {Yagi}, \citenamefont {Yunes}, \citenamefont {Zhao},\ and\ \citenamefont {Zhu}}]{Zhang:2019iim}%
  \BibitemOpen
  \bibfield  {author} {\bibinfo {author} {\bibfnamefont {C.}~\bibnamefont {Zhang}}, \bibinfo {author} {\bibfnamefont {X.}~\bibnamefont {Zhao}}, \bibinfo {author} {\bibfnamefont {A.}~\bibnamefont {Wang}}, \bibinfo {author} {\bibfnamefont {B.}~\bibnamefont {Wang}}, \bibinfo {author} {\bibfnamefont {K.}~\bibnamefont {Yagi}}, \bibinfo {author} {\bibfnamefont {N.}~\bibnamefont {Yunes}}, \bibinfo {author} {\bibfnamefont {W.}~\bibnamefont {Zhao}},\ and\ \bibinfo {author} {\bibfnamefont {T.}~\bibnamefont {Zhu}},\ }\bibfield  {title} {\bibinfo {title} {{Gravitational waves from the quasicircular inspiral of compact binaries in Einstein-aether theory}},\ }\href {https://doi.org/10.1103/PhysRevD.104.069905} {\bibfield  {journal} {\bibinfo  {journal} {Phys. Rev. D}\ }\textbf {\bibinfo {volume} {101}},\ \bibinfo {pages} {044002} (\bibinfo {year} {2020})},\ \bibinfo {note} {[Erratum: Phys.Rev.D 104, 069905(E) (2021)]},\ \Eprint {https://arxiv.org/abs/1911.10278} {arXiv:1911.10278 [gr-qc]} \BibitemShut {NoStop}%
\bibitem [{\citenamefont {Blanchet}(2014)}]{Blanchet:2013haa}%
  \BibitemOpen
  \bibfield  {author} {\bibinfo {author} {\bibfnamefont {L.}~\bibnamefont {Blanchet}},\ }\bibfield  {title} {\bibinfo {title} {{Gravitational Radiation from Post-Newtonian Sources and Inspiralling Compact Binaries}},\ }\href {https://doi.org/10.12942/lrr-2014-2} {\bibfield  {journal} {\bibinfo  {journal} {Living Rev. Rel.}\ }\textbf {\bibinfo {volume} {17}},\ \bibinfo {pages} {2} (\bibinfo {year} {2014})},\ \Eprint {https://arxiv.org/abs/1310.1528} {arXiv:1310.1528 [gr-qc]} \BibitemShut {NoStop}%
\bibitem [{\citenamefont {Yagi}\ \emph {et~al.}(2014{\natexlab{a}})\citenamefont {Yagi}, \citenamefont {Blas}, \citenamefont {Yunes},\ and\ \citenamefont {Barausse}}]{Yagi:2013qpa}%
  \BibitemOpen
  \bibfield  {author} {\bibinfo {author} {\bibfnamefont {K.}~\bibnamefont {Yagi}}, \bibinfo {author} {\bibfnamefont {D.}~\bibnamefont {Blas}}, \bibinfo {author} {\bibfnamefont {N.}~\bibnamefont {Yunes}},\ and\ \bibinfo {author} {\bibfnamefont {E.}~\bibnamefont {Barausse}},\ }\bibfield  {title} {\bibinfo {title} {{Strong Binary Pulsar Constraints on Lorentz Violation in Gravity}},\ }\href {https://doi.org/10.1103/PhysRevLett.112.161101} {\bibfield  {journal} {\bibinfo  {journal} {Phys. Rev. Lett.}\ }\textbf {\bibinfo {volume} {112}},\ \bibinfo {pages} {161101} (\bibinfo {year} {2014}{\natexlab{a}})},\ \Eprint {https://arxiv.org/abs/1307.6219} {arXiv:1307.6219 [gr-qc]} \BibitemShut {NoStop}%
\bibitem [{\citenamefont {Perkins}\ \emph {et~al.}(2021)\citenamefont {Perkins}, \citenamefont {Nair}, \citenamefont {Silva},\ and\ \citenamefont {Yunes}}]{Perkins:2021mhb}%
  \BibitemOpen
  \bibfield  {author} {\bibinfo {author} {\bibfnamefont {S.~E.}\ \bibnamefont {Perkins}}, \bibinfo {author} {\bibfnamefont {R.}~\bibnamefont {Nair}}, \bibinfo {author} {\bibfnamefont {H.~O.}\ \bibnamefont {Silva}},\ and\ \bibinfo {author} {\bibfnamefont {N.}~\bibnamefont {Yunes}},\ }\bibfield  {title} {\bibinfo {title} {{Improved gravitational-wave constraints on higher-order curvature theories of gravity}},\ }\href {https://doi.org/10.1103/PhysRevD.104.024060} {\bibfield  {journal} {\bibinfo  {journal} {Phys. Rev. D}\ }\textbf {\bibinfo {volume} {104}},\ \bibinfo {pages} {024060} (\bibinfo {year} {2021})},\ \Eprint {https://arxiv.org/abs/2104.11189} {arXiv:2104.11189 [gr-qc]} \BibitemShut {NoStop}%
\bibitem [{\citenamefont {Yagi}\ and\ \citenamefont {Yunes}(2016)}]{Yagi:2015pkc}%
  \BibitemOpen
  \bibfield  {author} {\bibinfo {author} {\bibfnamefont {K.}~\bibnamefont {Yagi}}\ and\ \bibinfo {author} {\bibfnamefont {N.}~\bibnamefont {Yunes}},\ }\bibfield  {title} {\bibinfo {title} {{Binary Love Relations}},\ }\href {https://doi.org/10.1088/0264-9381/33/13/13LT01} {\bibfield  {journal} {\bibinfo  {journal} {Class. Quant. Grav.}\ }\textbf {\bibinfo {volume} {33}},\ \bibinfo {pages} {13LT01} (\bibinfo {year} {2016})},\ \Eprint {https://arxiv.org/abs/1512.02639} {arXiv:1512.02639 [gr-qc]} \BibitemShut {NoStop}%
\bibitem [{\citenamefont {Yagi}\ and\ \citenamefont {Yunes}(2017{\natexlab{a}})}]{Yagi:2016qmr}%
  \BibitemOpen
  \bibfield  {author} {\bibinfo {author} {\bibfnamefont {K.}~\bibnamefont {Yagi}}\ and\ \bibinfo {author} {\bibfnamefont {N.}~\bibnamefont {Yunes}},\ }\bibfield  {title} {\bibinfo {title} {{Approximate Universal Relations among Tidal Parameters for Neutron Star Binaries}},\ }\href {https://doi.org/10.1088/1361-6382/34/1/015006} {\bibfield  {journal} {\bibinfo  {journal} {Class. Quant. Grav.}\ }\textbf {\bibinfo {volume} {34}},\ \bibinfo {pages} {015006} (\bibinfo {year} {2017}{\natexlab{a}})},\ \Eprint {https://arxiv.org/abs/1608.06187} {arXiv:1608.06187 [gr-qc]} \BibitemShut {NoStop}%
\bibitem [{\citenamefont {Carson}\ \emph {et~al.}(2019)\citenamefont {Carson}, \citenamefont {Chatziioannou}, \citenamefont {Haster}, \citenamefont {Yagi},\ and\ \citenamefont {Yunes}}]{Carson:2019rjx}%
  \BibitemOpen
  \bibfield  {author} {\bibinfo {author} {\bibfnamefont {Z.}~\bibnamefont {Carson}}, \bibinfo {author} {\bibfnamefont {K.}~\bibnamefont {Chatziioannou}}, \bibinfo {author} {\bibfnamefont {C.-J.}\ \bibnamefont {Haster}}, \bibinfo {author} {\bibfnamefont {K.}~\bibnamefont {Yagi}},\ and\ \bibinfo {author} {\bibfnamefont {N.}~\bibnamefont {Yunes}},\ }\bibfield  {title} {\bibinfo {title} {{Equation-of-state insensitive relations after GW170817}},\ }\href {https://doi.org/10.1103/PhysRevD.99.083016} {\bibfield  {journal} {\bibinfo  {journal} {Phys. Rev. D}\ }\textbf {\bibinfo {volume} {99}},\ \bibinfo {pages} {083016} (\bibinfo {year} {2019})},\ \Eprint {https://arxiv.org/abs/1903.03909} {arXiv:1903.03909 [gr-qc]} \BibitemShut {NoStop}%
\bibitem [{\citenamefont {Yagi}\ and\ \citenamefont {Yunes}(2013{\natexlab{a}})}]{Yagi:2013bca}%
  \BibitemOpen
  \bibfield  {author} {\bibinfo {author} {\bibfnamefont {K.}~\bibnamefont {Yagi}}\ and\ \bibinfo {author} {\bibfnamefont {N.}~\bibnamefont {Yunes}},\ }\bibfield  {title} {\bibinfo {title} {{I-Love-Q}},\ }\href {https://doi.org/10.1126/science.1236462} {\bibfield  {journal} {\bibinfo  {journal} {Science}\ }\textbf {\bibinfo {volume} {341}},\ \bibinfo {pages} {365} (\bibinfo {year} {2013}{\natexlab{a}})},\ \Eprint {https://arxiv.org/abs/1302.4499} {arXiv:1302.4499 [gr-qc]} \BibitemShut {NoStop}%
\bibitem [{\citenamefont {Yagi}\ and\ \citenamefont {Yunes}(2013{\natexlab{b}})}]{Yagi:2013awa}%
  \BibitemOpen
  \bibfield  {author} {\bibinfo {author} {\bibfnamefont {K.}~\bibnamefont {Yagi}}\ and\ \bibinfo {author} {\bibfnamefont {N.}~\bibnamefont {Yunes}},\ }\bibfield  {title} {\bibinfo {title} {{I-Love-Q Relations in Neutron Stars and their Applications to Astrophysics, Gravitational Waves and Fundamental Physics}},\ }\href {https://doi.org/10.1103/PhysRevD.88.023009} {\bibfield  {journal} {\bibinfo  {journal} {Phys. Rev. D}\ }\textbf {\bibinfo {volume} {88}},\ \bibinfo {pages} {023009} (\bibinfo {year} {2013}{\natexlab{b}})},\ \Eprint {https://arxiv.org/abs/1303.1528} {arXiv:1303.1528 [gr-qc]} \BibitemShut {NoStop}%
\bibitem [{\citenamefont {Maselli}\ \emph {et~al.}(2013)\citenamefont {Maselli}, \citenamefont {Cardoso}, \citenamefont {Ferrari}, \citenamefont {Gualtieri},\ and\ \citenamefont {Pani}}]{Maselli:2013mva}%
  \BibitemOpen
  \bibfield  {author} {\bibinfo {author} {\bibfnamefont {A.}~\bibnamefont {Maselli}}, \bibinfo {author} {\bibfnamefont {V.}~\bibnamefont {Cardoso}}, \bibinfo {author} {\bibfnamefont {V.}~\bibnamefont {Ferrari}}, \bibinfo {author} {\bibfnamefont {L.}~\bibnamefont {Gualtieri}},\ and\ \bibinfo {author} {\bibfnamefont {P.}~\bibnamefont {Pani}},\ }\bibfield  {title} {\bibinfo {title} {{Equation-of-state-independent relations in neutron stars}},\ }\href {https://doi.org/10.1103/PhysRevD.88.023007} {\bibfield  {journal} {\bibinfo  {journal} {Phys. Rev. D}\ }\textbf {\bibinfo {volume} {88}},\ \bibinfo {pages} {023007} (\bibinfo {year} {2013})},\ \Eprint {https://arxiv.org/abs/1304.2052} {arXiv:1304.2052 [gr-qc]} \BibitemShut {NoStop}%
\bibitem [{\citenamefont {Jacobson}\ and\ \citenamefont {Mattingly}(2001)}]{Jacobson:2000xp}%
  \BibitemOpen
  \bibfield  {author} {\bibinfo {author} {\bibfnamefont {T.}~\bibnamefont {Jacobson}}\ and\ \bibinfo {author} {\bibfnamefont {D.}~\bibnamefont {Mattingly}},\ }\bibfield  {title} {\bibinfo {title} {{Gravity with a dynamical preferred frame}},\ }\href {https://doi.org/10.1103/PhysRevD.64.024028} {\bibfield  {journal} {\bibinfo  {journal} {Phys. Rev. D}\ }\textbf {\bibinfo {volume} {64}},\ \bibinfo {pages} {024028} (\bibinfo {year} {2001})},\ \Eprint {https://arxiv.org/abs/gr-qc/0007031} {arXiv:gr-qc/0007031} \BibitemShut {NoStop}%
\bibitem [{\citenamefont {Jacobson}(2014)}]{Jacobson:2013xta}%
  \BibitemOpen
  \bibfield  {author} {\bibinfo {author} {\bibfnamefont {T.}~\bibnamefont {Jacobson}},\ }\bibfield  {title} {\bibinfo {title} {{Undoing the twist: The Ho\v{r}ava limit of Einstein-aether theory}},\ }\href {https://doi.org/10.1103/PhysRevD.89.081501} {\bibfield  {journal} {\bibinfo  {journal} {Phys. Rev. D}\ }\textbf {\bibinfo {volume} {89}},\ \bibinfo {pages} {081501(R)} (\bibinfo {year} {2014})},\ \Eprint {https://arxiv.org/abs/1310.5115} {arXiv:1310.5115 [gr-qc]} \BibitemShut {NoStop}%
\bibitem [{\citenamefont {Jacobson}\ and\ \citenamefont {Mattingly}(2004)}]{Jacobson:2004ts}%
  \BibitemOpen
  \bibfield  {author} {\bibinfo {author} {\bibfnamefont {T.}~\bibnamefont {Jacobson}}\ and\ \bibinfo {author} {\bibfnamefont {D.}~\bibnamefont {Mattingly}},\ }\bibfield  {title} {\bibinfo {title} {{Einstein-Aether waves}},\ }\href {https://doi.org/10.1103/PhysRevD.70.024003} {\bibfield  {journal} {\bibinfo  {journal} {Phys. Rev. D}\ }\textbf {\bibinfo {volume} {70}},\ \bibinfo {pages} {024003} (\bibinfo {year} {2004})},\ \Eprint {https://arxiv.org/abs/gr-qc/0402005} {arXiv:gr-qc/0402005} \BibitemShut {NoStop}%
\bibitem [{\citenamefont {Foster}(2007)}]{Foster:2007gr}%
  \BibitemOpen
  \bibfield  {author} {\bibinfo {author} {\bibfnamefont {B.~Z.}\ \bibnamefont {Foster}},\ }\bibfield  {title} {\bibinfo {title} {{Strong field effects on binary systems in Einstein-aether theory}},\ }\href {https://doi.org/10.1103/PhysRevD.76.084033} {\bibfield  {journal} {\bibinfo  {journal} {Phys. Rev. D}\ }\textbf {\bibinfo {volume} {76}},\ \bibinfo {pages} {084033} (\bibinfo {year} {2007})},\ \Eprint {https://arxiv.org/abs/0706.0704} {arXiv:0706.0704 [gr-qc]} \BibitemShut {NoStop}%
\bibitem [{\citenamefont {Foster}\ and\ \citenamefont {Jacobson}(2006)}]{Foster:2005dk}%
  \BibitemOpen
  \bibfield  {author} {\bibinfo {author} {\bibfnamefont {B.~Z.}\ \bibnamefont {Foster}}\ and\ \bibinfo {author} {\bibfnamefont {T.}~\bibnamefont {Jacobson}},\ }\bibfield  {title} {\bibinfo {title} {{Post-Newtonian parameters and constraints on Einstein-aether theory}},\ }\href {https://doi.org/10.1103/PhysRevD.73.064015} {\bibfield  {journal} {\bibinfo  {journal} {Phys. Rev. D}\ }\textbf {\bibinfo {volume} {73}},\ \bibinfo {pages} {064015} (\bibinfo {year} {2006})},\ \Eprint {https://arxiv.org/abs/gr-qc/0509083} {arXiv:gr-qc/0509083} \BibitemShut {NoStop}%
\bibitem [{\citenamefont {Veitch}\ \emph {et~al.}(2015)\citenamefont {Veitch}, \citenamefont {Raymond}, \citenamefont {Farr}, \citenamefont {Farr}, \citenamefont {Graff}, \citenamefont {Vitale} \emph {et~al.}}]{Veitch:2014wba}%
  \BibitemOpen
  \bibfield  {author} {\bibinfo {author} {\bibfnamefont {J.}~\bibnamefont {Veitch}}, \bibinfo {author} {\bibfnamefont {V.}~\bibnamefont {Raymond}}, \bibinfo {author} {\bibfnamefont {B.}~\bibnamefont {Farr}}, \bibinfo {author} {\bibfnamefont {W.}~\bibnamefont {Farr}}, \bibinfo {author} {\bibfnamefont {P.}~\bibnamefont {Graff}}, \bibinfo {author} {\bibfnamefont {S.}~\bibnamefont {Vitale}}, \emph {et~al.},\ }\bibfield  {title} {\bibinfo {title} {{Parameter estimation for compact binaries with ground-based gravitational-wave observations using the LALInference software library}},\ }\href {https://doi.org/10.1103/PhysRevD.91.042003} {\bibfield  {journal} {\bibinfo  {journal} {Phys. Rev. D}\ }\textbf {\bibinfo {volume} {91}},\ \bibinfo {pages} {042003} (\bibinfo {year} {2015})},\ \Eprint {https://arxiv.org/abs/1409.7215} {arXiv:1409.7215 [gr-qc]} \BibitemShut {NoStop}%
\bibitem [{\citenamefont {Abbott}\ \emph {et~al.}(2019)\citenamefont {Abbott} \emph {et~al.}}]{LIGOScientific:2018hze}%
  \BibitemOpen
  \bibfield  {author} {\bibinfo {author} {\bibfnamefont {B.~P.}\ \bibnamefont {Abbott}} \emph {et~al.} (\bibinfo {collaboration} {LIGO Scientific, Virgo}),\ }\bibfield  {title} {\bibinfo {title} {{Properties of the binary neutron star merger GW170817}},\ }\href {https://doi.org/10.1103/PhysRevX.9.011001} {\bibfield  {journal} {\bibinfo  {journal} {Phys. Rev. X}\ }\textbf {\bibinfo {volume} {9}},\ \bibinfo {pages} {011001} (\bibinfo {year} {2019})},\ \Eprint {https://arxiv.org/abs/1805.11579} {arXiv:1805.11579 [gr-qc]} \BibitemShut {NoStop}%
\bibitem [{\citenamefont {Abbott}\ \emph {et~al.}(2020{\natexlab{a}})\citenamefont {Abbott} \emph {et~al.}}]{LIGOScientific:2020aai}%
  \BibitemOpen
  \bibfield  {author} {\bibinfo {author} {\bibfnamefont {B.~P.}\ \bibnamefont {Abbott}} \emph {et~al.} (\bibinfo {collaboration} {LIGO Scientific, Virgo}),\ }\bibfield  {title} {\bibinfo {title} {{GW190425: Observation of a Compact Binary Coalescence with Total Mass $\sim 3.4 M_{\odot}$}},\ }\href {https://doi.org/10.3847/2041-8213/ab75f5} {\bibfield  {journal} {\bibinfo  {journal} {Astrophys. J. Lett.}\ }\textbf {\bibinfo {volume} {892}},\ \bibinfo {pages} {L3} (\bibinfo {year} {2020}{\natexlab{a}})},\ \Eprint {https://arxiv.org/abs/2001.01761} {arXiv:2001.01761 [astro-ph.HE]} \BibitemShut {NoStop}%
\bibitem [{\citenamefont {Chatziioannou}\ \emph {et~al.}(2018)\citenamefont {Chatziioannou}, \citenamefont {Haster},\ and\ \citenamefont {Zimmerman}}]{Chatziioannou:2018vzf}%
  \BibitemOpen
  \bibfield  {author} {\bibinfo {author} {\bibfnamefont {K.}~\bibnamefont {Chatziioannou}}, \bibinfo {author} {\bibfnamefont {C.-J.}\ \bibnamefont {Haster}},\ and\ \bibinfo {author} {\bibfnamefont {A.}~\bibnamefont {Zimmerman}},\ }\bibfield  {title} {\bibinfo {title} {{Measuring the neutron star tidal deformability with equation-of-state-independent relations and gravitational waves}},\ }\href {https://doi.org/10.1103/PhysRevD.97.104036} {\bibfield  {journal} {\bibinfo  {journal} {Phys. Rev. D}\ }\textbf {\bibinfo {volume} {97}},\ \bibinfo {pages} {104036} (\bibinfo {year} {2018})},\ \Eprint {https://arxiv.org/abs/1804.03221} {arXiv:1804.03221 [gr-qc]} \BibitemShut {NoStop}%
\bibitem [{\citenamefont {Foster}(2006)}]{Foster:2006az}%
  \BibitemOpen
  \bibfield  {author} {\bibinfo {author} {\bibfnamefont {B.~Z.}\ \bibnamefont {Foster}},\ }\bibfield  {title} {\bibinfo {title} {{Radiation damping in Einstein-aether theory}},\ }\href {https://doi.org/10.1103/PhysRevD.75.129904} {\bibfield  {journal} {\bibinfo  {journal} {Phys. Rev. D}\ }\textbf {\bibinfo {volume} {73}},\ \bibinfo {pages} {104012} (\bibinfo {year} {2006})},\ \bibinfo {note} {[Erratum: Phys.Rev.D 75, 129904(E) (2007)]},\ \Eprint {https://arxiv.org/abs/gr-qc/0602004} {arXiv:gr-qc/0602004} \BibitemShut {NoStop}%
\bibitem [{\citenamefont {Yagi}\ \emph {et~al.}(2014{\natexlab{b}})\citenamefont {Yagi}, \citenamefont {Blas}, \citenamefont {Barausse},\ and\ \citenamefont {Yunes}}]{Yagi:2013ava}%
  \BibitemOpen
  \bibfield  {author} {\bibinfo {author} {\bibfnamefont {K.}~\bibnamefont {Yagi}}, \bibinfo {author} {\bibfnamefont {D.}~\bibnamefont {Blas}}, \bibinfo {author} {\bibfnamefont {E.}~\bibnamefont {Barausse}},\ and\ \bibinfo {author} {\bibfnamefont {N.}~\bibnamefont {Yunes}},\ }\bibfield  {title} {\bibinfo {title} {{Constraints on Einstein-\AE{}ther theory and Ho\v{r}ava gravity from binary pulsar observations}},\ }\href {https://doi.org/10.1103/PhysRevD.89.084067} {\bibfield  {journal} {\bibinfo  {journal} {Phys. Rev. D}\ }\textbf {\bibinfo {volume} {89}},\ \bibinfo {pages} {084067} (\bibinfo {year} {2014}{\natexlab{b}})},\ \bibinfo {note} {[Erratum: Phys.Rev.D 90, 069902(E) (2014), Erratum: Phys.Rev.D 90, 069901(E) (2014)]},\ \Eprint {https://arxiv.org/abs/1311.7144} {arXiv:1311.7144 [gr-qc]} \BibitemShut {NoStop}%
\bibitem [{\citenamefont {Chatziioannou}\ \emph {et~al.}(2012)\citenamefont {Chatziioannou}, \citenamefont {Yunes},\ and\ \citenamefont {Cornish}}]{Chatziioannou:2012rf}%
  \BibitemOpen
  \bibfield  {author} {\bibinfo {author} {\bibfnamefont {K.}~\bibnamefont {Chatziioannou}}, \bibinfo {author} {\bibfnamefont {N.}~\bibnamefont {Yunes}},\ and\ \bibinfo {author} {\bibfnamefont {N.}~\bibnamefont {Cornish}},\ }\bibfield  {title} {\bibinfo {title} {{Model-Independent Test of General Relativity: An Extended post-Einsteinian Framework with Complete Polarization Content}},\ }\href {https://doi.org/10.1103/PhysRevD.86.022004} {\bibfield  {journal} {\bibinfo  {journal} {Phys. Rev. D}\ }\textbf {\bibinfo {volume} {86}},\ \bibinfo {pages} {022004} (\bibinfo {year} {2012})},\ \bibinfo {note} {[Erratum: Phys.Rev.D 95, 129901(E) (2017)]},\ \Eprint {https://arxiv.org/abs/1204.2585} {arXiv:1204.2585 [gr-qc]} \BibitemShut {NoStop}%
\bibitem [{\citenamefont {Schumacher}\ \emph {et~al.}(2023)\citenamefont {Schumacher}, \citenamefont {Yunes},\ and\ \citenamefont {Yagi}}]{Schumacher:2023jxq}%
  \BibitemOpen
  \bibfield  {author} {\bibinfo {author} {\bibfnamefont {K.}~\bibnamefont {Schumacher}}, \bibinfo {author} {\bibfnamefont {N.}~\bibnamefont {Yunes}},\ and\ \bibinfo {author} {\bibfnamefont {K.}~\bibnamefont {Yagi}},\ }\bibfield  {title} {\bibinfo {title} {{Gravitational Wave Polarizations with Different Propagation Speeds}},\ }\href@noop {} {\  (\bibinfo {year} {2023})},\ \Eprint {https://arxiv.org/abs/2308.05589} {arXiv:2308.05589 [gr-qc]} \BibitemShut {NoStop}%
\bibitem [{\citenamefont {Poisson}\ and\ \citenamefont {Will}(2014)}]{PoissonAndWill}%
  \BibitemOpen
  \bibfield  {author} {\bibinfo {author} {\bibfnamefont {E.}~\bibnamefont {Poisson}}\ and\ \bibinfo {author} {\bibfnamefont {C.}~\bibnamefont {Will}},\ }\href@noop {} {\emph {\bibinfo {title} {Gravity}}}\ (\bibinfo  {publisher} {Cambridge University Press},\ \bibinfo {year} {2014})\BibitemShut {NoStop}%
\bibitem [{\citenamefont {Yunes}\ and\ \citenamefont {Siemens}(2013)}]{Yunes:2013dva}%
  \BibitemOpen
  \bibfield  {author} {\bibinfo {author} {\bibfnamefont {N.}~\bibnamefont {Yunes}}\ and\ \bibinfo {author} {\bibfnamefont {X.}~\bibnamefont {Siemens}},\ }\bibfield  {title} {\bibinfo {title} {{Gravitational-Wave Tests of General Relativity with Ground-Based Detectors and Pulsar Timing-Arrays}},\ }\href {https://doi.org/10.12942/lrr-2013-9} {\bibfield  {journal} {\bibinfo  {journal} {Living Rev. Rel.}\ }\textbf {\bibinfo {volume} {16}},\ \bibinfo {pages} {9} (\bibinfo {year} {2013})},\ \Eprint {https://arxiv.org/abs/1304.3473} {arXiv:1304.3473 [gr-qc]} \BibitemShut {NoStop}%
\bibitem [{\citenamefont {Husa}\ \emph {et~al.}(2016)\citenamefont {Husa}, \citenamefont {Khan}, \citenamefont {Hannam}, \citenamefont {P\"urrer}, \citenamefont {Ohme}, \citenamefont {Forteza},\ and\ \citenamefont {Boh\'e}}]{Husa:2015iqa}%
  \BibitemOpen
  \bibfield  {author} {\bibinfo {author} {\bibfnamefont {S.}~\bibnamefont {Husa}}, \bibinfo {author} {\bibfnamefont {S.}~\bibnamefont {Khan}}, \bibinfo {author} {\bibfnamefont {M.}~\bibnamefont {Hannam}}, \bibinfo {author} {\bibfnamefont {M.}~\bibnamefont {P\"urrer}}, \bibinfo {author} {\bibfnamefont {F.}~\bibnamefont {Ohme}}, \bibinfo {author} {\bibfnamefont {X.~J.}\ \bibnamefont {Forteza}},\ and\ \bibinfo {author} {\bibfnamefont {A.}~\bibnamefont {Boh\'e}},\ }\bibfield  {title} {\bibinfo {title} {{Frequency-domain gravitational waves from nonprecessing black-hole binaries. I. New numerical waveforms and anatomy of the signal}},\ }\href {https://doi.org/10.1103/PhysRevD.93.044006} {\bibfield  {journal} {\bibinfo  {journal} {Phys. Rev. D}\ }\textbf {\bibinfo {volume} {93}},\ \bibinfo {pages} {044006} (\bibinfo {year} {2016})},\ \Eprint {https://arxiv.org/abs/1508.07250} {arXiv:1508.07250 [gr-qc]} \BibitemShut {NoStop}%
\bibitem [{\citenamefont {Dietrich}\ \emph {et~al.}(2019{\natexlab{a}})\citenamefont {Dietrich}, \citenamefont {Samajdar}, \citenamefont {Khan}, \citenamefont {Johnson-McDaniel}, \citenamefont {Dudi},\ and\ \citenamefont {Tichy}}]{Dietrich:2019kaq}%
  \BibitemOpen
  \bibfield  {author} {\bibinfo {author} {\bibfnamefont {T.}~\bibnamefont {Dietrich}}, \bibinfo {author} {\bibfnamefont {A.}~\bibnamefont {Samajdar}}, \bibinfo {author} {\bibfnamefont {S.}~\bibnamefont {Khan}}, \bibinfo {author} {\bibfnamefont {N.~K.}\ \bibnamefont {Johnson-McDaniel}}, \bibinfo {author} {\bibfnamefont {R.}~\bibnamefont {Dudi}},\ and\ \bibinfo {author} {\bibfnamefont {W.}~\bibnamefont {Tichy}},\ }\bibfield  {title} {\bibinfo {title} {{Improving the NRTidal model for binary neutron star systems}},\ }\href {https://doi.org/10.1103/PhysRevD.100.044003} {\bibfield  {journal} {\bibinfo  {journal} {Phys. Rev. D}\ }\textbf {\bibinfo {volume} {100}},\ \bibinfo {pages} {044003} (\bibinfo {year} {2019}{\natexlab{a}})},\ \Eprint {https://arxiv.org/abs/1905.06011} {arXiv:1905.06011 [gr-qc]} \BibitemShut {NoStop}%
\bibitem [{\citenamefont {Khan}\ \emph {et~al.}(2016)\citenamefont {Khan}, \citenamefont {Husa}, \citenamefont {Hannam}, \citenamefont {Ohme}, \citenamefont {P\"urrer}, \citenamefont {Forteza},\ and\ \citenamefont {Boh\'e}}]{Khan:2015jqa}%
  \BibitemOpen
  \bibfield  {author} {\bibinfo {author} {\bibfnamefont {S.}~\bibnamefont {Khan}}, \bibinfo {author} {\bibfnamefont {S.}~\bibnamefont {Husa}}, \bibinfo {author} {\bibfnamefont {M.}~\bibnamefont {Hannam}}, \bibinfo {author} {\bibfnamefont {F.}~\bibnamefont {Ohme}}, \bibinfo {author} {\bibfnamefont {M.}~\bibnamefont {P\"urrer}}, \bibinfo {author} {\bibfnamefont {X.~J.}\ \bibnamefont {Forteza}},\ and\ \bibinfo {author} {\bibfnamefont {A.}~\bibnamefont {Boh\'e}},\ }\bibfield  {title} {\bibinfo {title} {{Frequency-domain gravitational waves from nonprecessing black-hole binaries. II. A phenomenological model for the advanced detector era}},\ }\href {https://doi.org/10.1103/PhysRevD.93.044007} {\bibfield  {journal} {\bibinfo  {journal} {Phys. Rev. D}\ }\textbf {\bibinfo {volume} {93}},\ \bibinfo {pages} {044007} (\bibinfo {year} {2016})},\ \Eprint {https://arxiv.org/abs/1508.07253} {arXiv:1508.07253 [gr-qc]} \BibitemShut {NoStop}%
\bibitem [{\citenamefont {Dietrich}\ \emph {et~al.}(2019{\natexlab{b}})\citenamefont {Dietrich}, \citenamefont {Khan}, \citenamefont {Dudi}, \citenamefont {Kapadia}, \citenamefont {Kumar}, \citenamefont {Nagar} \emph {et~al.}}]{Dietrich:2018uni}%
  \BibitemOpen
  \bibfield  {author} {\bibinfo {author} {\bibfnamefont {T.}~\bibnamefont {Dietrich}}, \bibinfo {author} {\bibfnamefont {S.}~\bibnamefont {Khan}}, \bibinfo {author} {\bibfnamefont {R.}~\bibnamefont {Dudi}}, \bibinfo {author} {\bibfnamefont {S.}~\bibnamefont {Kapadia}}, \bibinfo {author} {\bibfnamefont {P.}~\bibnamefont {Kumar}}, \bibinfo {author} {\bibfnamefont {A.}~\bibnamefont {Nagar}}, \emph {et~al.},\ }\bibfield  {title} {\bibinfo {title} {{Matter imprints in waveform models for neutron star binaries: Tidal and self-spin effects}},\ }\href {https://doi.org/10.1103/PhysRevD.99.024029} {\bibfield  {journal} {\bibinfo  {journal} {Phys. Rev. D}\ }\textbf {\bibinfo {volume} {99}},\ \bibinfo {pages} {024029} (\bibinfo {year} {2019}{\natexlab{b}})},\ \Eprint {https://arxiv.org/abs/1804.02235} {arXiv:1804.02235 [gr-qc]} \BibitemShut {NoStop}%
\bibitem [{\citenamefont {Wade}\ \emph {et~al.}(2014)\citenamefont {Wade}, \citenamefont {Creighton}, \citenamefont {Ochsner}, \citenamefont {Lackey}, \citenamefont {Farr}, \citenamefont {Littenberg},\ and\ \citenamefont {Raymond}}]{Wade:2014vqa}%
  \BibitemOpen
  \bibfield  {author} {\bibinfo {author} {\bibfnamefont {L.}~\bibnamefont {Wade}}, \bibinfo {author} {\bibfnamefont {J.~D.~E.}\ \bibnamefont {Creighton}}, \bibinfo {author} {\bibfnamefont {E.}~\bibnamefont {Ochsner}}, \bibinfo {author} {\bibfnamefont {B.~D.}\ \bibnamefont {Lackey}}, \bibinfo {author} {\bibfnamefont {B.~F.}\ \bibnamefont {Farr}}, \bibinfo {author} {\bibfnamefont {T.~B.}\ \bibnamefont {Littenberg}},\ and\ \bibinfo {author} {\bibfnamefont {V.}~\bibnamefont {Raymond}},\ }\bibfield  {title} {\bibinfo {title} {{Systematic and statistical errors in a bayesian approach to the estimation of the neutron-star equation of state using advanced gravitational wave detectors}},\ }\href {https://doi.org/10.1103/PhysRevD.89.103012} {\bibfield  {journal} {\bibinfo  {journal} {Phys. Rev. D}\ }\textbf {\bibinfo {volume} {89}},\ \bibinfo {pages} {103012} (\bibinfo {year} {2014})},\ \Eprint {https://arxiv.org/abs/1402.5156} {arXiv:1402.5156 [gr-qc]} \BibitemShut {NoStop}%
\bibitem [{\citenamefont {Gao}\ \emph {et~al.}(2016)\citenamefont {Gao}, \citenamefont {Zhang},\ and\ \citenamefont {L\"u}}]{Gao:2015xle}%
  \BibitemOpen
  \bibfield  {author} {\bibinfo {author} {\bibfnamefont {H.}~\bibnamefont {Gao}}, \bibinfo {author} {\bibfnamefont {B.}~\bibnamefont {Zhang}},\ and\ \bibinfo {author} {\bibfnamefont {H.-J.}\ \bibnamefont {L\"u}},\ }\bibfield  {title} {\bibinfo {title} {{Constraints on binary neutron star merger product from short GRB observations}},\ }\href {https://doi.org/10.1103/PhysRevD.93.044065} {\bibfield  {journal} {\bibinfo  {journal} {Phys. Rev. D}\ }\textbf {\bibinfo {volume} {93}},\ \bibinfo {pages} {044065} (\bibinfo {year} {2016})},\ \Eprint {https://arxiv.org/abs/1511.00753} {arXiv:1511.00753 [astro-ph.HE]} \BibitemShut {NoStop}%
\bibitem [{\citenamefont {Kawaguchi}\ \emph {et~al.}(2018)\citenamefont {Kawaguchi}, \citenamefont {Kiuchi}, \citenamefont {Kyutoku}, \citenamefont {Sekiguchi}, \citenamefont {Shibata},\ and\ \citenamefont {Taniguchi}}]{Kawaguchi:2018gvj}%
  \BibitemOpen
  \bibfield  {author} {\bibinfo {author} {\bibfnamefont {K.}~\bibnamefont {Kawaguchi}}, \bibinfo {author} {\bibfnamefont {K.}~\bibnamefont {Kiuchi}}, \bibinfo {author} {\bibfnamefont {K.}~\bibnamefont {Kyutoku}}, \bibinfo {author} {\bibfnamefont {Y.}~\bibnamefont {Sekiguchi}}, \bibinfo {author} {\bibfnamefont {M.}~\bibnamefont {Shibata}},\ and\ \bibinfo {author} {\bibfnamefont {K.}~\bibnamefont {Taniguchi}},\ }\bibfield  {title} {\bibinfo {title} {{Frequency-domain gravitational waveform models for inspiraling binary neutron stars}},\ }\href {https://doi.org/10.1103/PhysRevD.97.044044} {\bibfield  {journal} {\bibinfo  {journal} {Phys. Rev. D}\ }\textbf {\bibinfo {volume} {97}},\ \bibinfo {pages} {044044} (\bibinfo {year} {2018})},\ \Eprint {https://arxiv.org/abs/1802.06518} {arXiv:1802.06518 [gr-qc]} \BibitemShut {NoStop}%
\bibitem [{\citenamefont {Garfinkle}\ and\ \citenamefont {Jacobson}(2011)}]{Garfinkle:2011iw}%
  \BibitemOpen
  \bibfield  {author} {\bibinfo {author} {\bibfnamefont {D.}~\bibnamefont {Garfinkle}}\ and\ \bibinfo {author} {\bibfnamefont {T.}~\bibnamefont {Jacobson}},\ }\bibfield  {title} {\bibinfo {title} {{A positive energy theorem for Einstein-aether and Ho\v{r}ava gravity}},\ }\href {https://doi.org/10.1103/PhysRevLett.107.191102} {\bibfield  {journal} {\bibinfo  {journal} {Phys. Rev. Lett.}\ }\textbf {\bibinfo {volume} {107}},\ \bibinfo {pages} {191102} (\bibinfo {year} {2011})},\ \Eprint {https://arxiv.org/abs/1108.1835} {arXiv:1108.1835 [gr-qc]} \BibitemShut {NoStop}%
\bibitem [{\citenamefont {Eling}(2006)}]{Eling:2005zq}%
  \BibitemOpen
  \bibfield  {author} {\bibinfo {author} {\bibfnamefont {C.}~\bibnamefont {Eling}},\ }\bibfield  {title} {\bibinfo {title} {{Energy in the Einstein-aether theory}},\ }\href {https://doi.org/10.1103/PhysRevD.80.129905} {\bibfield  {journal} {\bibinfo  {journal} {Phys. Rev. D}\ }\textbf {\bibinfo {volume} {73}},\ \bibinfo {pages} {084026} (\bibinfo {year} {2006})},\ \bibinfo {note} {[Erratum: Phys.Rev.D 80, 129905(E) (2009)]},\ \Eprint {https://arxiv.org/abs/gr-qc/0507059} {arXiv:gr-qc/0507059} \BibitemShut {NoStop}%
\bibitem [{\citenamefont {Sarbach}\ \emph {et~al.}(2019)\citenamefont {Sarbach}, \citenamefont {Barausse},\ and\ \citenamefont {Preciado-L\'opez}}]{Sarbach:2019yso}%
  \BibitemOpen
  \bibfield  {author} {\bibinfo {author} {\bibfnamefont {O.}~\bibnamefont {Sarbach}}, \bibinfo {author} {\bibfnamefont {E.}~\bibnamefont {Barausse}},\ and\ \bibinfo {author} {\bibfnamefont {J.~A.}\ \bibnamefont {Preciado-L\'opez}},\ }\bibfield  {title} {\bibinfo {title} {{Well-posed Cauchy formulation for Einstein-\ae{}ther theory}},\ }\href {https://doi.org/10.1088/1361-6382/ab2e13} {\bibfield  {journal} {\bibinfo  {journal} {Class. Quant. Grav.}\ }\textbf {\bibinfo {volume} {36}},\ \bibinfo {pages} {165007} (\bibinfo {year} {2019})},\ \Eprint {https://arxiv.org/abs/1902.05130} {arXiv:1902.05130 [gr-qc]} \BibitemShut {NoStop}%
\bibitem [{\citenamefont {Mattingly}\ and\ \citenamefont {Jacobson}(2002)}]{Mattingly:2001yd}%
  \BibitemOpen
  \bibfield  {author} {\bibinfo {author} {\bibfnamefont {D.}~\bibnamefont {Mattingly}}\ and\ \bibinfo {author} {\bibfnamefont {T.}~\bibnamefont {Jacobson}},\ }\bibfield  {title} {\bibinfo {title} {{Relativistic gravity with a dynamical preferred frame}},\ }in\ \href {https://doi.org/10.1142/9789812778123_0042} {\emph {\bibinfo {booktitle} {{2nd Meeting on CPT and Lorentz Symmetry}}}}\ (\bibinfo {year} {2002})\ pp.\ \bibinfo {pages} {331--335},\ \Eprint {https://arxiv.org/abs/gr-qc/0112012} {arXiv:gr-qc/0112012} \BibitemShut {NoStop}%
\bibitem [{\citenamefont {Adam}\ \emph {et~al.}(2022)\citenamefont {Adam}, \citenamefont {Figueras}, \citenamefont {Jacobson},\ and\ \citenamefont {Wiseman}}]{Adam:2021vsk}%
  \BibitemOpen
  \bibfield  {author} {\bibinfo {author} {\bibfnamefont {A.}~\bibnamefont {Adam}}, \bibinfo {author} {\bibfnamefont {P.}~\bibnamefont {Figueras}}, \bibinfo {author} {\bibfnamefont {T.}~\bibnamefont {Jacobson}},\ and\ \bibinfo {author} {\bibfnamefont {T.}~\bibnamefont {Wiseman}},\ }\bibfield  {title} {\bibinfo {title} {{Rotating black holes in Einstein-aether theory}},\ }\href {https://doi.org/10.1088/1361-6382/ac5053} {\bibfield  {journal} {\bibinfo  {journal} {Class. Quant. Grav.}\ }\textbf {\bibinfo {volume} {39}},\ \bibinfo {pages} {125001} (\bibinfo {year} {2022})},\ \Eprint {https://arxiv.org/abs/2108.00005} {arXiv:2108.00005 [gr-qc]} \BibitemShut {NoStop}%
\bibitem [{\citenamefont {O'Reilly}\ \emph {et~al.}(2020)\citenamefont {O'Reilly}, \citenamefont {Branchesi}, \citenamefont {Haino},\ and\ \citenamefont {Gemme}}]{ligo_SN_forecast}%
  \BibitemOpen
  \bibfield  {author} {\bibinfo {author} {\bibfnamefont {B.}~\bibnamefont {O'Reilly}}, \bibinfo {author} {\bibfnamefont {M.}~\bibnamefont {Branchesi}}, \bibinfo {author} {\bibfnamefont {S.}~\bibnamefont {Haino}},\ and\ \bibinfo {author} {\bibfnamefont {G.}~\bibnamefont {Gemme}},\ }\href {https://dcc.ligo.org/LIGO-T2000012/public} {\emph {\bibinfo {title} {{LIGO Document T2000012-v1}}}},\ \bibinfo {type} {Tech. Rep.}\ (\bibinfo {year} {2020})\BibitemShut {NoStop}%
\bibitem [{\citenamefont {Gelman}\ and\ \citenamefont {Rubin}(1992)}]{Gelman:1992zz}%
  \BibitemOpen
  \bibfield  {author} {\bibinfo {author} {\bibfnamefont {A.}~\bibnamefont {Gelman}}\ and\ \bibinfo {author} {\bibfnamefont {D.~B.}\ \bibnamefont {Rubin}},\ }\bibfield  {title} {\bibinfo {title} {{Inference from Iterative Simulation Using Multiple Sequences}},\ }\href {https://doi.org/10.1214/ss/1177011136} {\bibfield  {journal} {\bibinfo  {journal} {Statist. Sci.}\ }\textbf {\bibinfo {volume} {7}},\ \bibinfo {pages} {457} (\bibinfo {year} {1992})}\BibitemShut {NoStop}%
\bibitem [{\citenamefont {Gelman}\ \emph {et~al.}(2013)\citenamefont {Gelman}, \citenamefont {Carlin}, \citenamefont {Stern}, \citenamefont {Dunson}, \citenamefont {Vehtari},\ and\ \citenamefont {Rubin}}]{Gelman2013}%
  \BibitemOpen
  \bibfield  {author} {\bibinfo {author} {\bibfnamefont {A.}~\bibnamefont {Gelman}}, \bibinfo {author} {\bibfnamefont {J.~B.}\ \bibnamefont {Carlin}}, \bibinfo {author} {\bibfnamefont {H.~S.}\ \bibnamefont {Stern}}, \bibinfo {author} {\bibfnamefont {D.~B.}\ \bibnamefont {Dunson}}, \bibinfo {author} {\bibfnamefont {A.}~\bibnamefont {Vehtari}},\ and\ \bibinfo {author} {\bibfnamefont {D.~B.}\ \bibnamefont {Rubin}},\ }\href {https://doi.org/10.1201/b16018} {\emph {\bibinfo {title} {Bayesian Data Analysis}}}\ (\bibinfo  {publisher} {Chapman and Hall/{CRC}, New York},\ \bibinfo {year} {2013})\BibitemShut {NoStop}%
\bibitem [{\citenamefont {Abbott}\ \emph {et~al.}(2021{\natexlab{c}})\citenamefont {Abbott} \emph {et~al.}}]{LIGOScientific:2019lzm}%
  \BibitemOpen
  \bibfield  {author} {\bibinfo {author} {\bibfnamefont {R.}~\bibnamefont {Abbott}} \emph {et~al.} (\bibinfo {collaboration} {LIGO Scientific, Virgo}),\ }\bibfield  {title} {\bibinfo {title} {{Open data from the first and second observing runs of Advanced LIGO and Advanced Virgo}},\ }\href {https://doi.org/10.1016/j.softx.2021.100658} {\bibfield  {journal} {\bibinfo  {journal} {SoftwareX}\ }\textbf {\bibinfo {volume} {13}},\ \bibinfo {pages} {100658} (\bibinfo {year} {2021}{\natexlab{c}})},\ \Eprint {https://arxiv.org/abs/1912.11716} {arXiv:1912.11716 [gr-qc]} \BibitemShut {NoStop}%
\bibitem [{\citenamefont {Abbott}\ \emph {et~al.}(2020{\natexlab{b}})\citenamefont {Abbott} \emph {et~al.}}]{Abbott2020}%
  \BibitemOpen
  \bibfield  {author} {\bibinfo {author} {\bibfnamefont {B.~P.}\ \bibnamefont {Abbott}} \emph {et~al.},\ }\bibfield  {title} {\bibinfo {title} {Prospects for observing and localizing gravitational-wave transients with advanced {LIGO}, advanced virgo and {KAGRA}},\ }\bibfield  {journal} {\bibinfo  {journal} {Living Reviews in Relativity}\ }\textbf {\bibinfo {volume} {23}},\ \href {https://doi.org/10.1007/s41114-020-00026-9} {10.1007/s41114-020-00026-9} (\bibinfo {year} {2020}{\natexlab{b}})\BibitemShut {NoStop}%
\bibitem [{\citenamefont {McKechan}\ \emph {et~al.}(2010)\citenamefont {McKechan}, \citenamefont {Robinson},\ and\ \citenamefont {Sathyaprakash}}]{McKechan:2010kp}%
  \BibitemOpen
  \bibfield  {author} {\bibinfo {author} {\bibfnamefont {D.~J.~A.}\ \bibnamefont {McKechan}}, \bibinfo {author} {\bibfnamefont {C.}~\bibnamefont {Robinson}},\ and\ \bibinfo {author} {\bibfnamefont {B.~S.}\ \bibnamefont {Sathyaprakash}},\ }\bibfield  {title} {\bibinfo {title} {{A tapering window for time-domain templates and simulated signals in the detection of gravitational waves from coalescing compact binaries}},\ }\href {https://doi.org/10.1088/0264-9381/27/8/084020} {\bibfield  {journal} {\bibinfo  {journal} {Class. Quant. Grav.}\ }\textbf {\bibinfo {volume} {27}},\ \bibinfo {pages} {084020} (\bibinfo {year} {2010})},\ \Eprint {https://arxiv.org/abs/1003.2939} {arXiv:1003.2939 [gr-qc]} \BibitemShut {NoStop}%
\bibitem [{\citenamefont {Boh\'e}\ \emph {et~al.}(2015)\citenamefont {Boh\'e}, \citenamefont {Faye}, \citenamefont {Marsat},\ and\ \citenamefont {Porter}}]{Bohe:2015ana}%
  \BibitemOpen
  \bibfield  {author} {\bibinfo {author} {\bibfnamefont {A.}~\bibnamefont {Boh\'e}}, \bibinfo {author} {\bibfnamefont {G.}~\bibnamefont {Faye}}, \bibinfo {author} {\bibfnamefont {S.}~\bibnamefont {Marsat}},\ and\ \bibinfo {author} {\bibfnamefont {E.~K.}\ \bibnamefont {Porter}},\ }\bibfield  {title} {\bibinfo {title} {{Quadratic-in-spin effects in the orbital dynamics and gravitational-wave energy flux of compact binaries at the 3PN order}},\ }\href {https://doi.org/10.1088/0264-9381/32/19/195010} {\bibfield  {journal} {\bibinfo  {journal} {Class. Quant. Grav.}\ }\textbf {\bibinfo {volume} {32}},\ \bibinfo {pages} {195010} (\bibinfo {year} {2015})},\ \Eprint {https://arxiv.org/abs/1501.01529} {arXiv:1501.01529 [gr-qc]} \BibitemShut {NoStop}%
\bibitem [{\citenamefont {Mishra}\ \emph {et~al.}(2016)\citenamefont {Mishra}, \citenamefont {Kela}, \citenamefont {Arun},\ and\ \citenamefont {Faye}}]{Mishra:2016whh}%
  \BibitemOpen
  \bibfield  {author} {\bibinfo {author} {\bibfnamefont {C.~K.}\ \bibnamefont {Mishra}}, \bibinfo {author} {\bibfnamefont {A.}~\bibnamefont {Kela}}, \bibinfo {author} {\bibfnamefont {K.~G.}\ \bibnamefont {Arun}},\ and\ \bibinfo {author} {\bibfnamefont {G.}~\bibnamefont {Faye}},\ }\bibfield  {title} {\bibinfo {title} {{Ready-to-use post-Newtonian gravitational waveforms for binary black holes with nonprecessing spins: An update}},\ }\href {https://doi.org/10.1103/PhysRevD.93.084054} {\bibfield  {journal} {\bibinfo  {journal} {Phys. Rev. D}\ }\textbf {\bibinfo {volume} {93}},\ \bibinfo {pages} {084054} (\bibinfo {year} {2016})},\ \Eprint {https://arxiv.org/abs/1601.05588} {arXiv:1601.05588 [gr-qc]} \BibitemShut {NoStop}%
\bibitem [{\citenamefont {Krishnendu}\ \emph {et~al.}(2017)\citenamefont {Krishnendu}, \citenamefont {Arun},\ and\ \citenamefont {Mishra}}]{Krishnendu:2017shb}%
  \BibitemOpen
  \bibfield  {author} {\bibinfo {author} {\bibfnamefont {N.~V.}\ \bibnamefont {Krishnendu}}, \bibinfo {author} {\bibfnamefont {K.~G.}\ \bibnamefont {Arun}},\ and\ \bibinfo {author} {\bibfnamefont {C.~K.}\ \bibnamefont {Mishra}},\ }\bibfield  {title} {\bibinfo {title} {{Testing the binary black hole nature of a compact binary coalescence}},\ }\href {https://doi.org/10.1103/PhysRevLett.119.091101} {\bibfield  {journal} {\bibinfo  {journal} {Phys. Rev. Lett.}\ }\textbf {\bibinfo {volume} {119}},\ \bibinfo {pages} {091101} (\bibinfo {year} {2017})},\ \Eprint {https://arxiv.org/abs/1701.06318} {arXiv:1701.06318 [gr-qc]} \BibitemShut {NoStop}%
\bibitem [{\citenamefont {Yagi}\ and\ \citenamefont {Yunes}(2017{\natexlab{b}})}]{Yagi:2016bkt}%
  \BibitemOpen
  \bibfield  {author} {\bibinfo {author} {\bibfnamefont {K.}~\bibnamefont {Yagi}}\ and\ \bibinfo {author} {\bibfnamefont {N.}~\bibnamefont {Yunes}},\ }\bibfield  {title} {\bibinfo {title} {{Approximate Universal Relations for Neutron Stars and Quark Stars}},\ }\href {https://doi.org/10.1016/j.physrep.2017.03.002} {\bibfield  {journal} {\bibinfo  {journal} {Phys. Rept.}\ }\textbf {\bibinfo {volume} {681}},\ \bibinfo {pages} {1} (\bibinfo {year} {2017}{\natexlab{b}})},\ \Eprint {https://arxiv.org/abs/1608.02582} {arXiv:1608.02582 [gr-qc]} \BibitemShut {NoStop}%
\end{thebibliography}%
\end{document}